\documentclass[%
 reprint,
 amsmath,amssymb,
 aps,
]{revtex4-2}

\usepackage{graphicx}
\usepackage{dcolumn}
\usepackage{bm}
\usepackage{hyperref}
\usepackage{xcolor}
\usepackage[mathlines]{lineno}

\usepackage[normalem]{ulem}

\usepackage{amsmath,amssymb}

\usepackage{comment}

\renewcommand{\arraystretch}{1.5}

\begin{document}

\preprint{APS/123-QED}

\title{A fast algorithm for 2D Rigidity Percolation}



\author{Nina Javerzat}
\thanks{All authors contributed equally; \\
\href{nina.javerzat@univ-grenoble-alpes.fr}{nina.javerzat@univ-grenoble-alpes.fr};
\href{daniele.notarmuzi@tuwien.ac.at}{daniele.notarmuzi@tuwien.ac.at};}
\affiliation{%
 Université Grenoble Alpes, CNRS, LIPhy, 38000 Grenoble, France
}%

\author{Daniele Notarmuzi\textsuperscript{*}}
\affiliation{Institut für Theoretische Physik, TU Wien, Wiedner Hauptstraße 8-10, A-1040 Wien, Austria}%







\date{\today}

\begin{abstract}
Rigidity Percolation is a crucial framework for describing rigidity transitions in amorphous systems.
We present a new, efficient algorithm to study central-force Rigidity Percolation in two dimensions. This algorithm combines the Pebble Game algorithm, the Newman-Ziff approach to Connectivity Percolation, as well as novel rigorous results in rigidity theory, to exactly identify rigid clusters over the full bond concentration range, in a time that scales as $N^{1.02}$ for a system of $N$ nodes. We perform extensive numerical simulations with systems larger than $500$ million nodes, far beyond the previous limitations. We obtain new, precise estimates for the critical exponents, $\nu=1.1694(8)$ and $D_f=1.8423(7)$, and locate the critical threshold at $p_c = 0.6602741(4)$. 
Besides opening the way to further accurate numerical studies of Rigidity Percolation, our work provides new rigorous theoretical insights on specific cluster merging mechanisms that distinguish it from the standard Connectivity Percolation problem.  
\end{abstract}

\maketitle


\section{\label{sec:intro} Introduction}

Percolation is one of the most versatile conceptual frameworks in statistical physics and describes a wealth of natural phenomena, from polymer gelation to the propagation of diseases. In standard --Connectivity--  Percolation (CP) 
the central objects are connectivity clusters, namely sets of mutually connected nodes. As more and more bonds are added to the network, clusters grow and coalesce, and a system-spanning (macroscopic) cluster eventually emerges at a critical value of the bond concentration. 

While CP theory provides an extensive framework to characterize the connectivity properties of networks \cite{stauffer_aharony, duminil_copin2018}, Rigidity Percolation (RP) addresses instead their \emph{mechanical stability}: is the network able to transmit stresses and sustain external loads? It was realized early that this question can be formulated  as a constraint-counting problem \cite{Maxwell1864, Laman1970}; the rigidity of a network depends only on its topology and is largely independent of the precise physical nature of bonds. In central-force networks, on which we focus in this work, nodes possess $d$ degrees of freedom in $d$ dimensions, while bonds introduce constraints by fixing the distances between pairs of nodes.
Rigid clusters are sets of mutually rigid bonds, that is, subnetworks in which all degrees of freedom are fixed by the constraints except for the $d(d+1)/2$ global ones. Floppy clusters instead, have unfixed degrees of freedom that allow for deformation along soft (floppy) modes. While at low bond concentration the system is typically macroscopically floppy (i.e. composed of small, non-rigidly connected rigid clusters), increasing the concentration above a critical threshold leads to the emergence of a macroscopic rigid cluster (see e.g. \cite{JacobsThorpe1995}). Rigidity Percolation has attracted much attention in Soft Matter, as a simple and versatile framework to model liquid to amorphous solid transitions in systems such as  colloidal gels~\cite{mehdi2019,Tsurusawa2019Direct,Fenton2023Minimal}, living tissues~\cite{Petridou2021,Hannezo2022Rigidity,Lenne2022Biological,Manning2024}, fiber networks~\cite{Broedersz2011,Bolton1990Rigidity} or granular packings~\cite{silke2016, Liu2019,Ellenbroek2015Rigidity,Berthier2019Rigidity}. Notably, it has recently been revealed how the RP critical exponents control the viscoelastic properties of colloidal gels in their solid phase~\cite{Bantawa2023, richard2025}. Moreover, recent works gather evidence for a larger role of rigid clusters in the physics of amorphous systems, as controlling e.g. the behaviour at the brittle-ductile transition~\cite{silke2019} or the large stress fluctuations at the shear thickening transition~\cite{emanuelaDST}. 
\\\\
Inquiring about mutual rigidity, rather than mutual connectivity, makes RP much more difficult than the usual CP problem. 
A distinguishing feature of RP is its non-locality: as we make precise in section \ref{sec:theorems}, the activation of a single bond can trigger the rigidification of
arbitrarily large spatial regions, by coalescence of numerous rigid clusters into a single one.
Likewise, cutting a single bond can lead to cascades of rigidity loss \cite{Moukarzel1999}. As a matter of fact, RP remains largely less understood than CP. Rigorous results are rather scarce: the existence of a critical bond concentration strictly above the CP threshold has been proven on the 2D triangular lattice \cite{holroyd98}; the uniqueness of the infinite rigid cluster has been proven in Ref.~\cite{Haggstrom2003}.
Exact results have been obtained on non-generic networks such as hierarchical and Erdös-Rényi graphs \cite{barre2009, jordan2022} (see also \cite{Barre2005, Barre2006} for connections with satisfiability problems).  
In most cases however, especially those relevant to Soft Matter applications, the characterization of the RP universality classes relies essentially on numerical simulations, either to determine the values of critical exponents \cite{JacobsThorpe1995,Liu2019}, or to explore fine aspects of the universality class such as conformal invariance \cite{ninaconf,ninasle}.
\\\\
Yet, identifying efficiently rigid clusters in large systems is not an easy task. 
Numerical studies were made possible thanks to the Pebble Game algorithm 
\cite{Jacobs1997An,JacobsThorpe1995} (later generalized to frictional systems \cite{silke2016,LESTER2018225}), a combinatorial implementation of Laman theorem for graph rigidity, that we briefly describe in section~\ref{sec:PebGame}. 
Using the Pebble Game algorithm, the time to identify rigid clusters at $p_c^{RP}$ scales with the system size $N$ as $N^{1.15}$~\cite{Jacobs1997An}. This allowed to
estimate the values of the
critical exponents for central-force RP, 
$\nu^{\rm RP}=1.21\pm0.06,\;\beta^{\rm RP}=0.175\pm0.02,\; D_f^{\rm RP}=1.86\pm0.02$, respectively the 
correlation length and order parameter exponents, and the fractal dimension of the spanning rigid cluster,
with a sufficient precision to claim the existence of a new universality class, distinct from random 
CP~\cite{JacobsThorpe1995}. To our knowledge, there has been no attempt to determine the central-force RP exponents with better 
precision since then, the barrier being the harsh computational cost 
of analyzing systems larger than $\sim10^6$ nodes (for comparison, accurate studies of the CP critical point involve systems with $O(10^8)$ nodes \cite{Ziff_2011}). 
\\\\
Classifying rigidity transitions as belonging or not to the (uncorrelated~\footnote{In our, as well as earlier work \cite{JacobsThorpe1995}, bonds are activated in a random, uncorrelated way. Recent work has examined the effect of local correlations \cite{mehdi2019} but to our knowledge the effect of long-range correlations in the bond activation on the RP universality class has not been addressed. Our algorithm can be generalized to incorporate such correlations and we plan this for future work.}) central-force RP universality class is crucial to achieve a better understanding of the variety of liquid-to-amorphous solid transitions~\cite{mehdi2019,sciadv2023,Petridou2021,Lois2008,Liu2019,latvakokko1,latvakokko2}, and to assess the robustness of the central-force RP universality class. Providing accurate numerical standards for the critical exponents, and, more generally, understanding the specificities of the RP problem as compared to the well-studied CP one, are therefore major goals. 
\\\\
In this paper we introduce a new algorithm for RP. It is based on the powerful Newman-Ziff (NZ) approach to CP\cite{Newman2000Efficient}, where bonds are activated one at a time, and the state of the system, namely the identification of clusters,
is fully known at each step.
Studying percolation à la Newman-Ziff allows notably to measure the quantities of interest (probability to percolate, size of the largest component) at each bond activation, namely for 
each value of the bond filling fraction (fraction of active bonds),
 thereby covering the complete phase diagram in a single run. 
To obtain a NZ-like algorithm for RP, we first shed new light on the specific cluster merging mechanisms at play, by providing rigorous results that predict the consequences of each new bond activation. Exploiting these results we construct an algorithm that scales with the system size as $N^{1.02}$, i.e. that covers the entire phase diagram in almost linear time.
Our algorithm pushes the size limitation to $N \gtrsim 10^8$, allowing for precise studies of the RP universality class, starting with accurate determination of the main critical exponents \cite{DN_prep}.

The paper is organized as follows. In section \ref{sec:PebGame} we briefly describe the standard Pebble Game as defined by Jacobs and Hendrickson \cite{Jacobs1997An, JacobsThorpe1995}, and explain in more details how to perform pebble searches, that are key to the implementation of our own algorithm. In section \ref{sec:NZ} we recall the Newman-Ziff algorithm for CP. In section \ref{sec:theorems} we present the three theorems that allow us to design a NZ-like algorithm for RP, namely that allow, with minimal computational cost, to deduce the new state of the system after each bond activation. The proofs of those theorems, as well as a brief review of 2D rigidity theory and additional new results, are presented in the Supplemental Material (SM), section 1.
We give an overview of the strategy behind our algorithm in section \ref{sec:overview}, while the implementation is detailed in section \ref{sec:implementation}. In section \ref{sec:wrapping} we adapt the (CP) Machta algorithm \cite{Machta1996} to detect the wrapping of rigid clusters.
Finally we discuss the performance of our complete algorithm in section \ref{sec:results}. We conclude with several perspectives in section \ref{sec:conclusion}.

\section{The Jacobs and Hendrickson approach}
\label{sec:PebGame}

The central challenge in studying RP is the identification of rigid clusters, namely of the maximal sets of mutually rigid bonds.
In 2D, an exact criterion is given by Laman theorem on graph rigidity \cite{Laman1970} (cf. SM section 1 for more details).
Jacobs and Hendrickson (JH)~\cite{Jacobs1997An} introduced a clever algorithm to implement efficiently Laman’s criterion, the Pebble Game: the degrees of freedom of the system are represented as ``pebbles'', assigned to the nodes. Pebbles are removed 
each time an independent bond (constraint) is added to the system. Note that, in this formulation, three pebbles can always be gathered across any bond, since the system always has three global degrees of freedom in 2D. Additional pebbles correspond to floppy modes.

\subsection{The Pebble Game} 
\label{sec:JH_algo}

The JH algorithm is divided into two stages, filling and clustering. All bonds are initially inactive and each node has two pebbles attached to it, representing its two degrees of freedom. In the filling stage, bonds are independently activated with probability $p$. Each new activated bond is classified as an independent or redundant constraint using \emph{pebble searches} (see next section): a new bond $uv$ represents an independent constraint if $u$ and $v$ were not mutually rigid, namely if four pebbles can
be gathered across $uv$.
If the two nodes were already mutually rigid, only three pebbles can be gathered, and the new bond is redundant. 

Once all the bonds are activated and classified, rigid clusters are identified. 
Each rigid cluster is constructed by starting from a given unlabeled bond $b$, and testing the mutual rigidity of the other (unlabeled) bonds of the system with $b$, by performing pebble searches. Each bond that is found to be mutually rigid with respect to $b$, is given the same label as $b$. This is conveniently implemented as a Breadth First Search that stops when only floppy, or already labeled bonds, can be found.
\\\\
By construction, one run of the JH algorithm corresponds to a single value of the bond 
concentration $p$. The overall cost of the JH algorithm is determined by the number of pebble 
searches to be performed in a run  –which depends on $N$– as well as the cost of each search –which also depends on $N$. JH reported a scaling of the algorithm that depends on $p$, 
scaling as $N^{1.15}$ near the critical point $p \approx p_c$, and
linearly near full concentration.  In practice this means that studying the phase diagram with 
the JH algorithm requires to run it for (as many as possible) different values of $p$, with a 
total computing time that depends on the number of values considered, most of which being typically concentrated near the critical point with cost $N^{1.15}$ each.


Our algorithm solves these two limitations at once. First, the whole phase diagram is obtained in a single run.
This comes from adopting a Newman-Ziff approach, which we describe in the next section.
Secondly, we make such approach efficient by exploiting rigidity theory to make a strategic use of pebbles searches, keeping their number low. 

\subsection{Pebble searches}\label{sec:pebble_searches}

\begin{figure*}[!ht]
\begin{center}
\includegraphics[width=\textwidth]{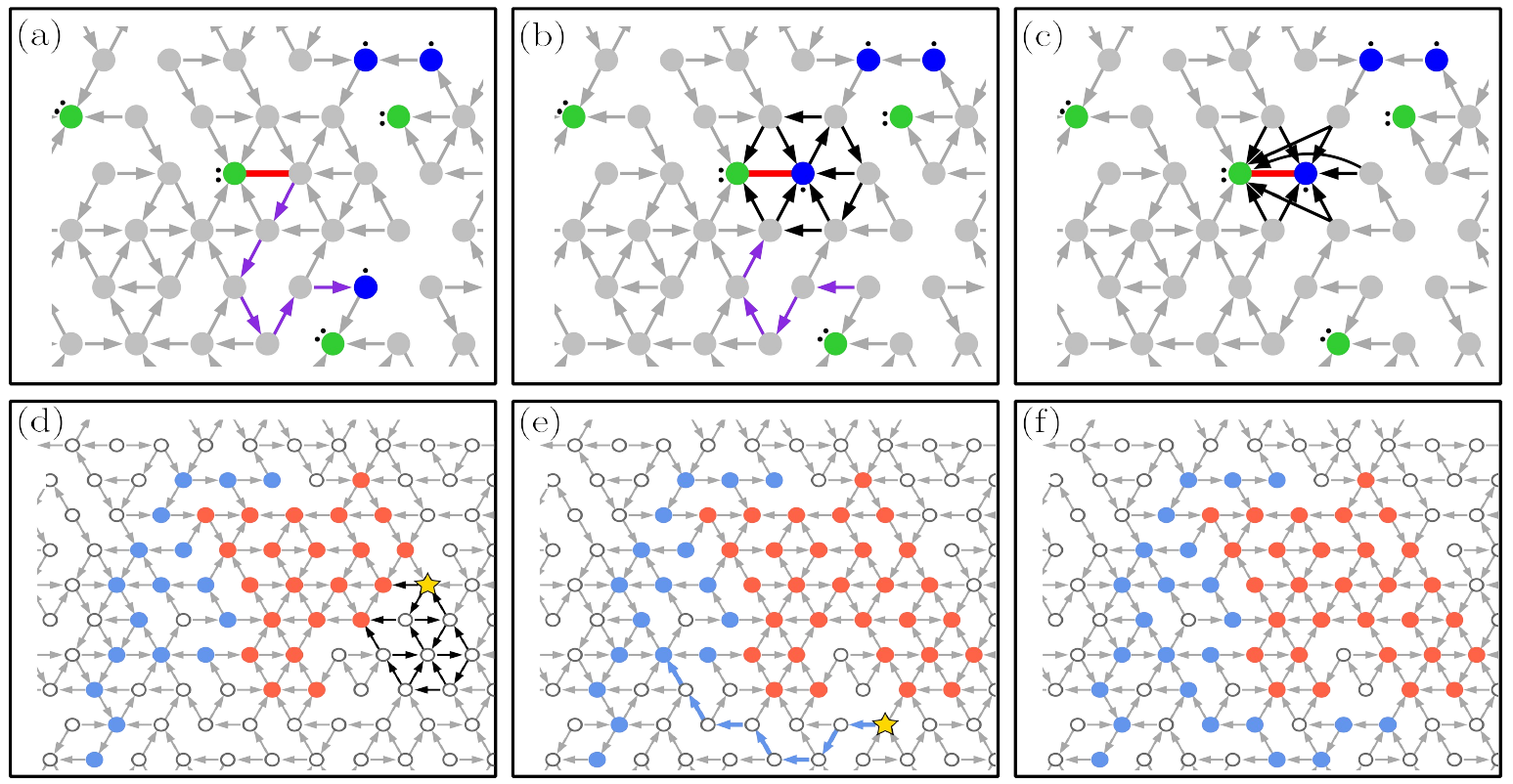}    
\end{center}
\caption{
Dynamics of pebble searches on the pebble graph.  
Top: searches of type I. Bottom: searches of type II. To ease visualization, bonds are not triangulated in the bottom panels.
(a) Green nodes have two pebbles, blue nodes have one, grey have zero. Pebbles are depicted as black
dots.
The red undirected bond is the
newly activated bond. A pebble search of type I starts at the grey end node of the new bond. The purple path
leads to a pebble.
(b) The path identified in (a) is reversed, the node where it started has now one pebble, and a second search is started. The search traverses the black edges without finding a pebble. 
(c) The edges traversed during the previous search are triangulated over the base made by the 
new bond.
(d) Red (blue) nodes have been marked rigid (floppy) during previous searches, while open circles represent unmarked nodes. A pebble search of type II starts at
the yellow star. The search traverses the black edges without finding a floppy node. 
Note that edges originating from rigid nodes are not walked.
(e) The nodes visited during the search of panel (d) are marked rigid. 
The visited edges are triangulated (not shown). A new search starts at the yellow star and, following the blue path, stops when
it hits a floppy node.
(f) The nodes traversed by the path identified in (e) are marked floppy.
}
\label{fig:PebSeraches}
\end{figure*}

Pebbles searches are used to check for mutual rigidity. These searches take place on the pebble graph, distinct from the physical lattice. Everytime an independent bond of the system is activated, a directed edge is added to the pebble graph and a pebble is removed from one of the end nodes, accounting from the removal of one degree of freedom. The edge is initially directed away from the node that loses a pebble.
For later use, we distinguish two types of pebble searches, whose dynamics is illustrated in Fig.~\ref{fig:PebSeraches}.

\textit{Type I searches --}
The purpose of pebble searches of type
I is to actually move the pebbles across the pebble graph, in order to keep track of the repartition of degrees of freedom. To gather a pebble at node $j$, a Breadth First Search (BFS)~\footnote{Depth First Search can be used as well. JH used BFS and we do the same in our work.} starting at $j$ is performed over the pebble graph, with the aim of finding
a path that leads from node $j$ to a node that has a pebble (Fig.~\ref{fig:PebSeraches}a). 
If a pebble is found, the direction of the pebble edges along the path is reversed and the
pebble is moved from the node where it has been found to node $j$ 
(Fig.~\ref{fig:PebSeraches}b). 

\textit{Type II searches -- } The purpose of pebble searches of type II is to test mutual 
rigidity without actually moving the pebbles across the pebble graph.
To check mutual rigidity of bonds $b_1$ and $b_2$, three pebbles (that are always available) are first gathered at $b_1$ using type I searches. These pebbles are temporarily frozen at their
location, i.e., they are made unavailable to the rest of the system, while the nodes of $b_1$ are marked as rigid. 
Then, a search is performed for a fourth pebble that could be gathered at the nodes of $b_2$:
if such pebble exists, bonds $b_1$ and $b_2$ are mutually floppy, otherwise they are mutually rigid. 
Note that, for $b_2$ to be rigid with respect to $b_1$, both its nodes must be rigid. If one node is
rigid and the other is floppy, the bond is overall floppy. This implies that, in contrast to CP, 
nodes in RP can belong to more than a rigid cluster. Each bond, however, belongs to a single rigid 
cluster. Nodes belonging to more than a rigid cluster are called pivots~\cite{Jacobs1997An}. 

The search proceeds as in the previous section, but the direction of pebble edges is never changed by type II searches, as pebbles are not actually moved. Moreover, if the search is successful (a pebble turns out to be available), 
all nodes along the path are marked as floppy (Fig.~\ref{fig:PebSeraches}e). Indeed, the found pebble could be gathered at any of them. On the contrary, after a failed search, all the visited nodes are marked rigid (Fig.~\ref{fig:PebSeraches}f): no pebble can be gathered at any of those visited nodes.

This marking of the nodes reduces the cost of subsequent searches, that can be terminated whenever a node, already marked as floppy during a previous search, is encountered. 
Moreover, edges of the pebble graph that start at nodes marked rigid during previous searches are not walked,
as they cannot lead to any pebble nor to a floppy node~\cite{Jacobs1997An}. 

\textit{Triangulation -- } 
Note that both types of searches can, in principle, result in a pebble not being found. 
When a pebble search fails, 
all the pebble edges traversed during the search are triangulated
over the bond that has three pebbles (Fig.~\ref{fig:PebSeraches}c). The triangulation
procedure is essential to keep pebble searches short~\cite{Jacobs1997An}.  


\section{The Newman-Ziff algorithm for CP}\label{sec:NZ}

In this section we recall briefly the NZ approach to CP \cite{NZ2001,Newman2000Efficient}. 
NZ simulated  percolation as a dynamical process, 
where bonds are activated one by one, and the state of the system 
is updated at each step, following simple 
rules. The algorithm allows therefore to generate instances of the system at every filling 
fraction,
 and does so in linear time. Measurements are then convoluted with the binomial
distribution to be transformed in functions of the bond concentration $p$~\cite{Newman2000Efficient}. The key idea in the NZ algorithm for CP, which we 
adapt to RP, is to describe each cluster as a directed 
tree: all nodes belonging to a 
 given connectivity cluster point to the same root in the corresponding tree. Each cluster is identified by its root, whose entry in the tree gives (minus) the size of the cluster. 

When a new bond $uv$ 
is activated, trees 
(clusters) must be updated accordingly. These updates are simple in CP, as only two events can happen. If $u$ and $v$ already belong to the same cluster (they point to the same root), the corresponding tree is unchanged. 
If $u$ and $v$ belong to two distinct clusters,
$\mathcal{C}_u$ and  $\mathcal{C}_v$  (they point to different roots), 
the new bond coalesces them into 
$\mathcal{C}_u \cup \mathcal{C}_v \cup\, \{uv\}$. Coalescence is performed by 
redirecting the root of the smallest cluster toward the root of the largest one.

The roots of $u$ and $v$ are identified by the so-called find operation, i.e.,
by traversing the tree from $u$ ($v$) to the corresponding root. 
The find operation is coupled with the so-called path compression: once
the root is found, the tree is compressed by redirecting all the traversed edges 
toward the identified root. This mechanism allows the find operation to have constant cost
even on large trees~\cite{Newman2000Efficient}. 


\section{Theorems for single bond activation}\label{sec:theorems}

In this paper we adapt the NZ strategy described in the previous section, to rigidity percolation. 
Namely, starting with an empty lattice and progressively activating the bonds, we are able to deduce the new state of the system after each new bond activation.
While this is relatively simple in CP, due to simple cluster events (cf. section \ref{sec:NZ}), the situation is more involved in RP, where the consequence of single bond activation is a priori not clear: is the new bond independent or redundant? Are new rigid clusters created or existing ones coalesced? If so, which ones and how?
In this section we provide theoretical results in 2D rigidity theory that allow us to identify the only three possible cluster events resulting from single bond activation, that we name Pivoting, Overconstraining, and Rigidification. The following results allow us to furthermore determine, without making use of pebble searches, (i) if an added bond is independent or redundant, (ii) what type of cluster event is triggered by a given bond activation. In addition, defining the pivotal class of a node as the number of distinct rigid clusters the node belongs to~\footnote{Inactive nodes have $\pi=0$
by definition.}, we also determine (iii) how the pivotal classes of nodes change. 
In section \ref{sec:overview} we explain how to combine these results with optimal pebble searches to obtain our final algorithm.
We start by giving basic principles of 2D rigidity, and refer to section 1 of the SM for further details.\\\\
\textit{Proposition 1 (Transitivity) -- } Denote $\mathcal{R}_1$, $\mathcal{R}_2$, $\mathcal{R}_3$ three rigid bodies.
\begin{enumerate}
    \item If $\mathcal{R}_1$ is rigid with respect to $\mathcal{R}_2$  and $\mathcal{R}_2$ is rigid with respect to $\mathcal{R}_3$, then $\mathcal{R}_1$ is rigid with respect to $\mathcal{R}_3$.
    \item If $\mathcal{R}_1$ is rigid with respect to $\mathcal{R}_2$  and $\mathcal{R}_2$ is floppy with respect to $\mathcal{R}_3$, then $\mathcal{R}_1$ is floppy with respect to $\mathcal{R}_3$.
\end{enumerate}
From now on, $\mathcal{R}$ denotes a rigid cluster, $\mathcal{C}$ a connectivity cluster and $\mathcal{C}_u$ is the connectivity cluster that contains node $u$.
\\\\
\noindent\textit{Proposition 2 (``Two-pivots rule'')~\cite{combinatorial_rigidity} --} In two dimensions, if two rigid clusters $\mathcal{R}_1$ and $\mathcal{R}_2$ share at least two pivots, then $\mathcal{R}_1\cup\mathcal{R}_2$ is a rigid cluster.
\\\\
In the following, $uv$ denotes a new activated bond. The proofs of the following statements are given in section 1B of the SM. Additional results on rigidity, which to our knowledge had not been proven before, are given in section 1A of the SM.\\\\
\noindent 
\textit{Theorem 1 (Pivoting) -- }
Assume $u\in\mathcal{C}_u$ and $v\in\mathcal{C}_v$ with $\mathcal{C}_u \neq \mathcal{C}_v$. Then,
\begin{enumerate}
    \item The bond $uv$ is independent.
    \item The activation of $uv$ creates a new rigid cluster composed of the bond $uv$ only.
    \item The pivotal class of $u$ and $v$ increases by 1. The pivotal class of any other node
     is unchanged.
\end{enumerate}

\noindent \textit{Theorem 2 (Overconstraining) -- } Assume that there exists a rigid cluster $\mathcal{R}$ such that $u,v\in\mathcal{R}$ before the activation of $uv$. Then,
\begin{enumerate}
    \item $uv$ is a redundant bond.
    \item   Upon activation of $uv$, $\mathcal{R}$ becomes $\mathcal{R}\cup \{uv\}$; all other rigid clusters remain identical.
     \item The pivotal classes of all nodes are unchanged.
\end{enumerate}

\noindent\textit{Theorem 3 (Rigidification) -- } Assume $\mathcal{C}_u = \mathcal{C}_v=\mathcal{C}$ and that, before the activation of $uv$, there is no rigid cluster that contains both $u$ and $v$. Then,
\begin{enumerate}
    \item $uv$ is an independent bond.
    \item The activation of $uv$ triggers a rigidification process resulting in the coalescence of $k+1$ rigid clusters $\mathcal{R}_1,\,\mathcal{R}_2,\cdots,\mathcal{R}_k$ and $\mathcal{R}_{k+1}=\{uv\}$ into a single rigid cluster $\mathcal{R}^+=\mathcal{R}_1\cup\mathcal{R}_2\cup\cdots\cup\mathcal{R}_k\cup \{uv\}$. The clusters $\mathcal{R}_1,\,\mathcal{R}_2,\cdots,\mathcal{R}_{k}$ all belong to $\mathcal{C}$, and each of them is connected to at least two other ones via at least two distinct pivots. 
    \item The pivotal class of these pivots decreases by at least one.
\end{enumerate}


\section{Overview of the algorithm}
\label{sec:overview}
In this section we present an algorithm that exploits the three theorems of section \ref{sec:theorems}. How we efficiently implement it in practice is
explained in the next section.

\subsection{Basic settings} The system is initially empty, i.e., there are no active bonds or nodes: each node has pivotal class 0 and carries two pebbles. The pebble graph is empty. 

For the hypotheses of the theorems to
be verifiable, the connectivity state of the network must be known at each new bond activation. 
We therefore implement the NZ algorithm for CP as done in Ref.~\cite{Newman2000Efficient}. 
The NZ algorithm scales linearly with system size, so that, 
with a very small additional time and memory cost, our algorithm produces the entire phase diagram
of the CP transition as well. 

To monitor the rigidity state of the system we adopt the same strategy as 
NZ~\cite{NZ2001} and represent 
rigid clusters as trees. For the sake of clarity, we indicate the trees that describe connectivity and rigid
clusters as $\mathcal{T}_{\rm{C}}$ and $\mathcal{T}_{\rm{R}}$ respectively. 
As nodes might belong to multiple rigid clusters, while bonds cannot, each node in a tree $\mathcal{T}_{\mathcal{R}}$ represents a bond in the corresponding
rigid cluster $\mathcal{R}$. 
Traversing a tree $\mathcal{T}_{\rm{R}}$ leads to a root bond that uniquely identifies the rigid cluster. 
Every time a root is identified by a find operation, the path compression is also performed
in order to keep a constant cost for each of these operations (cf. Section~\ref{sec:NZ}).
\subsection{Stages of the algorithm}
As for the NZ algorithm for CP, our goal is to deduce the new state of the system after each new bond activation, namely, to update the rigid clusters trees from the knowledge of the system state prior to the bond activation.
To this aim, we proceed as follows.

\begin{enumerate}
    \item Select a random bond $uv$ to activate.
    \item Identify $\mathcal{C}_u$ and $\mathcal{C}_v$ by finding their roots 
    \item If $\mathcal{C}_u \neq \mathcal{C}_v$, merge them, perform a pivoting step (cf. Section~\ref{subsec:pivoting}) and go back to step 1.
    \item If instead $\mathcal{C}_u = \mathcal{C}_v$, find all roots of the rigid clusters to which $u$ and $v$ belong~\footnote{Note that
    $\mathcal{C}_u = \mathcal{C}_v$ happens only if both $u$ and $v$ are active, hence
    each of them belongs to at least one rigid cluster.}.
    \item If a rigid cluster $\mathcal{R}$ containing both $u$ and $v$ does not already exists, i.e., 
    if a common root has not been 
    identified by the previous step, perform      
    a rigidification step (cf. Section~\ref{subsec:rigid_idea})  and go back to step 1.
    \item Otherwise, perform an overconstraining step (cf. Section~\ref{subsec:overconstr}) and go back to step 1.
\end{enumerate}
The algorithm ends when all the bonds have been activated, hence covering the entire filling fraction range. 
\\
The occurrence of pivoting, rigidification or overconstraining events typically depends on the bond concentration, as we show in Fig.~\ref{fig:RP_stages}. We now describe steps 3, 5, 6 in more details.

\subsection{Pivoting}
\label{subsec:pivoting}

At low concentration, and in particular below the CP transition, 
the activation of new bonds typically leads to pivoting, as shown in Fig.~\ref{fig:RP_stages}.

A pivoting step requires (i) to increase the pivotal class of both $u$ and $v$, 
(ii) to initialize a new $\mathcal{T}_{\rm{R}}$ tree with size one and root $uv$,
(iii) to update the pebble graph  by covering $uv$ with a pebble. 
Note that, by Theorem 1, the bond is already
known to be independent, and having one pebble to cover it is therefore sufficient.
Hence, if at least one node between $u$
and $v$ has a pebble, we simply use it without searching for it.
If both have zero pebbles, a pebble search of type I is performed.
Note that this search never fails.
As the size of $\mathcal{C}_u$ and $\mathcal{C}_v$ are known, the search is performed 
across the smallest of the two to make it faster. 

A pivoting step has constant cost if the pebble search is not performed,
otherwise the cost is that of the only search performed.  Overall, the cost of a pivoting step is quite low, see SM 2.

\subsection{Rigidification}
\label{subsec:rigid_idea}
Above the CP transition, the probability of having a macroscopic connectivity cluster increases sharply, so that the probability of activating a bond within the same connectivity cluster is large. However, while $p<p_{RP}$, rigid clusters are small (connectivity clusters have numerous floppy modes), leading to a large probability of a rigidification event (see Fig.~\ref{fig:RP_stages}). Namely, the new independent bond will connect a pair of nodes belonging to the same connectivity cluster, but different rigid clusters, and its activation will lead to the coalescence of a variable number of distinct rigid clusters into a single one.
 The construction of the new rigid cluster is at the core of our algorithm.
\\
The most obvious strategy to construct this rigid cluster is that of the JH algorithm, i.e.,
starting with all nodes unmarked except $u$ and $v$ that are marked rigid, use a Breadth First
Search over the neighbours of nodes marked rigid. In this way, the rigid cluster is constructed by 
identifying a surrounding shell of nodes marked floppy.
However, the implementation of this approach
results in an algorithm that we observe to be quadratic (see SM 3 A).
\\\\
The key of our approach is first to realize that
it is sufficient to test mutual rigidity of at most one bond per existing rigid cluster, with respect to the newly activated bond. We conveniently choose these bonds as the roots of the rigid clusters.
Indeed, by transitivity (Proposition 1), if the root of a rigid cluster $\mathcal{R}$ is rigid (floppy) with respect to $uv$, then all the bonds in $\mathcal{R}$ are also rigid (floppy) with respect to $uv$, and do not need to be checked.  
To build the rigid cluster, one could therefore test mutual rigidity between $uv$ and all the roots of the rigid clusters belonging to the connectivity cluster $\mathcal{C}$.
While testing the roots in an arbitrary order does not result in satisfactory scaling (see SM 3 B), the approach becomes highly efficient when the roots are tested by following paths in a \emph{Pivot Network}, as we now explain.
\\\\
To construct the rigid cluster resulting from rigidification, we start a Breadth First Search from $uv$
over roots of rigid clusters that could potentially coalesce into the forming one. To identify them, we introduce the Pivot Network
$\mathcal{P}$, a graph whose nodes represent the rigid clusters existing in the system. If two rigid clusters share a pivot node in the physical system, they share an edge in the Pivot Network. Note that, from the ``two-pivots rule'' (Proposition 2), two distinct rigid clusters can share at most one pivot. In practice, the nodes of the Pivot Network correspond to the root bonds of the physical system, that uniquely identify each rigid cluster. The goal of the rigidification step is to identify the $k$ coalescing clusters/roots of Theorem 3, in order to construct $\mathcal{R}^+=\mathcal{R}_1\cup\mathcal{R}_2\cup\cdots\cup\mathcal{R}_k\cup \{uv\}$. Initially, $\mathcal{R}^+$ is composed of the new activated bond $uv$ only and 
both $u$ and $v$ are pivots of it. The BFS over the Pivot Network is performed
by identifying the roots connected to $\mathcal{R}^+$ in the Pivot Network, and by checking the mutual
rigidity of each of these roots with respect to $uv$.
For each root $r$ found to be rigid with respect to $uv$, the corresponding
rigid cluster is coalesced into $\mathcal{R}^+$, i.e., $\mathcal{R}^+\to\mathcal{R}^+\cup\mathcal{R}_{r}$. 
The pivots of each coalesced rigid cluster $\mathcal{R}_{r}$ become pivots of $\mathcal{R}^+$.
The BFS then proceeds by further checking the new roots connected to $\mathcal{R}^+$ through these pivots.
As shown in Fig.~\ref{fig:rigidification_fig}, this iterative procedure builds the rigid cluster resulting from rigidification by walking the paths
in the Pivot Network that connect all the rigid clusters coalescing into $\mathcal{R}^+$. The BFS stops when a surrounding shell of floppy clusters has been identified.
\\
Practically, the cost of the algorithm is dominated by the cost of the rigidification steps (whose cost generally depends
on the bond concentration, see next section).

\subsection{Overconstraining}
\label{subsec:overconstr}
\begin{figure}[!ht]
\includegraphics[width=\columnwidth]{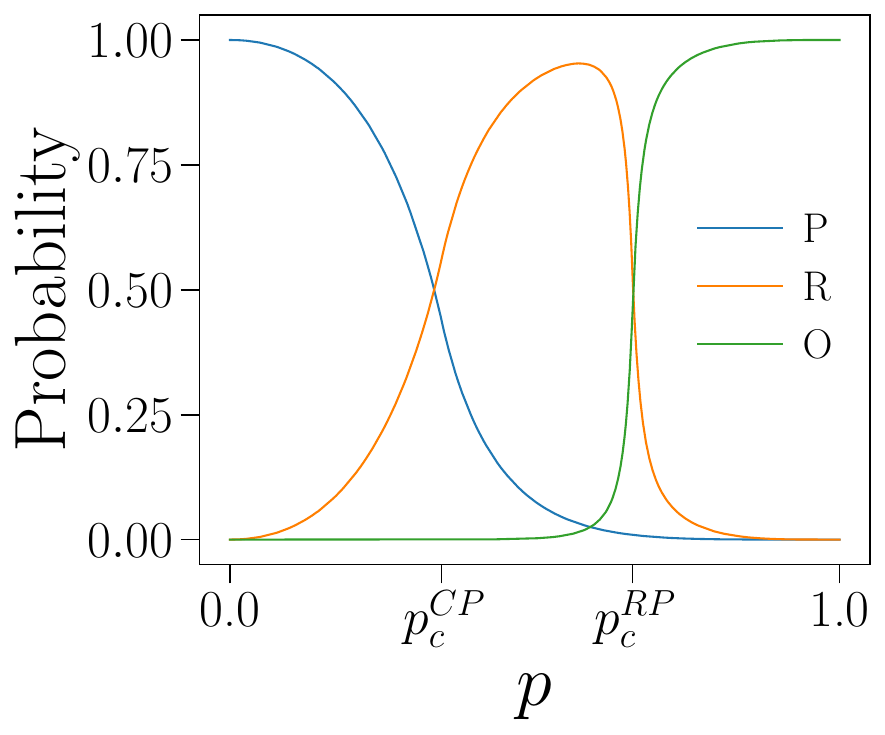}    
\caption{Stages of the rigidity transition. The blue, orange and green curves represent
the probability, as a function of the bond concentration, that the activation of a new bond 
results respectively in a pivoting event, a rigidification event or an overconstraining event. We indicate the critical bond concentrations $p_c^{CP}=2 \sin (\pi /18)$~\cite{sykes64} and $p_c^{RP}\approx 0.6602$~\cite{Jacobs1997An, JacobsThorpe1995} of the CP and RP
transition respectively.
Curves are shown for $L=2^{10}$ and are observed not to depend on $L$. 
}
\label{fig:RP_stages}
\end{figure}

At high bond concentration, and in particular right above the RP transition, the existence of a macroscopic 
rigid cluster leads to a sharp increase of the probability of overconstraining events, namely of connecting
two nodes belonging to the same rigid cluster (see Fig.~\ref{fig:RP_stages}).

An overconstraining step only requires to update the $\mathcal{T}_{\rm{R}}$ of the rigid cluster
to which both $u$ and $v$ belong by directing $uv$ to its root.
This operation clearly has constant cost.

\section{Implementation of the algorithm}
\label{sec:implementation}
We provide an explicit C++ implementation of the algorithm presented in the previous section for RP on the triangular lattice \cite{GitProject}.
The generalization to any two-dimensional graph, including off-lattice settings relevant for particle systems, is straightforward.

\subsection{Data structures}
The system is a triangular lattice of linear size $L$, where 
$i=0,\cdots,N-1$ integers are used to label nodes and $b=0,\cdots,M-1$ integers are used to label
bonds, with $N=L^2$ and $M=3N$.
$\mathcal{T}_{\rm{C}}$ and $\mathcal{T}_{\rm{R}}$ are two arrays of length $N$ and $M$
respectively. 
$\mathcal{T}_{\rm{C}}$ and $\mathcal{T}_{\rm{R}}$ are treated using the-non recursive
implementation of path compression of Ref.~\cite{NZ2001}.

The system is represented by a standard adjacency list, the pebble graph is a $2N$
array where entries $2i$ and $2i+1$ are the pebble neighbors of node $i$. Negative values correspond to inactive pebble edges.
We further define several other arrays of length $N$, the use of which is detailed in SM 4, 
including notably an array $\pi$, which stores the pivotal class of each node.
The Pivot Network $\mathcal{P}$ is implemented as an array of sets with $M$ entries.
C++ sets are balanced binary search trees that guarantee that no duplicates exist
and, for a set with $n$ elements, they have $\mathcal{O}(\log n)$ cost of insertion, deletion
and search~\footnote{We preferred sets to unordered sets as we observed the former to perform better
than the latter. Indeed, the complexity of basic operations on sets is guaranteed to be $\mathcal{O}(\log n)$. In contrast, while unordered sets offer $\mathcal{O}(1)$ complexity, their performance can degrade to $\mathcal{O}(n)$ due to rehashing of the underlying hash table.}.
Entries corresponding to root bonds  contain 
the pivots of the corresponding rigid cluster. All other entries are empty sets.
If the activation of $uv$ triggers a pivoting step, both $u$ and $v$ are
inserted in the entry of bond $uv$ if their pivotal class becomes larger than one. 
Since in this case $u$ and $v$ become pivots of the other rigid cluster(s) they belong to, they must be inserted in the corresponding entries of $\mathcal{P}$ as well. We identify these rigid clusters by applying find operations on the active bonds adjacent to $uv$.
The rigidification step is the more involved one, and we describe it in the next section. The Pivot Network is not affected by overconstraining.

\subsection{Rigidification step}\label{subsec:rigidification_step}

The Pivot Network is used to efficiently perform the BFS over the rigid clusters. 
We start the BFS with $\mathcal{R}=uv$ and iterate over its pivots, which are stored in the corresponding
entry of $\mathcal{P}$.
Every time two rigid clusters coalesce, $\mathcal{P}$  must be updated accordingly.

The first part of the rigidification step requires to update the pebble graph. To this aim, we
(i) gather four pebbles at the new bond by performing pebble searches of type I. Note that
we know the bond to be independent, so these searches never fail (implying
that we never fail pebble searches of type I); (ii) Use one of the pebbles
to cover the edge and update the pebble graph; (iii) Initialize a new $\mathcal{T}_{\rm{R}}$ tree 
with size one and root $uv$, just like in a pivoting step; (iv) Increase the pivotal
class of $u$ and $v$ and update $\mathcal{P}$ accordingly.
Finally, (v) freeze the remaining three
pebbles at their location.

Then we proceed with the BFS, which starts by inserting $u$ and $v$ in a queue of pivots.
Every time a pivot is inserted in this queue, we label it as enqueued using a dedicated array (see SM 4).
All nodes are initially unmarked, except for $u$ and $v$ that are marked rigid.
The rigidification step proceeds as follows:
\begin{enumerate}
    \item If the queue of pivots is empty, terminate the rigidification step. 
    Otherwise  pick a pivot $p$ and remove it from the queue. 
     \item Iterate over the bonds of $p$ .
     For each bond, find its root $ij$. Once all the roots
    are processed as detailed below, go back to step 1 and process the next pivot.
    \item 
    Consider the mark on node $i$.\\
    (i) If $i$ is marked rigid \textrightarrow go to step 4.\\
    (ii) If $i$ is marked floppy \textrightarrow go back to step 2. \\
    (iii) If $i$ is unmarked, perform a pebble search of type II to mark it and then operate according to     cases (i) and (ii).
    \item 
    Consider the mark on node $j$ and operate like in step 3.\\
    (i) If $j$ is marked rigid \textrightarrow go to step 5.\\
    (ii) If $j$ is marked floppy \textrightarrow go back to step 2. \\
    (iii) If $j$ is unmarked, perform a pebble search of type II to mark it and then operate according to 
    cases (i) and (ii).
\end{enumerate}

Note that the next steps are executed only if both $i$ and $j$ have been marked as rigid. This means, by transitivity, that the whole rigid cluster rooted by the bond $ij$, which we denote $\mathcal{R}_{ij}$, is mutually rigid with respect to the new bond and must be coalesced into the forming cluster $\mathcal{R}^+$. 
To this end, we first identify the largest among $\mathcal{R}^+$ and $\mathcal{R}_{ij}$. 
Let's assume, without loss of generality, 
that $\mathcal{R}_{ij}\equiv\mathcal{R}_{\rm large}$ is larger than $\mathcal{R}^+\equiv\mathcal{R}_{\rm small}$. 
We perform the following steps.
\begin{enumerate} 
    \setcounter{enumi}{4}
    \item Update the trees $\mathcal{T}_{\rm{R}}$: redirect the root $r_{\rm small}$ of $\mathcal{R}_{\rm small}$ to the root $r_{\rm large}$ of $\mathcal{R}_{\rm large}$ and update the entry of
    $r_{\rm large}$ in $\mathcal{T}_{\rm{R}}$ to keep track of the updated cluster size. 
    This step is the analog of a standard coalescence step in
    the NZ algorithm for CP (see sec.~\ref{sec:NZ}).
    \item Remove $p$ from the set of pivots of $\mathcal{R}_{\rm small}$, stored in $\mathcal{P}_{r_{\rm small}}$
    (after coalescence, $p$ is not anymore a pivot between $\mathcal{R}_{\rm large}$ and $\mathcal{R}_{\rm small}$)
    and decrease its pivotal class by one, $\pi_p\to \pi_p -1$. 
    If this results in $\pi_p=1$ ($p$ belongs only to one cluster and is not a pivot anymore) remove $p$ 
    from $\mathcal{P}_{r_{\rm large}}$  as well.
    \item Iterate over the remaining pivots $p'$ of $\mathcal{R}_{\rm small}$.
    If $p'$ is already in 
   $\mathcal{P}_{r_{\rm large}}$,  $\pi_{p'}\to \pi_{p'}-1$. If $\pi_{p'}$ becomes 1, remove $p'$ from $\mathcal{P}_{r_{\rm large}}$.
    If $p'$ is not yet in $\mathcal{P}_{r_{\rm large}}$, insert it and go to step 8.
    In any case, remove $p'$ from $\mathcal{P}_{r_{\rm small}}$.
    \item If $p'$ is not labeled as enqueued,
    insert it in the queue of pivots and label it as enqueued.
    This is absolutely crucial to prevent the insertion of the same pivot in the queue more than once. 
\end{enumerate}

\begin{figure*}[!ht]
\begin{center}
\includegraphics[width=\textwidth]{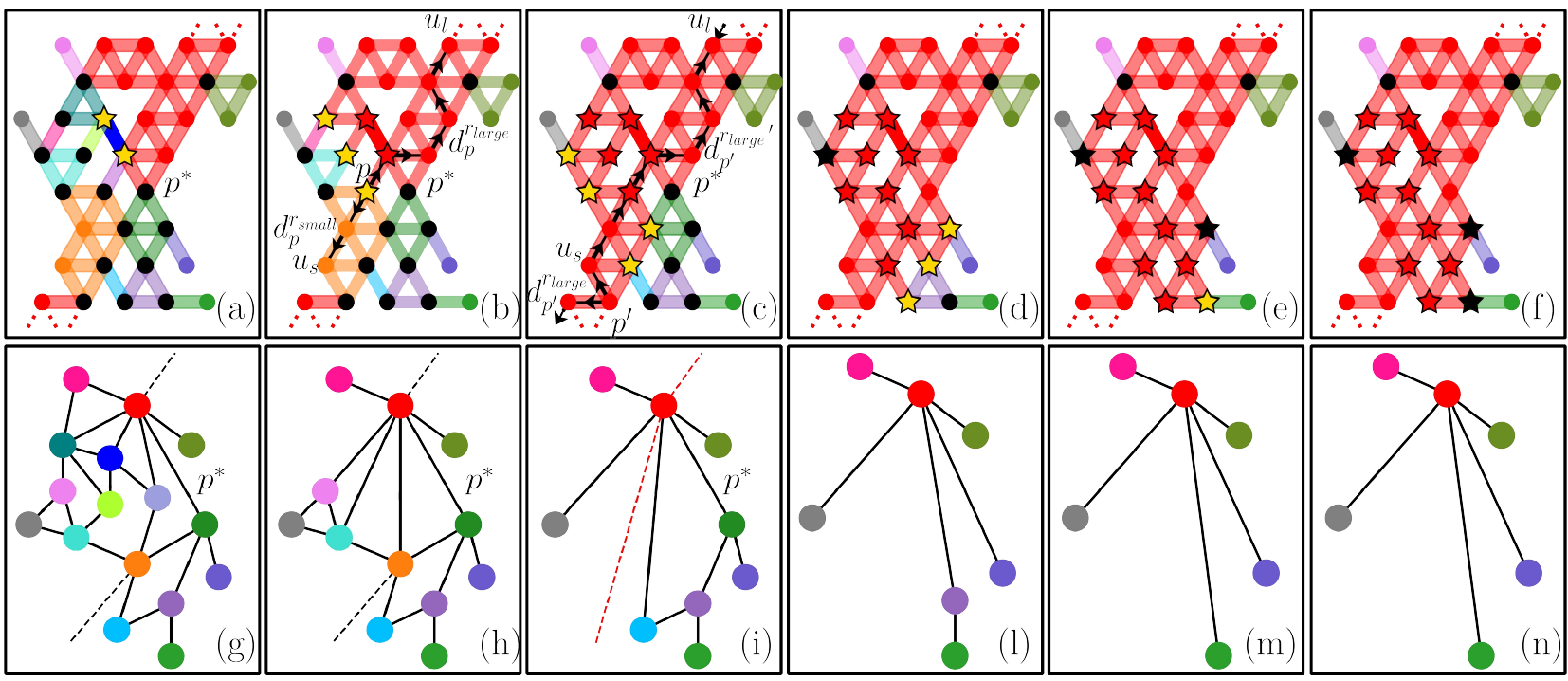}    
\end{center}
\caption{
Example of a rigidification process as implemented by our algorithm. The different steps are illustrated both on the triangular lattice (top row) and on the corresponding Pivot Network (bottom row). 
Each color indicates a different rigid cluster.
Top row: pivots are indicated with black dots. At each step, enqueued pivots are highlighted as stars.
(a) The blue bond is the newly activated one: its two end nodes,
marked with yellow stars, are initially enqueued in the queue of pivots. 
 Each rigid cluster is represented by a different color. Nodes with the same color of their bonds belong to that rigid cluster only, black nodes are pivots. 
 (b) The system state after having checked the rigid clusters led to by the pivots enqueued in panel (a).
Red stars indicate pivots that have been processed. The arrows illustrate the Machta displacements $d_p^{r_{\rm large}}$ and  $\,d_p^{r_{\rm small}}$, from pivot $p$ to the large and small roots.
 (c) The system state after having checked the rigid clusters led to by the pivots enqueued in panel (b). The displacement $d_{u_s}^{r_{\rm large}}$ has been updated according to Eq.~(\ref{eq:update_us}). The arrows show the displacements $d_{p'}^{r_{\rm large}}$ and $d_{p'}^{r_{\rm large}\,'}$ given by Eq.~(\ref{eq:machta_piv_update}), whose difference allows to detect that wrapping has occurred along the vertical axis. Note that the pivot $p^*$ is never enqueued, but that the green cluster is nonetheless reached via another pivot (yellow star), cf. discussion at the end of sec.~\ref{sec:implementation}. 
 (d) The system state after having checked the rigid clusters led to by the pivots enqueued in panel (c). Black stars indicate processed pivots leading to floppy clusters.
(e) The system state after having checked the rigid clusters led to by the pivots enqueued in panel (d).
(f) The final lattice configuration.
Bottom row: the pivot network corresponding to each panel of the top row. Nodes correspond to rigid clusters, edges represent pivots. 
Note that pivots with pivotal class three form triangles in this network; we indicate again the non-enqueued pivot $p^*$. 
Note that as rigid clusters coalesce, nodes disappear and the network is rearranged. Also, note that the bond representing pivot $p^*$ disappears from panel (i) to panel (l).
If it had larger pivotal class, it would still be an edge in the Pivot Network.}
\label{fig:rigidification_fig}
\end{figure*}

This process ends only when the queue is empty, i.e., when all the pivots have been
processed.
This protocol implements the rigidification strategy described in section \ref{subsec:rigid_idea} and in Fig.~\ref{fig:rigidification_fig}:
processing all pivots means that we attempt to build all possible paths in the Pivot Network that, starting from a 
rigid cluster that coalesces into $\mathcal{R}^+$, connect it to other existing rigid clusters.
The queue is empty when the only remaining paths lead to floppy clusters.
\\\\
Note that, as long as the rigidification process is not over, $\mathcal{R}_{\rm small}$ and $\mathcal{R}_{\rm large}$ may share several pivots, whose pivotal classes get updated at step 7. As can be seen in Fig.~\ref{fig:rigidification_fig} (the green and red clusters in panel (c)), such pivots arise from previous coalescence events, and do not contradict Proposition 2.\\
\\
Moreover, note that we only enqueue pivots of $\mathcal{R}_{\rm small}$ (step 7), namely the BFS progresses always from the pivots of the smallest cluster without enqueuing the ones of the largest cluster. This implies that, at the end of the process, the pivots of exactly one of the $\mathcal{R}_1,\cdots\,\mathcal{R}_{k+1}$, say $\bar{\mathcal{R}}$, have not been enqueued.
To see this, consider a coalescence between $\mathcal{R}_1$ and $\mathcal{R}_2$ and then a coalescence
between $\mathcal{R}_1 \cup \mathcal{R}_2$ and $\mathcal{R}_3$. After the first coalescence, 
the non-enqueued pivots belong to one rigid cluster only (say $\mathcal{R}_1$). After the second event,
if $\mathcal{R}_{\rm large} = \mathcal{R}_3$, then the pivots of $\mathcal{R}_1$ gets enqueued
and the ones of $\mathcal{R}_3$ are the only ones not enqueued, while in the opposite case the
only non-enqueued ones remain those of $\mathcal{R}_1$.
The same reasoning applies to all the successive coalescences.
Ignoring the pivots of $\bar{\mathcal{R}}$ still leads to the correct construction of $\mathcal{R}^+$. Namely, none of the $\mathcal{R}_1,\cdots\,\mathcal{R}_{k+1}$ is ``forgotten'' during rigidification. Indeed, by Theorem 3, any rigid cluster that shares a pivot with $\bar{\mathcal{R}}$ and that coalesces into $\mathcal{R}^+$
shares at least one other pivot with at least another coalescing cluster, 
from which it will be eventually reached by the BFS. In other words, clusters that share a pivot with $\bar{\mathcal{R}}$ but with none of
the other coalescing ones are surely floppy.
This is illustrated in Fig.~\ref{fig:rigidification_fig}, where the pivot of the large (red) cluster labeled $p^*$ is never enqueued. The green rigid cluster connected to the large one via $p^*$ is nonetheless reached from another pivot, and coalesced.
\\\\
Not enqueuing all pivots is crucial to make the algorithm fast. Indeed, when a large rigid
cluster exists, a very large number of pivots that lead to floppy clusters would be enqueued,
making the iteration of step 1 exceedingly long. This is especially true above the RP transition,
where the macroscopic rigid cluster has $\mathcal{O}(N)$ pivots: enqueuing them 
 would result in a quadratic algorithm.

\section{Detecting wrapping}\label{sec:wrapping}
\subsection{The Machta algorithm in CP}\label{sec:Machta_CP}
In systems with doubly periodic boundary conditions, the criterion for percolation is the wrapping of a cluster around one, the other, or both periodic boundaries. This corresponds to the existence of a non-contractible loop inside a cluster. A clever way to detect wrapping in CP is the Machta algorithm \cite{Machta1996}, that we recall briefly before generalizing it to the more involved RP problem (we refer to \cite{NZ2001, Machta1996} for a more detailed explanation of the Machta algorithm). The key difference between a contractible loop (non-wrapping) and a non-contractible one (wrapping) is the algebraic distance traveled as one goes around the loop. The essence of Machta algorithm is to exploit this fact, by keeping track of the algebraic displacements from each node to a given reference node inside its cluster, conveniently taken as the root. When a new activated bond connects two nodes of a same cluster, the difference of their respective displacements to the root is computed: wrapping has occurred if the difference of displacements is of the order of the system's length.
In the original Machta algorithm for CP, two variables are assigned to each node, that give the $x$ and $y$ displacements along a path inside the cluster, from that node to its parent in the tree.
Machta displacements are updated in sync with the tree, whenever roots get redirected and when path compression is performed. Wrapping is checked only when connecting two nodes that are found to belong to the same cluster, as this is the only cluster event that might lead to wrapping. 

\subsection{A Machta algorithm for RP}\label{sec:Machta_RP}
In RP, a first complication comes from the fact that nodes may belong to more than one rigid cluster. For each node, we therefore keep track of the displacements associated to each rigid cluster it belongs to. Namely, if node $u$ belongs to $k$ rigid clusters, so that it is shared by bonds $b_1,\,b_2,\cdots,b_k$, whose respective parents 
in the trees are $a_1=(u_1,v_1),\,a_2=(u_2,v_2),\cdots,a_k=(u_k,v_k)$,
we compute the Machta displacements $d_u^{a_i}\equiv \left\{\delta_x(u\to u_i),\delta_y(u\to u_i)\right\}$, for $\quad i=1,\cdots,k$. We conventionally set the reference point as the first node ($u_i$) of the parent bond. 
Two arrays of length $N$ are used to store the displacements along the two dimensions of the lattice.
In our C++ implementation, each entry in the array contains a map. So, for example,
to store the horizontal displacements of $u$
we use the array $dx$ such that $dx[u]$ is a map and the displacement $\delta_x(u\to u_i)$
is stored as the value with key $a_i$, i.e, $dx[u][a_i]$.

The second notable difference with CP is the type of cluster events that may lead to wrapping. In RP, wrapping may occur both during overconstraining and rigidification. Below we detail the update of displacements and wrapping detection, when activating the new bond $b=(u,v)$.
\\\\
\textit{Pivoting --} Bond $b$ is a rigid cluster of size one (cf. sec.~\ref{sec:theorems}), and the root of its own tree. The displacement of its first node is therefore set as $d_u^b=\left\{0,0\right\}$ (reference node), while $d_v^b=\left\{\delta_x(v\to u) \right\}$.
\\\\
\textit{Overconstraining -- } This case is akin to CP: adding a new bond to an existing rigid cluster may ``close it'' around the periodic boundary(ies). Therefore, when activating $b=(u,v)$ leads to overconstraining, we simply compare the total displacements to the root for $u$ and $v$.
\\\\
\textit{Rigidification -- } Each time a rigid cluster is coalesced into the forming one (as described in section \ref{subsec:rigidification_step}), the root of the smallest cluster, $r_{\rm small}=(u_s,v_s)$ is redirected to the root of the largest, $r_{\rm large}=(u_l,v_l)$. Denoting $p$ the pivot node between $\mathcal{R}_{\rm small}$ and $\mathcal{R}_{\rm large}$, we update the displacements as
\begin{equation}\label{eq:update_us}
\begin{aligned}
    &d_{u_s}^{r_{\rm large}}\to d_{u_s}^{r_{\rm large}\,'} = -d_{p}^{r_{\rm small}} + d_p^{r_{\rm large}}\\
    &d_{v_s}^{r_{\rm large}} \to d_{v_s}^{r_{\rm large}\,'} = d_{v_s}^{r_{\rm small}} + d_{u_s}^{r_{\rm large}}.
\end{aligned}
\end{equation}
An example update of $d_{u_s}^{r_{\rm large}}$ is shown in Fig.~\ref{fig:rigidification_fig} (panels (b) and (c)).
After coalescing $\mathcal{R}_{\rm small}$, some of its pivots are added to the pivots of $\mathcal{R}_{\rm large}$ (step 7 in sec.\ref{subsec:rigidification_step}). We update their  displacements accordingly, 
\begin{equation}
\label{eq:machta_piv_update}
\forall p' \in \mathcal{P}_{r_{\rm small}},\quad d_{p'}^{r_{\rm large}\,'} = d_{p'}^{r_{\rm small}} + d_{u_s}^{ r_{\rm large}}.
\end{equation}
To check if the coalescence step has created a non-contractible loop, we compute, for each node $p'$ that is a pivot between small and large, the difference $d_{p'}^{r_{\rm large}\,'} - d_{p'}^{r_{\rm large}}$. 
Namely, while in CP and overconstraining the closure of the loop is checked at the new bond, during rigidification it is more convenient to check it at pivots. This difference is the algebraic distance along the closed loop. It is zero if the rigid cluster does not (yet) wrap, non-zero if wrapping has occurred. An example of such wrapping detection is given in Fig.~\ref{fig:rigidification_fig} (panel (c)).
Note that the nodes of the small root may be pivots too. Therefore, we also check wrapping by computing $d_{u_s}^{r_{\rm large}\,'} - d_{u_s}^{r_{\rm large}}$ and $d_{v_s}^{r_{\rm large}\,'} - d_{v_s}^{r_{\rm large}}$.

\section{Results}
\label{sec:results}
\subsection{Performance of the algorithm}
\begin{figure*}[!ht]
\begin{center}
\includegraphics[width=\textwidth]{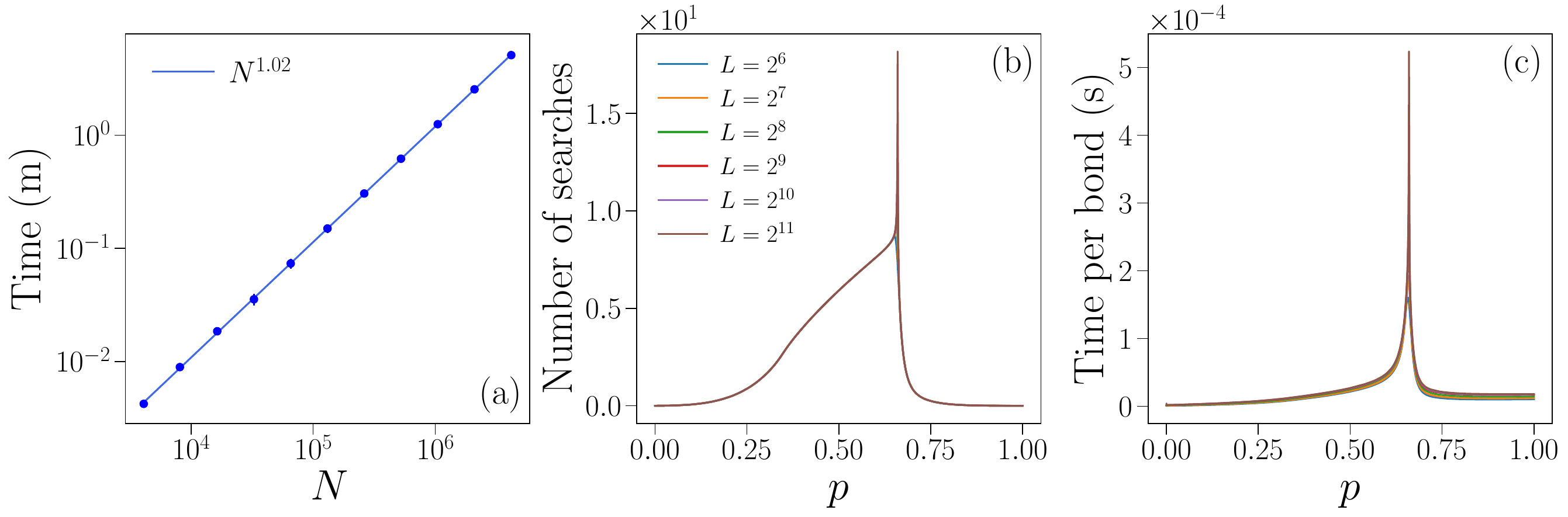}    
\end{center}
\caption{
Performance of the algorithm. 
(a) The average time (in minutes) to complete a simulation as a function of the system size. 
The blue
solid line is the best fit to the log-log of the data and shows the scaling $N^{1.02}$.
Error bars are smaller than the markers' size.
(b) Average number of pebble searches (of any type) needed at each bond activation, as function of the bond concentration $p$, for different system sizes $N=L^2$.
(c) Average time (in seconds) of each bond activation as function of $p$, for the same system sizes as in panel (b).
 }
\label{fig:results_fig}
\end{figure*}

Fig.~\ref{fig:results_fig} shows the performance of the algorithm presented
in the previous sections, including the implementation of wrapping detection, on the triangular 
lattice. Quantities that depend on the filling fraction are convoluted as in \cite{Newman2000Efficient} and shown as
a function of $p$. Averages are made over $10^5$ for the smallest sizes, and over about $600$ for the largest one.
The performance of the algorithm, quantified by the average time to complete
a simulation, is observed to scale as $N^{1.02}$ (panel (a)). A striking feature of the 
algorithm is the extremely low number of pebble searches needed to update the system's
state after the activation of a bond, shown as a function of the bond concentration
$p$ (panel (b)). This function peaks near $p_{c}^{RP}$ with a maximum that is of the order of 10
even at large system sizes. 
The average time of each bond activation
as a function of the bond concentration is shown on panel (c). Despite this function
also shows a peak near the rigidity transition, the fundamental observation here is that
the time per bond is not proportional to the number of pebble searches.
Therefore,
contrary to rigidification protocols based on the JH approach (cf. SM 3 A), 
the total execution time cannot be deduced from the number of searches. For example, for $p_c^{CP} \lesssim p < p_c^{RP}$ the number of searches grows much more rapidly than the time per bond, which remains 
low. A deeper investigation is required to understand the scaling of the algorithm. 

\begin{figure}
    \centering
    \includegraphics[scale=.5]{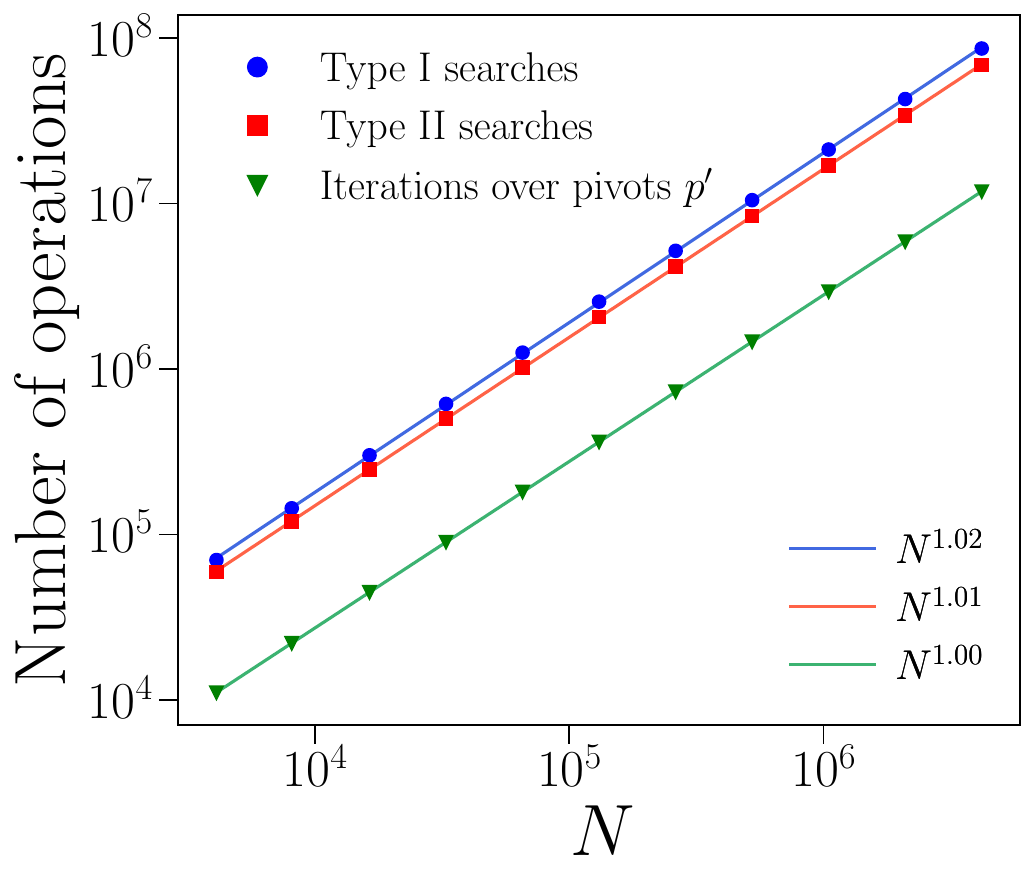}
    \caption{Average number of nodes visited during type I searches (blue circles,
    scaling as $N^{1.02}$), average number of nodes visited during type II searches (red squares,
    scaling as $N^{1.01}$) and average number of pivots $p'$ over wich we iterate
    during step 7 of the algorithm (green triangles, scaling as $N^{1.00}$). Data are
    shown as a function of the system size. 
    }
    \label{fig:comp_complex}
\end{figure}

To this aim, we first note that Fig.~\ref{fig:results_fig}a shows the performance of our code,
which is only a proxy of the computational complexity of the algorithm and, in particular, sets an upper bound on it (see SM 5). The computational complexity
is truly quantified by the number of operations performed. To estimate it, we focus on the rigidification step, which is the main contributor to the cost of the algorithm.
The observation is that there are mainly three operations that determine how expensive a
rigidification step is. The first operation is the preliminary search of four pebbles
using type I searches. The second operation is the repeated application of type II 
searches to mark nodes. The last expensive operation is the iteration of pivots $p'$, at step 7 of the algorithm. Note that the number of pivots
$p$ in step 1 is at most equal to the number of pivots $p'$, as only a fraction of the
$p'$ gets enqueued and hence iterated over in step 1 (with the exception of $u$ and $v$,
which represent however an $N$-independent cost). To quantify how expensive pebble searches are,
we count the total number of nodes visited during each search and we further count
the number of pivots $p'$. Results are shown in Fig.~\ref{fig:comp_complex}.

Data show that the main cost of the algorithm is due to pebble searches of type I,
despite they are less numerous than pebble searches of type II (which, however, are only
slightly less expensive than the former)
The number of pivots $p'$
represents instead a minor contribution to the cost of the algorithm. Importantly, pebbles
searches of type I are estimated to have a cost scaling as $N^{1.02}$, i.e., the
same scaling inferred for the performance. This suggests that the observed performance
well quantifies the computational complexity and that its (extremely small) superlinearity
is an intrinsic feature of the algorithm, i.e., that it is implementation independent.
Overall, we conclude that the computational complexity of the algorithm presented in this work
is $N^{1.02}$, i.e., the phase diagram of the RP transition can be computed in a time
that --to any practical purpose-- is linear in the system size (see SM 5).

\subsection{Estimation of the RP critical exponents and critical threshold}
We apply our algorithm to the determination of the main critical exponents -- the correlation length exponent $\nu$, the fractal dimension $D_f$, the susceptibility exponent $\gamma$ and the order parameter exponent $\beta$.
We also determine the critical point $p_c$.
We perform simulations for system sizes ranging from $L=2^6=64$ to $L=2^{14.5}=23170$, and choose the numbers of samples so as to activate at least $7 \cdot 10^{11}$ bonds per value of $L$ (see SM 6).
For each run of the algorithm, and for both the connectivity and rigid clusters, we store: the number of bonds $m_x$ and $m_{xy}$ at which wrapping happens for the first time, respectively along the horizontal and along both horizontal and vertical directions; the size $S_{xy}$ of the $xy$-wrapping cluster right at wrapping;
the maximum value $\chi_{\rm max}$ of the susceptibility.

We then perform scaling analysis to determine $D_f$, $\nu$, $\gamma/\nu$ and $p_c$, for both the RP and CP transitions, the latter being used to benchmark our scaling analysis.
The values of all exponents are reported in Table \ref{tab:exponents}, giving a comparison of the CP and RP universality classes. These values are validated
by the excellent collapse of the rescaled scaling functions (order parameter, susceptibility
and wrapping probability) shown in Fig.~\ref{fig:collapse}. Overall, our results strengthen the long-standing hypothesis~\cite{JacobsThorpe1995} that
2d CP and central-force RP belong to different universality classes.

\subsubsection{Fractal dimension $D_f$}\label{sec:Df}
The fractal dimension $D_f$ is estimated from the (rescaled) size of the wrapping cluster $S_{xy}/(3L^2)$, expected to scale as  $S_{xy}\sim L^{D_f-2}$~\cite{stauffer_aharony}. For CP, a simple power-law fit gives an estimate in excellent agreement with the exact value $91/48$, reported in Figure~\ref{fig:exponents} (a). For RP, deviations to scaling are present at small sizes, so that we choose to fit $S_{xy}^{\rm RP}$ as a power-law, but only for sufficiently large $L$. The number of datapoints removed is chosen as the one maximizing the goodness of fit, and indicated in Fig.~\ref{fig:exponents}. Note that we also tried to fit the data of $S_{xy}$ (and similarly for $\Delta$, $\chi_{\rm max}$ and $p_c$) to the form $\mathcal{O}(L) = A\,L^{y}\left(1+B/L^{w}\right)$, which incorporates corrections to scaling. For all observables, such fits lead values of the exponent $y$ consistent with those obtained by removing the smallest sizes from a power-law fit.

\subsubsection{Susceptibility exponent $\gamma$}
The maximum of the susceptibility is known to scale as $\chi_{max}\sim L^{\gamma/\nu}$~\cite{stauffer_aharony}, from which we estimate the exponent $\gamma/\nu$. As for the size of the wrapping cluster $S_{xy}$ (cf. section \ref{sec:Df}), the data for CP is perfectly fitted by a power-law, while RP shows significant deviations to scaling at small sizes, see Figure~\ref{fig:exponents} (b). We adopt the same strategy as in section \ref{sec:Df} and fit only
data with sufficiently large $L>L_{\min}$, the value of $L_{\rm min}$ being chosen as maximizing the goodness of fit. The susceptibility exponent $\gamma^{\rm RP}$ is then computed using the value of $\nu^{\rm RP}$ obtained independently in section \ref{sec:nu}, and reported in Table~\ref{tab:exponents}.

\subsubsection{Correlation length exponent $\nu$}\label{sec:nu}
The measurements of $m_x$ and $m_{xy}$ allows to 
compute the wrapping probabilities $R_x$ (along the horizontal axis), $R_{xy}$ (along both axes), $R_e$ (along either axis) and $R_1$ (along one axis), see~\cite{NZ2001} and Fig.~S6.
For each wrapping probability, we compute the width of the transition region $\Delta(L)$, as explained in SM 6 A. For each wrapping type, $\Delta(L)$ is expected to scale as $\Delta(L)\sim L^{-1/\nu}$ \cite{stauffer_aharony}. 

\begin{figure*}
    \centering
\includegraphics[width=\linewidth]{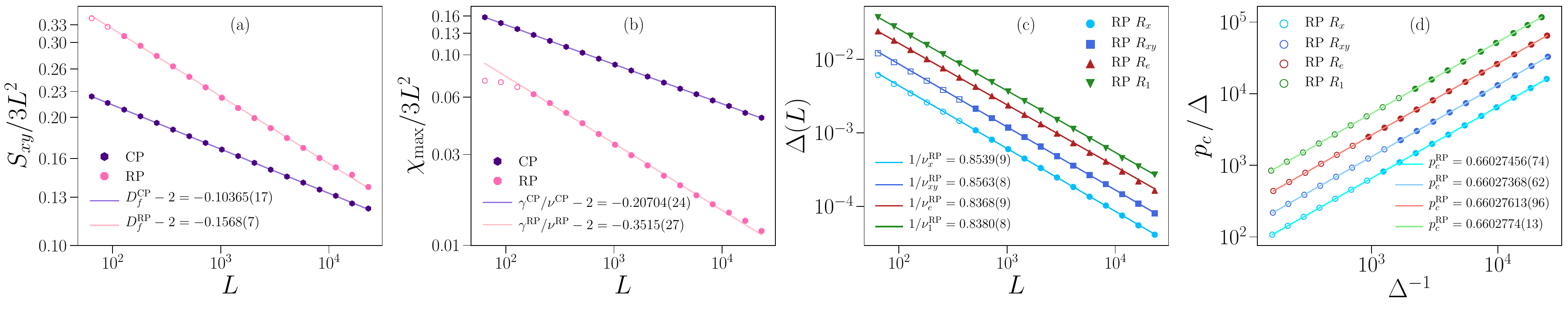}    
\caption{
Scaling analysis for RP.
(a), (b) Size $S_{xy}/(3L^2)$ of the wrapping cluster, and maximum $\chi_{\rm max}$ of the susceptibility, both rescaled by the total number of bonds, as function of $L$, shown for CP and RP. The lines are power-law fits of the data marked with filled symbols.
(c) Widths $\Delta(L)$ of the transition region for each wrapping type $x,\, xy,\, e,\,1$, as functions of the system size $L$. The slopes of the solid lines are reported in the legend and are obtained from power-law fits of the data marked with filled symbols.
For the sake of readability, data from $R_{xy}$, $R_e$ and $R_1$ have been shifted by a factor 2, 4 and 6 respectively.
(d) Finite-size critical thresholds $p_c/\Delta$, for each wrapping type, as functions of $\Delta^{-1}$. The slopes of the solid lines, reported in the legend, give the corresponding values of $p_c^{\rm RP}$. Data has been shifted and shown in log-log scale to ease readability.
}
\label{fig:exponents}
\end{figure*}

For CP, simple power-law fits of $\Delta(L)$ yield exponents in excellent agreement with the exact value $\nu^{\rm CP}=4/3$, for all four wrapping types. For RP, small deviations to scaling are visible in the residual plots, especially for $R_x$ and $R_{xy}$, while they appear suppressed in $R_e$ and $R_1$. We adopt the same strategy as in section \ref{sec:Df} and fit the data for $\Delta(L)$ as power-laws, discarding the smallest sizes as indicated in Figure~\ref{fig:exponents} (c). The corresponding values of $1/\nu^{\rm RP}$ are reported in Figure~\ref{fig:exponents} (c). Note that not all four wrapping probabilities are independent \cite{Newman2000Efficient}. We find that the estimates from $R_{x}$ and $R_{xy}$ lead to the best collapse for all four wrapping probabilities, and therefore compute the final value of $\nu^{\rm RP}$, reported in Table~\ref{tab:exponents}, as the average of $\nu^{\rm RP}_x$ and $\nu^{\rm RP}_{xy}$.



\subsubsection{Critical threshold $p_c$}

The wrapping probabilities $R_i$ allow to further compute, for each wrapping type, the size-dependent transition thresholds $p_c(L)$, namely the bond concentration at which the emergence of a wrapping cluster is expected on a finite lattice of size $L$ (cf. SM 6 A). In general we expect $p_c(L)-p_c \propto L^{-1/\nu}\propto \Delta(L)$, where $p_c=p_c(L\to\infty)$. For RP, pronounced deviations to scaling are visible at small sizes, namely $p_c^{\rm RP}(\Delta) = p_c^{\rm RP}+ A \,\Delta^{\rm RP}\left(1+B\,\Delta^{w}\right)$ for some positive exponent $w$. We consider the quantity $p_c^{\rm RP}/\Delta^{\rm RP}$ and fit
\begin{equation}\label{eq:pc_fit}
y(x)\equiv \frac{p_c^{\rm RP}(\Delta^{\rm RP})}{\Delta^{\rm RP}} = p_c\;x+ A
\end{equation}
for $L>L_{\rm min}$ sufficiently large, namely when $x=1/\Delta^{\rm RP}$ is sufficiently large that the correction term $\propto x^{-w}$ becomes negligible. For each wrapping type the value of $L_{\rm min}$ is chosen to maximize the goodness of fit.
 All four wrapping types yield excellent fits, as shown in Fig.~\ref{fig:exponents} (d). Our final estimate of $p_c^{\rm RP}$ reported in Table~\ref{tab:exponents} is the average of the estimates obtained from $R_x$ and $R_{xy}$, that have the lowest uncertainties.

\begin{figure*}
    \centering
    \includegraphics[width=\textwidth]{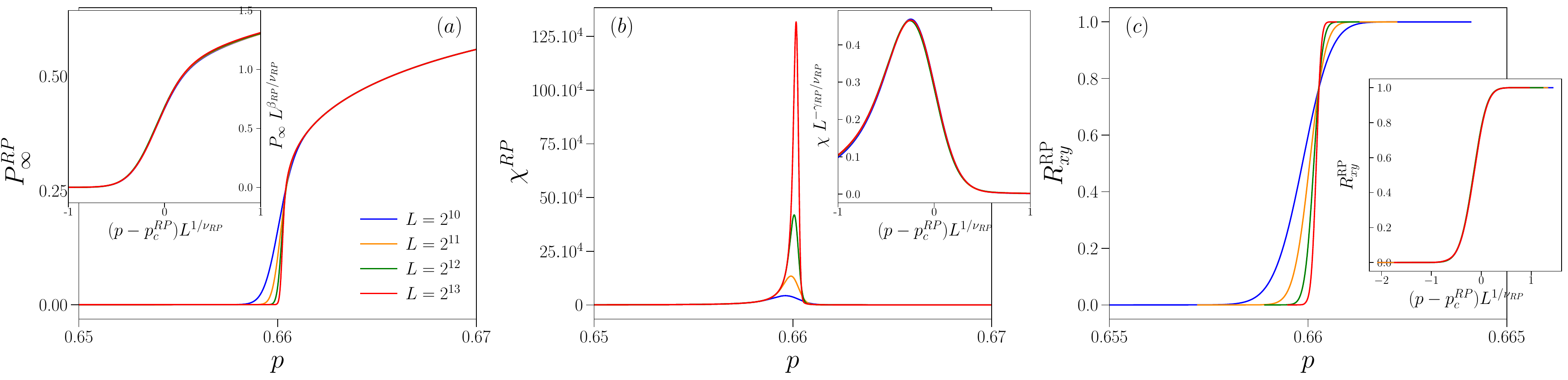}
\caption{Scaling functions for RP (main plots), and their collapse (insets) when rescaled using the values of the critical threshold and exponents in Table \ref{tab:exponents}. (a) Order parameter $P_{\infty}$ (b) Susceptibility $\chi$ (c) Wrapping probability $R_{xy}$.}
\label{fig:collapse}
\end{figure*}

\begin{table}
    \begin{tabular}{c|c|c}
     & CP & RP \\ \hline
     $D_f$ & $91/48\approx 1.895833$ & $1.8423(7)$\\
     $\nu$ & $4/3\approx 1.333333$  & $1.1694(8)$ \\
     $\gamma$ &  $43/18\approx 2.388889$ & $1.928(3)$ \\
     $\beta$ & $5/36\approx 0.138889$  & $0.1844(8)$ \\
     $p_c$ & $2\sin(\pi/18)\approx 0.347296$ & $0.6602741(4)$
    \end{tabular}
    \caption{CP and RP universality classes, and critical thresholds. The value of $\beta$ is computed using hyperscaling as $\beta = \nu(2-D_f)$.}
    \label{tab:exponents}
\end{table}

\section{Conclusion}\label{sec:conclusion}

In this work we have presented novel theoretical results 
on two dimensional central-force Rigidity Percolation
and have 
exploited them to derive an exact algorithm that computes the whole phase diagram of 
the model
in a time that is almost linear in the system size.
A C++ implementation of the algorithm is provided in~\cite{GitProject}.
We have assessed the complexity intrinsic to the algorithm on the triangular lattice, 
where each node has exactly six neighbors.
The rigidity transition, however, is of interest
also on different topologies~\cite{barre2009, jordan2022}. Our algorithm can be straightforwardly implemented on 
any system where rigidity is ruled by Laman theorem.
In particular, it can be readily used to analyze the rigidity of off-lattice configurations obtained from simulations of particle systems.
On networks with a broad degree distribution, where the average degree
might grow significantly with $N$~\cite{boguna2009Langevin}, the
repeated application of the find operation 
would introduce a supplementary, $N$-dependent cost. In any other network as,
e.g., Erd\H{o}s--R\'{e}nyi graphs we expect our algorithm to perform well. The actual
performance might depend on fine details of the system such as the structure of the
pebble graph (pebble searches might be more or less expensive than on the triangular
lattice) and on the geometry of rigid clusters (number of pivots they have).
\\\\\
Our algorithm pushes the size limitation encountered in previous works \cite{JacobsThorpe1995, mehdi2019, ninaconf, ninasle}, allowing us to characterize accurately the universality class: we have performed scaling analysis on systems up to $536\,848\,900$ nodes, and provide new estimates of the main critical exponents, $\nu$, $\beta$, $D_f$ and $\gamma$, as well as the critical threshold $p_c$.
The consistency of these values is validated by the excellent collapse of the rescaled scaling functions, shown in Fig.~\ref{fig:collapse}.
\\\\
On the theoretical side, our results provide new material to the long-standing question of the relationship between CP and RP \cite{deGennes1976,Moukarzel1999, Liu2019}, by showing explicitly that different cluster merging mechanisms are at play in the two problems. RP features notably non-local ``rigidification'' events that lead to the coalescence of several (rigid) clusters, a phenomenon absent in CP. In contrast with the standard Pebble Game, we construct the system in a dynamic way, making it straightforward to study accurately the frequency
and spatial extension of rigidification events~\footnote{While concluding the first version of this work we became aware of the preprint \cite{lu2024}, that proposes a similar dynamic approach to RP. The algorithm has not been published, but is reported to perform slightly worse than the standard Pebble Game.}. It would also be very interesting to generalize this approach to study rigidity loss \cite{ellenbroek2015}, in particular large-scale fluidization events whereby the removal of a single bond
 makes a macroscopic rigid cluster collapse into smaller clusters connected by non-rigid hinges. 

Although we focused here on networks with central forces, it is important to examine if some of our theoretical conclusions generalize to systems with non-central interactions. This could allow to combine generalized pebbles games \cite{silke2016, LESTER2018225} with the NZ approach we presented to develop faster algorithm for RP with, for instance, frictional interactions. Furthermore, our exact results may be exploited to inform non-exact, neural network-based algorithms for rigidity transitions~\cite{Head2025}. Finally, the exact determination of rigid clusters in three (or more) dimensions remains a crucial bottleneck \cite{emanuelaDST,richard2025}. Very recent results \cite{kim2024} have opened the possibility of designing exact algorithms for rigidity analysis in any dimension, and we can hope to make these highly efficient by incorporating the insights of the present work.

\textbf{Note added to the second version:} While finalizing the second version of this work --which completes our original study by determining the critical exponents, we became aware of the preprint \cite{lu2026}, which also reports new estimates for the RP critical exponents. To the best of our knowledge (the code not yet being publicly available), the algorithm described in \cite{lu2026} incorporates several strategies that we introduced in the first version of our work, but the system sizes considered in their analysis do not exceed 
$L=8192$. We observe that the reported values of the critical threshold and the correlation length exponent are in good agreement with our results. The larger value of the fractal dimension, $D_f=1.850(2)$, may be explained by the fact that it is obtained from a different set of observables, wrapping not being implemented in their algorithm.

\section{Acknowledgments}

This research was funded in whole by the Austrian Science Fund (FWF) under Project. No. Y-1163-N27. For open access purposes, the author has applied a CC BY public copyright license to any author-accepted manuscript version arising from this submission.
DN acknowledges 
computation time at the Vienna Scientific Cluster (VSC). NJ has been supported by a grant from MIAI@Grenoble Alpes (ANR-19-P3IA-0003). Part of the computations presented in this paper were performed using the GRICAD infrastructure (https://gricad.univ-grenoble-alpes.fr), which is supported by Grenoble research communities.
\\\\
The authors are grateful to Lorenzo Rovigatti, Silvano Ferrari, Emanuela Bianchi and Claudio Castellano for insightful discussions and valuable comments on the manuscript. We especially thank Hugo Vanneuville and Vincent Beffara for their precious help in finalizing  the proofs given in the SM. Finally, we thank the organizers of the 2024 Network Dynamics workshop in Les Houches, where this project started.

\section{Code availability}
The code is available at \cite{GitProject}.



\bibliography{biblio}

\begin{thebibliography}{10}%
\makeatletter
\providecommand \@ifxundefined [1]{%
 \@ifx{#1\undefined}
}%
\providecommand \@ifnum [1]{%
 \ifnum #1\expandafter \@firstoftwo
 \else \expandafter \@secondoftwo
 \fi
}%
\providecommand \@ifx [1]{%
 \ifx #1\expandafter \@firstoftwo
 \else \expandafter \@secondoftwo
 \fi
}%
\providecommand \natexlab [1]{#1}%
\providecommand \enquote  [1]{``#1''}%
\providecommand \bibnamefont  [1]{#1}%
\providecommand \bibfnamefont [1]{#1}%
\providecommand \citenamefont [1]{#1}%
\providecommand \href@noop [0]{\@secondoftwo}%
\providecommand \href [0]{\begingroup \@sanitize@url \@href}%
\providecommand \@href[1]{\@@startlink{#1}\@@href}%
\providecommand \@@href[1]{\endgroup#1\@@endlink}%
\providecommand \@sanitize@url [0]{\catcode `\\12\catcode `\$12\catcode `\&12\catcode `\#12\catcode `\^12\catcode `\_12\catcode `\%12\relax}%
\providecommand \@@startlink[1]{}%
\providecommand \@@endlink[0]{}%
\providecommand \url  [0]{\begingroup\@sanitize@url \@url }%
\providecommand \@url [1]{\endgroup\@href {#1}{\urlprefix }}%
\providecommand \urlprefix  [0]{URL }%
\providecommand \Eprint [0]{\href }%
\providecommand \doibase [0]{https://doi.org/}%
\providecommand \selectlanguage [0]{\@gobble}%
\providecommand \bibinfo  [0]{\@secondoftwo}%
\providecommand \bibfield  [0]{\@secondoftwo}%
\providecommand \translation [1]{[#1]}%
\providecommand \BibitemOpen [0]{}%
\providecommand \bibitemStop [0]{}%
\providecommand \bibitemNoStop [0]{.\EOS\space}%
\providecommand \EOS [0]{\spacefactor3000\relax}%
\providecommand \BibitemShut  [1]{\csname bibitem#1\endcsname}%
\let\auto@bib@innerbib\@empty
\bibitem [{\citenamefont {Graver}\ \emph {et~al.}(1993)\citenamefont {Graver}, \citenamefont {Servatius},\ and\ \citenamefont {Servatius}}]{combinatorial_rigidity}%
  \BibitemOpen
  \bibfield  {author} {\bibinfo {author} {\bibfnamefont {J.}~\bibnamefont {Graver}}, \bibinfo {author} {\bibfnamefont {B.}~\bibnamefont {Servatius}},\ and\ \bibinfo {author} {\bibfnamefont {H.}~\bibnamefont {Servatius}},\ }\href {https://books.google.fr/books?id=Q0dECQAAQBAJ} {\emph {\bibinfo {title} {Combinatorial Rigidity}}},\ Graduate studies in mathematics\ (\bibinfo  {publisher} {American Mathematical Society},\ \bibinfo {year} {1993})\BibitemShut {NoStop}%
\bibitem [{\citenamefont {Holroyd}(1998)}]{holroyd98}%
  \BibitemOpen
  \bibfield  {author} {\bibinfo {author} {\bibfnamefont {A.~E.}\ \bibnamefont {Holroyd}},\ }\bibfield  {title} {\bibinfo {title} {{Existence and uniqueness of infinite components in generic rigidity percolation}},\ }\href {https://doi.org/10.1214/aoap/1028903458} {\bibfield  {journal} {\bibinfo  {journal} {The Annals of Applied Probability}\ }\textbf {\bibinfo {volume} {8}},\ \bibinfo {pages} {944 } (\bibinfo {year} {1998})}\BibitemShut {NoStop}%
\bibitem [{\citenamefont {Gluck}(1975)}]{gluck}%
  \BibitemOpen
  \bibfield  {author} {\bibinfo {author} {\bibfnamefont {H.}~\bibnamefont {Gluck}},\ }\bibfield  {title} {\bibinfo {title} {Almost all simply connected closed surfaces are rigid},\ }in\ \href@noop {} {\emph {\bibinfo {booktitle} {Geometric Topology}}},\ \bibinfo {editor} {edited by\ \bibinfo {editor} {\bibfnamefont {L.~C.}\ \bibnamefont {Glaser}}\ and\ \bibinfo {editor} {\bibfnamefont {T.~B.}\ \bibnamefont {Rushing}}}\ (\bibinfo  {publisher} {Springer Berlin Heidelberg},\ \bibinfo {address} {Berlin, Heidelberg},\ \bibinfo {year} {1975})\ pp.\ \bibinfo {pages} {225--239}\BibitemShut {NoStop}%
\bibitem [{\citenamefont {Pollaczek-Geiringer}(1927)}]{Geiringer}%
  \BibitemOpen
  \bibfield  {author} {\bibinfo {author} {\bibfnamefont {H.}~\bibnamefont {Pollaczek-Geiringer}},\ }\bibfield  {title} {\bibinfo {title} {Über die gliederung ebener fachwerke},\ }\href {https://doi.org/https://doi.org/10.1002/zamm.19270070107} {\bibfield  {journal} {\bibinfo  {journal} {ZAMM - Journal of Applied Mathematics and Mechanics / Zeitschrift für Angewandte Mathematik und Mechanik}\ }\textbf {\bibinfo {volume} {7}},\ \bibinfo {pages} {58} (\bibinfo {year} {1927})},\ \Eprint {https://arxiv.org/abs/https://onlinelibrary.wiley.com/doi/pdf/10.1002/zamm.19270070107} {https://onlinelibrary.wiley.com/doi/pdf/10.1002/zamm.19270070107} \BibitemShut {NoStop}%
\bibitem [{\citenamefont {Laman}(1970)}]{Laman1970}%
  \BibitemOpen
  \bibfield  {author} {\bibinfo {author} {\bibfnamefont {G.}~\bibnamefont {Laman}},\ }\bibfield  {title} {\bibinfo {title} {On graphs and rigidity of plane skeletal structures},\ }\href {https://doi.org/10.1007/BF01534980} {\bibfield  {journal} {\bibinfo  {journal} {Journal of Engineering Mathematics}\ }\textbf {\bibinfo {volume} {4}},\ \bibinfo {pages} {331} (\bibinfo {year} {1970})}\BibitemShut {NoStop}%
\bibitem [{\citenamefont {Jacobs}\ and\ \citenamefont {Thorpe}(1995)}]{JacobsThorpe1995}%
  \BibitemOpen
  \bibfield  {author} {\bibinfo {author} {\bibfnamefont {D.~J.}\ \bibnamefont {Jacobs}}\ and\ \bibinfo {author} {\bibfnamefont {M.~F.}\ \bibnamefont {Thorpe}},\ }\bibfield  {title} {\bibinfo {title} {Generic rigidity percolation: The pebble game},\ }\href {https://doi.org/10.1103/PhysRevLett.75.4051} {\bibfield  {journal} {\bibinfo  {journal} {Phys. Rev. Lett.}\ }\textbf {\bibinfo {volume} {75}},\ \bibinfo {pages} {4051} (\bibinfo {year} {1995})}\BibitemShut {NoStop}%
\bibitem [{\citenamefont {Jacobs}\ and\ \citenamefont {Hendrickson}(1997)}]{Jacobs1997An}%
  \BibitemOpen
  \bibfield  {author} {\bibinfo {author} {\bibfnamefont {D.~J.}\ \bibnamefont {Jacobs}}\ and\ \bibinfo {author} {\bibfnamefont {B.}~\bibnamefont {Hendrickson}},\ }\bibfield  {title} {\bibinfo {title} {An algorithm for two-dimensional rigidity percolation: The pebble game},\ }\href {https://doi.org/https://doi.org/10.1006/jcph.1997.5809} {\bibfield  {journal} {\bibinfo  {journal} {Journal of Computational Physics}\ }\textbf {\bibinfo {volume} {137}},\ \bibinfo {pages} {346} (\bibinfo {year} {1997})}\BibitemShut {NoStop}%
\bibitem [{\citenamefont {Notarmuzi}\ and\ \citenamefont {Javerzat}(2025)}]{GitProject}%
  \BibitemOpen
  \bibfield  {author} {\bibinfo {author} {\bibfnamefont {D.}~\bibnamefont {Notarmuzi}}\ and\ \bibinfo {author} {\bibfnamefont {N.}~\bibnamefont {Javerzat}},\ }\href {https://doi.org/10.5281/zenodo.15584520} {\bibinfo {title} {10.5281/zenodo.15584520}},\ \bibinfo {howpublished} {\url{https://github.com/NinaJaverzat/NZRP/tree/main}} (\bibinfo {year} {2025}),\ \bibinfo {note} {simulation code}\BibitemShut {NoStop}%
\bibitem [{\citenamefont {Newman}\ and\ \citenamefont {Ziff}(2000)}]{Newman2000Efficient}%
  \BibitemOpen
  \bibfield  {author} {\bibinfo {author} {\bibfnamefont {M.~E.~J.}\ \bibnamefont {Newman}}\ and\ \bibinfo {author} {\bibfnamefont {R.~M.}\ \bibnamefont {Ziff}},\ }\bibfield  {title} {\bibinfo {title} {Efficient monte carlo algorithm and high-precision results for percolation},\ }\href {https://doi.org/10.1103/PhysRevLett.85.4104} {\bibfield  {journal} {\bibinfo  {journal} {Phys. Rev. Lett.}\ }\textbf {\bibinfo {volume} {85}},\ \bibinfo {pages} {4104} (\bibinfo {year} {2000})}\BibitemShut {NoStop}%
\bibitem [{\citenamefont {Newman}\ and\ \citenamefont {Ziff}(2001)}]{NZ2001}%
  \BibitemOpen
  \bibfield  {author} {\bibinfo {author} {\bibfnamefont {M.~E.~J.}\ \bibnamefont {Newman}}\ and\ \bibinfo {author} {\bibfnamefont {R.~M.}\ \bibnamefont {Ziff}},\ }\bibfield  {title} {\bibinfo {title} {Fast monte carlo algorithm for site or bond percolation},\ }\href {https://doi.org/10.1103/PhysRevE.64.016706} {\bibfield  {journal} {\bibinfo  {journal} {Phys. Rev. E}\ }\textbf {\bibinfo {volume} {64}},\ \bibinfo {pages} {016706} (\bibinfo {year} {2001})}\BibitemShut {NoStop}%
\end{thebibliography}%


\begin{thebibliography}{70}%
\makeatletter
\providecommand \@ifxundefined [1]{%
 \@ifx{#1\undefined}
}%
\providecommand \@ifnum [1]{%
 \ifnum #1\expandafter \@firstoftwo
 \else \expandafter \@secondoftwo
 \fi
}%
\providecommand \@ifx [1]{%
 \ifx #1\expandafter \@firstoftwo
 \else \expandafter \@secondoftwo
 \fi
}%
\providecommand \natexlab [1]{#1}%
\providecommand \enquote  [1]{``#1''}%
\providecommand \bibnamefont  [1]{#1}%
\providecommand \bibfnamefont [1]{#1}%
\providecommand \citenamefont [1]{#1}%
\providecommand \href@noop [0]{\@secondoftwo}%
\providecommand \href [0]{\begingroup \@sanitize@url \@href}%
\providecommand \@href[1]{\@@startlink{#1}\@@href}%
\providecommand \@@href[1]{\endgroup#1\@@endlink}%
\providecommand \@sanitize@url [0]{\catcode `\\12\catcode `\$12\catcode `\&12\catcode `\#12\catcode `\^12\catcode `\_12\catcode `\%12\relax}%
\providecommand \@@startlink[1]{}%
\providecommand \@@endlink[0]{}%
\providecommand \url  [0]{\begingroup\@sanitize@url \@url }%
\providecommand \@url [1]{\endgroup\@href {#1}{\urlprefix }}%
\providecommand \urlprefix  [0]{URL }%
\providecommand \Eprint [0]{\href }%
\providecommand \doibase [0]{https://doi.org/}%
\providecommand \selectlanguage [0]{\@gobble}%
\providecommand \bibinfo  [0]{\@secondoftwo}%
\providecommand \bibfield  [0]{\@secondoftwo}%
\providecommand \translation [1]{[#1]}%
\providecommand \BibitemOpen [0]{}%
\providecommand \bibitemStop [0]{}%
\providecommand \bibitemNoStop [0]{.\EOS\space}%
\providecommand \EOS [0]{\spacefactor3000\relax}%
\providecommand \BibitemShut  [1]{\csname bibitem#1\endcsname}%
\let\auto@bib@innerbib\@empty
\bibitem [{\citenamefont {Stauffer}\ and\ \citenamefont {Aharony}(1992)}]{stauffer_aharony}%
  \BibitemOpen
  \bibfield  {author} {\bibinfo {author} {\bibfnamefont {D.}~\bibnamefont {Stauffer}}\ and\ \bibinfo {author} {\bibfnamefont {A.}~\bibnamefont {Aharony}},\ }\href {https://www.taylorfrancis.com/books/mono/10.1201/9781315274386/introduction-percolation-theory-ammon-aharony-dietrich-stauffer} {\emph {\bibinfo {title} {Introduction To Percolation Theory}}}\ (\bibinfo  {publisher} {Taylor I\& Francis},\ \bibinfo {year} {1992})\BibitemShut {NoStop}%
\bibitem [{\citenamefont {Duminil-Copin}(2019)}]{duminil_copin2018}%
  \BibitemOpen
  \bibfield  {author} {\bibinfo {author} {\bibfnamefont {H.}~\bibnamefont {Duminil-Copin}},\ }\bibinfo {title} {Sixty years of percolation},\ in\ \href {https://doi.org/10.1142/9789813272880\_0162} {\emph {\bibinfo {booktitle} {Proceedings of the International Congress of Mathematicians (ICM 2018)}}}\ (\bibinfo  {publisher} {World Scientific},\ \bibinfo {year} {2019})\ pp.\ \bibinfo {pages} {2829--2856},\ \Eprint {https://arxiv.org/abs/https://arxiv.org/abs/1712.04651} {https://arxiv.org/abs/1712.04651} \BibitemShut {NoStop}%
\bibitem [{\citenamefont {and}(1864)}]{Maxwell1864}%
  \BibitemOpen
  \bibfield  {author} {\bibinfo {author} {\bibfnamefont {J.~C.~M.}\ \bibnamefont {and}},\ }\bibfield  {title} {\bibinfo {title} {L. on the calculation of the equilibrium and stiffness of frames},\ }\href {https://doi.org/10.1080/14786446408643668} {\bibfield  {journal} {\bibinfo  {journal} {The London, Edinburgh, and Dublin Philosophical Magazine and Journal of Science}\ }\textbf {\bibinfo {volume} {27}},\ \bibinfo {pages} {294} (\bibinfo {year} {1864})},\ \Eprint {https://arxiv.org/abs/https://doi.org/10.1080/14786446408643668} {https://doi.org/10.1080/14786446408643668} \BibitemShut {NoStop}%
\bibitem [{\citenamefont {Laman}(1970)}]{Laman1970}%
  \BibitemOpen
  \bibfield  {author} {\bibinfo {author} {\bibfnamefont {G.}~\bibnamefont {Laman}},\ }\bibfield  {title} {\bibinfo {title} {On graphs and rigidity of plane skeletal structures},\ }\href {https://doi.org/10.1007/BF01534980} {\bibfield  {journal} {\bibinfo  {journal} {Journal of Engineering Mathematics}\ }\textbf {\bibinfo {volume} {4}},\ \bibinfo {pages} {331} (\bibinfo {year} {1970})}\BibitemShut {NoStop}%
\bibitem [{\citenamefont {Jacobs}\ and\ \citenamefont {Thorpe}(1995)}]{JacobsThorpe1995}%
  \BibitemOpen
  \bibfield  {author} {\bibinfo {author} {\bibfnamefont {D.~J.}\ \bibnamefont {Jacobs}}\ and\ \bibinfo {author} {\bibfnamefont {M.~F.}\ \bibnamefont {Thorpe}},\ }\bibfield  {title} {\bibinfo {title} {Generic rigidity percolation: The pebble game},\ }\href {https://doi.org/10.1103/PhysRevLett.75.4051} {\bibfield  {journal} {\bibinfo  {journal} {Phys. Rev. Lett.}\ }\textbf {\bibinfo {volume} {75}},\ \bibinfo {pages} {4051} (\bibinfo {year} {1995})}\BibitemShut {NoStop}%
\bibitem [{\citenamefont {Zhang}\ \emph {et~al.}(2019)\citenamefont {Zhang}, \citenamefont {Zhang}, \citenamefont {Bouzid}, \citenamefont {Rocklin}, \citenamefont {Del~Gado},\ and\ \citenamefont {Mao}}]{mehdi2019}%
  \BibitemOpen
  \bibfield  {author} {\bibinfo {author} {\bibfnamefont {S.}~\bibnamefont {Zhang}}, \bibinfo {author} {\bibfnamefont {L.}~\bibnamefont {Zhang}}, \bibinfo {author} {\bibfnamefont {M.}~\bibnamefont {Bouzid}}, \bibinfo {author} {\bibfnamefont {D.~Z.}\ \bibnamefont {Rocklin}}, \bibinfo {author} {\bibfnamefont {E.}~\bibnamefont {Del~Gado}},\ and\ \bibinfo {author} {\bibfnamefont {X.}~\bibnamefont {Mao}},\ }\bibfield  {title} {\bibinfo {title} {Correlated rigidity percolation and colloidal gels},\ }\href {https://doi.org/10.1103/PhysRevLett.123.058001} {\bibfield  {journal} {\bibinfo  {journal} {Phys. Rev. Lett.}\ }\textbf {\bibinfo {volume} {123}},\ \bibinfo {pages} {058001} (\bibinfo {year} {2019})}\BibitemShut {NoStop}%
\bibitem [{\citenamefont {Tsurusawa}\ \emph {et~al.}(2019)\citenamefont {Tsurusawa}, \citenamefont {Leocmach}, \citenamefont {Russo},\ and\ \citenamefont {Tanaka}}]{Tsurusawa2019Direct}%
  \BibitemOpen
  \bibfield  {author} {\bibinfo {author} {\bibfnamefont {H.}~\bibnamefont {Tsurusawa}}, \bibinfo {author} {\bibfnamefont {M.}~\bibnamefont {Leocmach}}, \bibinfo {author} {\bibfnamefont {J.}~\bibnamefont {Russo}},\ and\ \bibinfo {author} {\bibfnamefont {H.}~\bibnamefont {Tanaka}},\ }\bibfield  {title} {\bibinfo {title} {Direct link between mechanical stability in gels and percolation of isostatic particles},\ }\href {https://doi.org/10.1126/sciadv.aav6090} {\bibfield  {journal} {\bibinfo  {journal} {Science Advances}\ }\textbf {\bibinfo {volume} {5}},\ \bibinfo {pages} {eaav6090} (\bibinfo {year} {2019})},\ \Eprint {https://arxiv.org/abs/https://www.science.org/doi/pdf/10.1126/sciadv.aav6090} {https://www.science.org/doi/pdf/10.1126/sciadv.aav6090} \BibitemShut {NoStop}%
\bibitem [{\citenamefont {Fenton}\ \emph {et~al.}(2023)\citenamefont {Fenton}, \citenamefont {Padmanabhan}, \citenamefont {Ryu}, \citenamefont {Nguyen}, \citenamefont {Zia},\ and\ \citenamefont {Helgeson}}]{Fenton2023Minimal}%
  \BibitemOpen
  \bibfield  {author} {\bibinfo {author} {\bibfnamefont {S.~M.}\ \bibnamefont {Fenton}}, \bibinfo {author} {\bibfnamefont {P.}~\bibnamefont {Padmanabhan}}, \bibinfo {author} {\bibfnamefont {B.~K.}\ \bibnamefont {Ryu}}, \bibinfo {author} {\bibfnamefont {T.~T.~D.}\ \bibnamefont {Nguyen}}, \bibinfo {author} {\bibfnamefont {R.~N.}\ \bibnamefont {Zia}},\ and\ \bibinfo {author} {\bibfnamefont {M.~E.}\ \bibnamefont {Helgeson}},\ }\bibfield  {title} {\bibinfo {title} {Minimal conditions for solidification and thermal processing of colloidal gels},\ }\href {https://doi.org/10.1073/pnas.2215922120} {\bibfield  {journal} {\bibinfo  {journal} {Proceedings of the National Academy of Sciences}\ }\textbf {\bibinfo {volume} {120}},\ \bibinfo {pages} {e2215922120} (\bibinfo {year} {2023})},\ \Eprint {https://arxiv.org/abs/https://www.pnas.org/doi/pdf/10.1073/pnas.2215922120} {https://www.pnas.org/doi/pdf/10.1073/pnas.2215922120} \BibitemShut {NoStop}%
\bibitem [{\citenamefont {Petridou}\ \emph {et~al.}(2021)\citenamefont {Petridou}, \citenamefont {Corominas-Murtra}, \citenamefont {Heisenberg},\ and\ \citenamefont {Hannezo}}]{Petridou2021}%
  \BibitemOpen
  \bibfield  {author} {\bibinfo {author} {\bibfnamefont {N.~I.}\ \bibnamefont {Petridou}}, \bibinfo {author} {\bibfnamefont {B.}~\bibnamefont {Corominas-Murtra}}, \bibinfo {author} {\bibfnamefont {C.-P.}\ \bibnamefont {Heisenberg}},\ and\ \bibinfo {author} {\bibfnamefont {E.}~\bibnamefont {Hannezo}},\ }\bibfield  {title} {\bibinfo {title} {Rigidity percolation uncovers a structural basis for embryonic tissue phase transitions},\ }\href {https://doi.org/https://doi.org/10.1016/j.cell.2021.02.017} {\bibfield  {journal} {\bibinfo  {journal} {Cell}\ }\textbf {\bibinfo {volume} {184}},\ \bibinfo {pages} {1914} (\bibinfo {year} {2021})}\BibitemShut {NoStop}%
\bibitem [{\citenamefont {Hannezo}\ and\ \citenamefont {Heisenberg}(2022)}]{Hannezo2022Rigidity}%
  \BibitemOpen
  \bibfield  {author} {\bibinfo {author} {\bibfnamefont {E.}~\bibnamefont {Hannezo}}\ and\ \bibinfo {author} {\bibfnamefont {C.-P.}\ \bibnamefont {Heisenberg}},\ }\bibfield  {title} {\bibinfo {title} {Rigidity transitions in development and disease},\ }\href {https://doi.org/10.1016/j.tcb.2021.12.006} {\bibfield  {journal} {\bibinfo  {journal} {Trends in Cell Biology}\ }\textbf {\bibinfo {volume} {32}},\ \bibinfo {pages} {433} (\bibinfo {year} {2022})},\ \bibinfo {note} {doi: 10.1016/j.tcb.2021.12.006}\BibitemShut {NoStop}%
\bibitem [{\citenamefont {Lenne}\ and\ \citenamefont {Trivedi}(2022{\natexlab{a}})}]{Lenne2022Biological}%
  \BibitemOpen
  \bibfield  {author} {\bibinfo {author} {\bibfnamefont {P.-F.}\ \bibnamefont {Lenne}}\ and\ \bibinfo {author} {\bibfnamefont {V.}~\bibnamefont {Trivedi}},\ }\bibfield  {title} {\bibinfo {title} {Sculpting tissues by phase transitions},\ }\href {https://doi.org/10.1038/s41467-022-28151-9} {\bibfield  {journal} {\bibinfo  {journal} {Nature Communications}\ }\textbf {\bibinfo {volume} {13}},\ \bibinfo {pages} {664} (\bibinfo {year} {2022}{\natexlab{a}})}\BibitemShut {NoStop}%
\bibitem [{\citenamefont {Manning}(2024)}]{Manning2024}%
  \BibitemOpen
  \bibfield  {author} {\bibinfo {author} {\bibfnamefont {M.~L.}\ \bibnamefont {Manning}},\ }\bibfield  {title} {\bibinfo {title} {Rigidity in mechanical biological networks},\ }\href {https://doi.org/10.1016/j.cub.2024.07.014} {\bibfield  {journal} {\bibinfo  {journal} {Current Biology}\ }\textbf {\bibinfo {volume} {34}},\ \bibinfo {pages} {R1024} (\bibinfo {year} {2024})}\BibitemShut {NoStop}%
\bibitem [{\citenamefont {Broedersz}\ \emph {et~al.}(2011)\citenamefont {Broedersz}, \citenamefont {Mao}, \citenamefont {Lubensky},\ and\ \citenamefont {MacKintosh}}]{Broedersz2011}%
  \BibitemOpen
  \bibfield  {author} {\bibinfo {author} {\bibfnamefont {C.~P.}\ \bibnamefont {Broedersz}}, \bibinfo {author} {\bibfnamefont {X.}~\bibnamefont {Mao}}, \bibinfo {author} {\bibfnamefont {T.~C.}\ \bibnamefont {Lubensky}},\ and\ \bibinfo {author} {\bibfnamefont {F.~C.}\ \bibnamefont {MacKintosh}},\ }\bibfield  {title} {\bibinfo {title} {Criticality and isostaticity in fibre networks},\ }\href {https://doi.org/10.1038/nphys2127} {\bibfield  {journal} {\bibinfo  {journal} {Nature Physics}\ }\textbf {\bibinfo {volume} {7}},\ \bibinfo {pages} {983} (\bibinfo {year} {2011})}\BibitemShut {NoStop}%
\bibitem [{\citenamefont {Bolton}\ and\ \citenamefont {Weaire}(1990)}]{Bolton1990Rigidity}%
  \BibitemOpen
  \bibfield  {author} {\bibinfo {author} {\bibfnamefont {F.}~\bibnamefont {Bolton}}\ and\ \bibinfo {author} {\bibfnamefont {D.}~\bibnamefont {Weaire}},\ }\bibfield  {title} {\bibinfo {title} {Rigidity loss transition in a disordered 2d froth},\ }\href {https://doi.org/10.1103/PhysRevLett.65.3449} {\bibfield  {journal} {\bibinfo  {journal} {Phys. Rev. Lett.}\ }\textbf {\bibinfo {volume} {65}},\ \bibinfo {pages} {3449} (\bibinfo {year} {1990})}\BibitemShut {NoStop}%
\bibitem [{\citenamefont {Henkes}\ \emph {et~al.}(2016)\citenamefont {Henkes}, \citenamefont {Quint}, \citenamefont {Fily},\ and\ \citenamefont {Schwarz}}]{silke2016}%
  \BibitemOpen
  \bibfield  {author} {\bibinfo {author} {\bibfnamefont {S.}~\bibnamefont {Henkes}}, \bibinfo {author} {\bibfnamefont {D.~A.}\ \bibnamefont {Quint}}, \bibinfo {author} {\bibfnamefont {Y.}~\bibnamefont {Fily}},\ and\ \bibinfo {author} {\bibfnamefont {J.~M.}\ \bibnamefont {Schwarz}},\ }\bibfield  {title} {\bibinfo {title} {Rigid cluster decomposition reveals criticality in frictional jamming},\ }\href {https://doi.org/10.1103/PhysRevLett.116.028301} {\bibfield  {journal} {\bibinfo  {journal} {Phys. Rev. Lett.}\ }\textbf {\bibinfo {volume} {116}},\ \bibinfo {pages} {028301} (\bibinfo {year} {2016})}\BibitemShut {NoStop}%
\bibitem [{\citenamefont {Liu}\ \emph {et~al.}(2019)\citenamefont {Liu}, \citenamefont {Henkes},\ and\ \citenamefont {Schwarz}}]{Liu2019}%
  \BibitemOpen
  \bibfield  {author} {\bibinfo {author} {\bibfnamefont {K.}~\bibnamefont {Liu}}, \bibinfo {author} {\bibfnamefont {S.}~\bibnamefont {Henkes}},\ and\ \bibinfo {author} {\bibfnamefont {J.~M.}\ \bibnamefont {Schwarz}},\ }\bibfield  {title} {\bibinfo {title} {Frictional rigidity percolation: A new universality class and its superuniversal connections through minimal rigidity proliferation},\ }\href {https://doi.org/10.1103/PhysRevX.9.021006} {\bibfield  {journal} {\bibinfo  {journal} {Phys. Rev. X}\ }\textbf {\bibinfo {volume} {9}},\ \bibinfo {pages} {021006} (\bibinfo {year} {2019})}\BibitemShut {NoStop}%
\bibitem [{\citenamefont {Ellenbroek}\ \emph {et~al.}(2015{\natexlab{a}})\citenamefont {Ellenbroek}, \citenamefont {Hagh}, \citenamefont {Kumar}, \citenamefont {Thorpe},\ and\ \citenamefont {van Hecke}}]{Ellenbroek2015Rigidity}%
  \BibitemOpen
  \bibfield  {author} {\bibinfo {author} {\bibfnamefont {W.~G.}\ \bibnamefont {Ellenbroek}}, \bibinfo {author} {\bibfnamefont {V.~F.}\ \bibnamefont {Hagh}}, \bibinfo {author} {\bibfnamefont {A.}~\bibnamefont {Kumar}}, \bibinfo {author} {\bibfnamefont {M.~F.}\ \bibnamefont {Thorpe}},\ and\ \bibinfo {author} {\bibfnamefont {M.}~\bibnamefont {van Hecke}},\ }\bibfield  {title} {\bibinfo {title} {Rigidity loss in disordered systems: Three scenarios},\ }\href {https://doi.org/10.1103/PhysRevLett.114.135501} {\bibfield  {journal} {\bibinfo  {journal} {Phys. Rev. Lett.}\ }\textbf {\bibinfo {volume} {114}},\ \bibinfo {pages} {135501} (\bibinfo {year} {2015}{\natexlab{a}})}\BibitemShut {NoStop}%
\bibitem [{\citenamefont {Berthier}\ \emph {et~al.}(2019{\natexlab{a}})\citenamefont {Berthier}, \citenamefont {Kollmer}, \citenamefont {Henkes}, \citenamefont {Liu}, \citenamefont {Schwarz},\ and\ \citenamefont {Daniels}}]{Berthier2019Rigidity}%
  \BibitemOpen
  \bibfield  {author} {\bibinfo {author} {\bibfnamefont {E.}~\bibnamefont {Berthier}}, \bibinfo {author} {\bibfnamefont {J.~E.}\ \bibnamefont {Kollmer}}, \bibinfo {author} {\bibfnamefont {S.~E.}\ \bibnamefont {Henkes}}, \bibinfo {author} {\bibfnamefont {K.}~\bibnamefont {Liu}}, \bibinfo {author} {\bibfnamefont {J.~M.}\ \bibnamefont {Schwarz}},\ and\ \bibinfo {author} {\bibfnamefont {K.~E.}\ \bibnamefont {Daniels}},\ }\bibfield  {title} {\bibinfo {title} {Rigidity percolation control of the brittle-ductile transition in disordered networks},\ }\href {https://doi.org/10.1103/PhysRevMaterials.3.075602} {\bibfield  {journal} {\bibinfo  {journal} {Phys. Rev. Mater.}\ }\textbf {\bibinfo {volume} {3}},\ \bibinfo {pages} {075602} (\bibinfo {year} {2019}{\natexlab{a}})}\BibitemShut {NoStop}%
\bibitem [{\citenamefont {Bantawa}\ \emph {et~al.}(2023)\citenamefont {Bantawa}, \citenamefont {Keshavarz}, \citenamefont {Geri}, \citenamefont {Bouzid}, \citenamefont {Divoux}, \citenamefont {McKinley},\ and\ \citenamefont {Del~Gado}}]{Bantawa2023}%
  \BibitemOpen
  \bibfield  {author} {\bibinfo {author} {\bibfnamefont {M.}~\bibnamefont {Bantawa}}, \bibinfo {author} {\bibfnamefont {B.}~\bibnamefont {Keshavarz}}, \bibinfo {author} {\bibfnamefont {M.}~\bibnamefont {Geri}}, \bibinfo {author} {\bibfnamefont {M.}~\bibnamefont {Bouzid}}, \bibinfo {author} {\bibfnamefont {T.}~\bibnamefont {Divoux}}, \bibinfo {author} {\bibfnamefont {G.~H.}\ \bibnamefont {McKinley}},\ and\ \bibinfo {author} {\bibfnamefont {E.}~\bibnamefont {Del~Gado}},\ }\bibfield  {title} {\bibinfo {title} {The hidden hierarchical nature of soft particulate gels},\ }\href {https://doi.org/10.1038/s41567-023-01988-7} {\bibfield  {journal} {\bibinfo  {journal} {Nature Physics}\ }\textbf {\bibinfo {volume} {19}},\ \bibinfo {pages} {1178} (\bibinfo {year} {2023})}\BibitemShut {NoStop}%
\bibitem [{\citenamefont {Richard}\ and\ \citenamefont {Bouzid}(2025)}]{richard2025}%
  \BibitemOpen
  \bibfield  {author} {\bibinfo {author} {\bibfnamefont {D.}~\bibnamefont {Richard}}\ and\ \bibinfo {author} {\bibfnamefont {M.}~\bibnamefont {Bouzid}},\ }\href {https://arxiv.org/abs/2504.19568} {\bibinfo {title} {How rigidity percolation and bending stiffness shape colloidal gel elasticity}} (\bibinfo {year} {2025}),\ \Eprint {https://arxiv.org/abs/2504.19568} {arXiv:2504.19568 [cond-mat.soft]} \BibitemShut {NoStop}%
\bibitem [{\citenamefont {Berthier}\ \emph {et~al.}(2019{\natexlab{b}})\citenamefont {Berthier}, \citenamefont {Kollmer}, \citenamefont {Henkes}, \citenamefont {Liu}, \citenamefont {Schwarz},\ and\ \citenamefont {Daniels}}]{silke2019}%
  \BibitemOpen
  \bibfield  {author} {\bibinfo {author} {\bibfnamefont {E.}~\bibnamefont {Berthier}}, \bibinfo {author} {\bibfnamefont {J.~E.}\ \bibnamefont {Kollmer}}, \bibinfo {author} {\bibfnamefont {S.~E.}\ \bibnamefont {Henkes}}, \bibinfo {author} {\bibfnamefont {K.}~\bibnamefont {Liu}}, \bibinfo {author} {\bibfnamefont {J.~M.}\ \bibnamefont {Schwarz}},\ and\ \bibinfo {author} {\bibfnamefont {K.~E.}\ \bibnamefont {Daniels}},\ }\bibfield  {title} {\bibinfo {title} {Rigidity percolation control of the brittle-ductile transition in disordered networks},\ }\href {https://doi.org/10.1103/PhysRevMaterials.3.075602} {\bibfield  {journal} {\bibinfo  {journal} {Phys. Rev. Mater.}\ }\textbf {\bibinfo {volume} {3}},\ \bibinfo {pages} {075602} (\bibinfo {year} {2019}{\natexlab{b}})}\BibitemShut {NoStop}%
\bibitem [{\citenamefont {Goyal}\ \emph {et~al.}(2024)\citenamefont {Goyal}, \citenamefont {Martys},\ and\ \citenamefont {Del~Gado}}]{emanuelaDST}%
  \BibitemOpen
  \bibfield  {author} {\bibinfo {author} {\bibfnamefont {A.}~\bibnamefont {Goyal}}, \bibinfo {author} {\bibfnamefont {N.~S.}\ \bibnamefont {Martys}},\ and\ \bibinfo {author} {\bibfnamefont {E.}~\bibnamefont {Del~Gado}},\ }\bibfield  {title} {\bibinfo {title} {Flow induced rigidity percolation in shear thickening suspensions},\ }\href {https://doi.org/10.1122/8.0000786} {\bibfield  {journal} {\bibinfo  {journal} {Journal of Rheology}\ }\textbf {\bibinfo {volume} {68}},\ \bibinfo {pages} {219} (\bibinfo {year} {2024})},\ \Eprint {https://arxiv.org/abs/https://pubs.aip.org/sor/jor/article-pdf/68/2/219/19504709/219\_1\_8.0000786.pdf} {https://pubs.aip.org/sor/jor/article-pdf/68/2/219/19504709/219\_1\_8.0000786.pdf} \BibitemShut {NoStop}%
\bibitem [{\citenamefont {Moukarzel}\ and\ \citenamefont {Duxbury}(1999)}]{Moukarzel1999}%
  \BibitemOpen
  \bibfield  {author} {\bibinfo {author} {\bibfnamefont {C.}~\bibnamefont {Moukarzel}}\ and\ \bibinfo {author} {\bibfnamefont {P.~M.}\ \bibnamefont {Duxbury}},\ }\bibfield  {title} {\bibinfo {title} {Comparison of rigidity and connectivity percolation in two dimensions},\ }\href {https://doi.org/10.1103/PhysRevE.59.2614} {\bibfield  {journal} {\bibinfo  {journal} {Phys. Rev. E}\ }\textbf {\bibinfo {volume} {59}},\ \bibinfo {pages} {2614} (\bibinfo {year} {1999})}\BibitemShut {NoStop}%
\bibitem [{\citenamefont {Holroyd}(1998)}]{holroyd98}%
  \BibitemOpen
  \bibfield  {author} {\bibinfo {author} {\bibfnamefont {A.~E.}\ \bibnamefont {Holroyd}},\ }\bibfield  {title} {\bibinfo {title} {{Existence and uniqueness of infinite components in generic rigidity percolation}},\ }\href {https://doi.org/10.1214/aoap/1028903458} {\bibfield  {journal} {\bibinfo  {journal} {The Annals of Applied Probability}\ }\textbf {\bibinfo {volume} {8}},\ \bibinfo {pages} {944 } (\bibinfo {year} {1998})}\BibitemShut {NoStop}%
\bibitem [{\citenamefont {H{\"a}ggstr{\"o}m}(2003)}]{Haggstrom2003}%
  \BibitemOpen
  \bibfield  {author} {\bibinfo {author} {\bibfnamefont {O.}~\bibnamefont {H{\"a}ggstr{\"o}m}},\ }\bibfield  {title} {\bibinfo {title} {Uniqueness of infinite rigid components in percolation models: the case of nonplanar lattices},\ }\href {https://doi.org/10.1007/s00440-003-0290-2} {\bibfield  {journal} {\bibinfo  {journal} {Probability Theory and Related Fields}\ }\textbf {\bibinfo {volume} {127}},\ \bibinfo {pages} {513} (\bibinfo {year} {2003})}\BibitemShut {NoStop}%
\bibitem [{\citenamefont {Barr\'e}(2009)}]{barre2009}%
  \BibitemOpen
  \bibfield  {author} {\bibinfo {author} {\bibfnamefont {J.}~\bibnamefont {Barr\'e}},\ }\bibfield  {title} {\bibinfo {title} {Hierarchical models of rigidity percolation},\ }\href {https://doi.org/10.1103/PhysRevE.80.061108} {\bibfield  {journal} {\bibinfo  {journal} {Phys. Rev. E}\ }\textbf {\bibinfo {volume} {80}},\ \bibinfo {pages} {061108} (\bibinfo {year} {2009})}\BibitemShut {NoStop}%
\bibitem [{\citenamefont {Jord\'{a}n}\ and\ \citenamefont {Tanigawa}(2022)}]{jordan2022}%
  \BibitemOpen
  \bibfield  {author} {\bibinfo {author} {\bibfnamefont {T.}~\bibnamefont {Jord\'{a}n}}\ and\ \bibinfo {author} {\bibfnamefont {S.-i.}\ \bibnamefont {Tanigawa}},\ }\bibfield  {title} {\bibinfo {title} {Rigidity of random subgraphs and eigenvalues of stiffness matrices},\ }\href {https://doi.org/10.1137/20M1349849} {\bibfield  {journal} {\bibinfo  {journal} {SIAM Journal on Discrete Mathematics}\ }\textbf {\bibinfo {volume} {36}},\ \bibinfo {pages} {2367} (\bibinfo {year} {2022})},\ \Eprint {https://arxiv.org/abs/https://doi.org/10.1137/20M1349849} {https://doi.org/10.1137/20M1349849} \BibitemShut {NoStop}%
\bibitem [{\citenamefont {Barr{\'e}}\ \emph {et~al.}(2005)\citenamefont {Barr{\'e}}, \citenamefont {Bishop}, \citenamefont {Lookman},\ and\ \citenamefont {Saxena}}]{Barre2005}%
  \BibitemOpen
  \bibfield  {author} {\bibinfo {author} {\bibfnamefont {J.}~\bibnamefont {Barr{\'e}}}, \bibinfo {author} {\bibfnamefont {A.~R.}\ \bibnamefont {Bishop}}, \bibinfo {author} {\bibfnamefont {T.}~\bibnamefont {Lookman}},\ and\ \bibinfo {author} {\bibfnamefont {A.}~\bibnamefont {Saxena}},\ }\bibfield  {title} {\bibinfo {title} {The cavity method for the rigidity transition},\ }\href {https://doi.org/10.1007/s10955-004-2709-2} {\bibfield  {journal} {\bibinfo  {journal} {Journal of Statistical Physics}\ }\textbf {\bibinfo {volume} {118}},\ \bibinfo {pages} {1057} (\bibinfo {year} {2005})}\BibitemShut {NoStop}%
\bibitem [{\citenamefont {Rivoire}\ and\ \citenamefont {Barr\'e}(2006)}]{Barre2006}%
  \BibitemOpen
  \bibfield  {author} {\bibinfo {author} {\bibfnamefont {O.}~\bibnamefont {Rivoire}}\ and\ \bibinfo {author} {\bibfnamefont {J.}~\bibnamefont {Barr\'e}},\ }\bibfield  {title} {\bibinfo {title} {Exactly solvable models of adaptive networks},\ }\href {https://doi.org/10.1103/PhysRevLett.97.148701} {\bibfield  {journal} {\bibinfo  {journal} {Phys. Rev. Lett.}\ }\textbf {\bibinfo {volume} {97}},\ \bibinfo {pages} {148701} (\bibinfo {year} {2006})}\BibitemShut {NoStop}%
\bibitem [{\citenamefont {Javerzat}\ and\ \citenamefont {Bouzid}(2023)}]{ninaconf}%
  \BibitemOpen
  \bibfield  {author} {\bibinfo {author} {\bibfnamefont {N.}~\bibnamefont {Javerzat}}\ and\ \bibinfo {author} {\bibfnamefont {M.}~\bibnamefont {Bouzid}},\ }\bibfield  {title} {\bibinfo {title} {Evidences of conformal invariance in 2d rigidity percolation},\ }\href {https://doi.org/10.1103/PhysRevLett.130.268201} {\bibfield  {journal} {\bibinfo  {journal} {Phys. Rev. Lett.}\ }\textbf {\bibinfo {volume} {130}},\ \bibinfo {pages} {268201} (\bibinfo {year} {2023})}\BibitemShut {NoStop}%
\bibitem [{\citenamefont {Javerzat}(2024)}]{ninasle}%
  \BibitemOpen
  \bibfield  {author} {\bibinfo {author} {\bibfnamefont {N.}~\bibnamefont {Javerzat}},\ }\bibfield  {title} {\bibinfo {title} {Schramm-loewner evolution in 2d rigidity percolation},\ }\href {https://doi.org/10.1103/PhysRevLett.132.018201} {\bibfield  {journal} {\bibinfo  {journal} {Phys. Rev. Lett.}\ }\textbf {\bibinfo {volume} {132}},\ \bibinfo {pages} {018201} (\bibinfo {year} {2024})}\BibitemShut {NoStop}%
\bibitem [{\citenamefont {Jacobs}\ and\ \citenamefont {Hendrickson}(1997)}]{Jacobs1997An}%
  \BibitemOpen
  \bibfield  {author} {\bibinfo {author} {\bibfnamefont {D.~J.}\ \bibnamefont {Jacobs}}\ and\ \bibinfo {author} {\bibfnamefont {B.}~\bibnamefont {Hendrickson}},\ }\bibfield  {title} {\bibinfo {title} {An algorithm for two-dimensional rigidity percolation: The pebble game},\ }\href {https://doi.org/https://doi.org/10.1006/jcph.1997.5809} {\bibfield  {journal} {\bibinfo  {journal} {Journal of Computational Physics}\ }\textbf {\bibinfo {volume} {137}},\ \bibinfo {pages} {346} (\bibinfo {year} {1997})}\BibitemShut {NoStop}%
\bibitem [{\citenamefont {Lester}\ and\ \citenamefont {Li}(2018)}]{LESTER2018225}%
  \BibitemOpen
  \bibfield  {author} {\bibinfo {author} {\bibfnamefont {D.}~\bibnamefont {Lester}}\ and\ \bibinfo {author} {\bibfnamefont {R.}~\bibnamefont {Li}},\ }\bibfield  {title} {\bibinfo {title} {The frictional pebble game: An algorithm for rigidity percolation in saturated frictional assemblies},\ }\href {https://doi.org/https://doi.org/10.1016/j.jcp.2018.05.016} {\bibfield  {journal} {\bibinfo  {journal} {Journal of Computational Physics}\ }\textbf {\bibinfo {volume} {369}},\ \bibinfo {pages} {225} (\bibinfo {year} {2018})}\BibitemShut {NoStop}%
\bibitem [{\citenamefont {Ziff}\ \emph {et~al.}(2011)\citenamefont {Ziff}, \citenamefont {Simmons},\ and\ \citenamefont {Kleban}}]{Ziff_2011}%
  \BibitemOpen
  \bibfield  {author} {\bibinfo {author} {\bibfnamefont {R.~M.}\ \bibnamefont {Ziff}}, \bibinfo {author} {\bibfnamefont {J.~J.~H.}\ \bibnamefont {Simmons}},\ and\ \bibinfo {author} {\bibfnamefont {P.}~\bibnamefont {Kleban}},\ }\bibfield  {title} {\bibinfo {title} {Factorization of correlations in two-dimensional percolation on the plane and torus},\ }\href {https://doi.org/10.1088/1751-8113/44/6/065002} {\bibfield  {journal} {\bibinfo  {journal} {Journal of Physics A: Mathematical and Theoretical}\ }\textbf {\bibinfo {volume} {44}},\ \bibinfo {pages} {065002} (\bibinfo {year} {2011})}\BibitemShut {NoStop}%
\bibitem [{Note1()}]{Note1}%
  \BibitemOpen
  \bibinfo {note} {In our, as well as earlier work \cite {JacobsThorpe1995}, bonds are activated in a random, uncorrelated way. Recent work has examined the effect of local correlations \cite {mehdi2019} but to our knowledge the effect of long-range correlations in the bond activation on the RP universality class has not been addressed. Our algorithm can be generalized to incorporate such correlations and we plan this for future work.}\BibitemShut {Stop}%
\bibitem [{\citenamefont {Dashti}\ \emph {et~al.}(2023)\citenamefont {Dashti}, \citenamefont {Saberi}, \citenamefont {Rahbari},\ and\ \citenamefont {Kurths}}]{sciadv2023}%
  \BibitemOpen
  \bibfield  {author} {\bibinfo {author} {\bibfnamefont {H.}~\bibnamefont {Dashti}}, \bibinfo {author} {\bibfnamefont {A.~A.}\ \bibnamefont {Saberi}}, \bibinfo {author} {\bibfnamefont {S.}~\bibnamefont {Rahbari}},\ and\ \bibinfo {author} {\bibfnamefont {J.}~\bibnamefont {Kurths}},\ }\bibfield  {title} {\bibinfo {title} {Emergence of rigidity percolation in flowing granular systems},\ }\href {https://doi.org/10.1126/sciadv.adh5586} {\bibfield  {journal} {\bibinfo  {journal} {Science Advances}\ }\textbf {\bibinfo {volume} {9}},\ \bibinfo {pages} {eadh5586} (\bibinfo {year} {2023})},\ \Eprint {https://arxiv.org/abs/https://www.science.org/doi/pdf/10.1126/sciadv.adh5586} {https://www.science.org/doi/pdf/10.1126/sciadv.adh5586} \BibitemShut {NoStop}%
\bibitem [{\citenamefont {Lois}\ \emph {et~al.}(2008)\citenamefont {Lois}, \citenamefont {Blawzdziewicz},\ and\ \citenamefont {O'Hern}}]{Lois2008}%
  \BibitemOpen
  \bibfield  {author} {\bibinfo {author} {\bibfnamefont {G.}~\bibnamefont {Lois}}, \bibinfo {author} {\bibfnamefont {J.}~\bibnamefont {Blawzdziewicz}},\ and\ \bibinfo {author} {\bibfnamefont {C.~S.}\ \bibnamefont {O'Hern}},\ }\bibfield  {title} {\bibinfo {title} {Jamming transition and new percolation universality classes in particulate systems with attraction},\ }\href {https://doi.org/10.1103/PhysRevLett.100.028001} {\bibfield  {journal} {\bibinfo  {journal} {Phys. Rev. Lett.}\ }\textbf {\bibinfo {volume} {100}},\ \bibinfo {pages} {028001} (\bibinfo {year} {2008})}\BibitemShut {NoStop}%
\bibitem [{\citenamefont {Latva-Kokko}\ \emph {et~al.}(2001)\citenamefont {Latva-Kokko}, \citenamefont {M\"akinen},\ and\ \citenamefont {Timonen}}]{latvakokko1}%
  \BibitemOpen
  \bibfield  {author} {\bibinfo {author} {\bibfnamefont {M.}~\bibnamefont {Latva-Kokko}}, \bibinfo {author} {\bibfnamefont {J.}~\bibnamefont {M\"akinen}},\ and\ \bibinfo {author} {\bibfnamefont {J.}~\bibnamefont {Timonen}},\ }\bibfield  {title} {\bibinfo {title} {Rigidity transition in two-dimensional random fiber networks},\ }\href {https://doi.org/10.1103/PhysRevE.63.046113} {\bibfield  {journal} {\bibinfo  {journal} {Phys. Rev. E}\ }\textbf {\bibinfo {volume} {63}},\ \bibinfo {pages} {046113} (\bibinfo {year} {2001})}\BibitemShut {NoStop}%
\bibitem [{\citenamefont {Latva-Kokko}\ and\ \citenamefont {Timonen}(2001)}]{latvakokko2}%
  \BibitemOpen
  \bibfield  {author} {\bibinfo {author} {\bibfnamefont {M.}~\bibnamefont {Latva-Kokko}}\ and\ \bibinfo {author} {\bibfnamefont {J.}~\bibnamefont {Timonen}},\ }\bibfield  {title} {\bibinfo {title} {Rigidity of random networks of stiff fibers in the low-density limit},\ }\href {https://doi.org/10.1103/PhysRevE.64.066117} {\bibfield  {journal} {\bibinfo  {journal} {Phys. Rev. E}\ }\textbf {\bibinfo {volume} {64}},\ \bibinfo {pages} {066117} (\bibinfo {year} {2001})}\BibitemShut {NoStop}%
\bibitem [{\citenamefont {Newman}\ and\ \citenamefont {Ziff}(2000)}]{Newman2000Efficient}%
  \BibitemOpen
  \bibfield  {author} {\bibinfo {author} {\bibfnamefont {M.~E.~J.}\ \bibnamefont {Newman}}\ and\ \bibinfo {author} {\bibfnamefont {R.~M.}\ \bibnamefont {Ziff}},\ }\bibfield  {title} {\bibinfo {title} {Efficient monte carlo algorithm and high-precision results for percolation},\ }\href {https://doi.org/10.1103/PhysRevLett.85.4104} {\bibfield  {journal} {\bibinfo  {journal} {Phys. Rev. Lett.}\ }\textbf {\bibinfo {volume} {85}},\ \bibinfo {pages} {4104} (\bibinfo {year} {2000})}\BibitemShut {NoStop}%
\bibitem [{\citenamefont {Notarmuzi}\ and\ \citenamefont {Javerzat}()}]{DN_prep}%
  \BibitemOpen
  \bibfield  {author} {\bibinfo {author} {\bibfnamefont {D.}~\bibnamefont {Notarmuzi}}\ and\ \bibinfo {author} {\bibfnamefont {N.}~\bibnamefont {Javerzat}},\ }\href@noop {} {\bibinfo {title} {in preparation}}\BibitemShut {NoStop}%
\bibitem [{\citenamefont {Machta}\ \emph {et~al.}(1996)\citenamefont {Machta}, \citenamefont {Choi}, \citenamefont {Lucke}, \citenamefont {Schweizer},\ and\ \citenamefont {Chayes}}]{Machta1996}%
  \BibitemOpen
  \bibfield  {author} {\bibinfo {author} {\bibfnamefont {J.}~\bibnamefont {Machta}}, \bibinfo {author} {\bibfnamefont {Y.~S.}\ \bibnamefont {Choi}}, \bibinfo {author} {\bibfnamefont {A.}~\bibnamefont {Lucke}}, \bibinfo {author} {\bibfnamefont {T.}~\bibnamefont {Schweizer}},\ and\ \bibinfo {author} {\bibfnamefont {L.~M.}\ \bibnamefont {Chayes}},\ }\bibfield  {title} {\bibinfo {title} {Invaded cluster algorithm for potts models},\ }\href {https://doi.org/10.1103/PhysRevE.54.1332} {\bibfield  {journal} {\bibinfo  {journal} {Phys. Rev. E}\ }\textbf {\bibinfo {volume} {54}},\ \bibinfo {pages} {1332} (\bibinfo {year} {1996})}\BibitemShut {NoStop}%
\bibitem [{Note2()}]{Note2}%
  \BibitemOpen
  \bibinfo {note} {Depth First Search can be used as well. JH used BFS and we do the same in our work.}\BibitemShut {Stop}%
\bibitem [{\citenamefont {Newman}\ and\ \citenamefont {Ziff}(2001)}]{NZ2001}%
  \BibitemOpen
  \bibfield  {author} {\bibinfo {author} {\bibfnamefont {M.~E.~J.}\ \bibnamefont {Newman}}\ and\ \bibinfo {author} {\bibfnamefont {R.~M.}\ \bibnamefont {Ziff}},\ }\bibfield  {title} {\bibinfo {title} {Fast monte carlo algorithm for site or bond percolation},\ }\href {https://doi.org/10.1103/PhysRevE.64.016706} {\bibfield  {journal} {\bibinfo  {journal} {Phys. Rev. E}\ }\textbf {\bibinfo {volume} {64}},\ \bibinfo {pages} {016706} (\bibinfo {year} {2001})}\BibitemShut {NoStop}%
\bibitem [{Note3()}]{Note3}%
  \BibitemOpen
  \bibinfo {note} {Inactive nodes have $\pi =0$ by definition.}\BibitemShut {Stop}%
\bibitem [{\citenamefont {Graver}\ \emph {et~al.}(1993)\citenamefont {Graver}, \citenamefont {Servatius},\ and\ \citenamefont {Servatius}}]{combinatorial_rigidity}%
  \BibitemOpen
  \bibfield  {author} {\bibinfo {author} {\bibfnamefont {J.}~\bibnamefont {Graver}}, \bibinfo {author} {\bibfnamefont {B.}~\bibnamefont {Servatius}},\ and\ \bibinfo {author} {\bibfnamefont {H.}~\bibnamefont {Servatius}},\ }\href {https://books.google.fr/books?id=Q0dECQAAQBAJ} {\emph {\bibinfo {title} {Combinatorial Rigidity}}},\ Graduate studies in mathematics\ (\bibinfo  {publisher} {American Mathematical Society},\ \bibinfo {year} {1993})\BibitemShut {NoStop}%
\bibitem [{Note4()}]{Note4}%
  \BibitemOpen
  \bibinfo {note} {Note that $\protect \mathcal {C}_u = \protect \mathcal {C}_v$ happens only if both $u$ and $v$ are active, hence each of them belongs to at least one rigid cluster.}\BibitemShut {Stop}%
\bibitem [{\citenamefont {Sykes}\ and\ \citenamefont {Essam}(1964)}]{sykes64}%
  \BibitemOpen
  \bibfield  {author} {\bibinfo {author} {\bibfnamefont {M.~F.}\ \bibnamefont {Sykes}}\ and\ \bibinfo {author} {\bibfnamefont {J.~W.}\ \bibnamefont {Essam}},\ }\bibfield  {title} {\bibinfo {title} {Exact critical percolation probabilities for site and bond problems in two dimensions},\ }\href {https://doi.org/10.1063/1.1704215} {\bibfield  {journal} {\bibinfo  {journal} {Journal of Mathematical Physics}\ }\textbf {\bibinfo {volume} {5}},\ \bibinfo {pages} {1117} (\bibinfo {year} {1964})},\ \Eprint {https://arxiv.org/abs/https://pubs.aip.org/aip/jmp/article-pdf/5/8/1117/19003400/1117\_1\_online.pdf} {https://pubs.aip.org/aip/jmp/article-pdf/5/8/1117/19003400/1117\_1\_online.pdf} \BibitemShut {NoStop}%
\bibitem [{\citenamefont {Notarmuzi}\ and\ \citenamefont {Javerzat}(2025)}]{GitProject}%
  \BibitemOpen
  \bibfield  {author} {\bibinfo {author} {\bibfnamefont {D.}~\bibnamefont {Notarmuzi}}\ and\ \bibinfo {author} {\bibfnamefont {N.}~\bibnamefont {Javerzat}},\ }\href {https://doi.org/10.5281/zenodo.15584520} {\bibinfo {title} {10.5281/zenodo.15584520}},\ \bibinfo {howpublished} {\url{https://github.com/NinaJaverzat/NZRP/tree/main}} (\bibinfo {year} {2025}),\ \bibinfo {note} {simulation code}\BibitemShut {NoStop}%
\bibitem [{Note5()}]{Note5}%
  \BibitemOpen
  \bibinfo {note} {We preferred sets to unordered sets as we observed the former to perform better than the latter. Indeed, the complexity of basic operations on sets is guaranteed to be $\protect \mathcal {O}(\log n)$. In contrast, while unordered sets offer $\protect \mathcal {O}(1)$ complexity, their performance can degrade to $\protect \mathcal {O}(n)$ due to rehashing of the underlying hash table.}\BibitemShut {Stop}%
\bibitem [{\citenamefont {Bogu\~n\'a}\ \emph {et~al.}(2009)\citenamefont {Bogu\~n\'a}, \citenamefont {Castellano},\ and\ \citenamefont {Pastor-Satorras}}]{boguna2009Langevin}%
  \BibitemOpen
  \bibfield  {author} {\bibinfo {author} {\bibfnamefont {M.}~\bibnamefont {Bogu\~n\'a}}, \bibinfo {author} {\bibfnamefont {C.}~\bibnamefont {Castellano}},\ and\ \bibinfo {author} {\bibfnamefont {R.}~\bibnamefont {Pastor-Satorras}},\ }\bibfield  {title} {\bibinfo {title} {Langevin approach for the dynamics of the contact process on annealed scale-free networks},\ }\href {https://doi.org/10.1103/PhysRevE.79.036110} {\bibfield  {journal} {\bibinfo  {journal} {Phys. Rev. E}\ }\textbf {\bibinfo {volume} {79}},\ \bibinfo {pages} {036110} (\bibinfo {year} {2009})}\BibitemShut {NoStop}%
\bibitem [{\citenamefont {{De Gennes, P.G.}}(1976)}]{deGennes1976}%
  \BibitemOpen
  \bibfield  {author} {\bibinfo {author} {\bibnamefont {{De Gennes, P.G.}}},\ }\bibfield  {title} {\bibinfo {title} {On a relation between percolation theory and the elasticity of gels},\ }\href {https://doi.org/10.1051/jphyslet:019760037010100} {\bibfield  {journal} {\bibinfo  {journal} {J. Physique Lett.}\ }\textbf {\bibinfo {volume} {37}},\ \bibinfo {pages} {1} (\bibinfo {year} {1976})}\BibitemShut {NoStop}%
\bibitem [{Note6()}]{Note6}%
  \BibitemOpen
  \bibinfo {note} {While concluding the first version of this work we became aware of the preprint \cite {lu2024}, that proposes a similar dynamic approach to RP. The algorithm has not been published, but is reported to perform slightly worse than the standard Pebble Game.}\BibitemShut {Stop}%
\bibitem [{\citenamefont {Ellenbroek}\ \emph {et~al.}(2015{\natexlab{b}})\citenamefont {Ellenbroek}, \citenamefont {Hagh}, \citenamefont {Kumar}, \citenamefont {Thorpe},\ and\ \citenamefont {van Hecke}}]{ellenbroek2015}%
  \BibitemOpen
  \bibfield  {author} {\bibinfo {author} {\bibfnamefont {W.~G.}\ \bibnamefont {Ellenbroek}}, \bibinfo {author} {\bibfnamefont {V.~F.}\ \bibnamefont {Hagh}}, \bibinfo {author} {\bibfnamefont {A.}~\bibnamefont {Kumar}}, \bibinfo {author} {\bibfnamefont {M.~F.}\ \bibnamefont {Thorpe}},\ and\ \bibinfo {author} {\bibfnamefont {M.}~\bibnamefont {van Hecke}},\ }\bibfield  {title} {\bibinfo {title} {Rigidity loss in disordered systems: Three scenarios},\ }\href {https://doi.org/10.1103/PhysRevLett.114.135501} {\bibfield  {journal} {\bibinfo  {journal} {Phys. Rev. Lett.}\ }\textbf {\bibinfo {volume} {114}},\ \bibinfo {pages} {135501} (\bibinfo {year} {2015}{\natexlab{b}})}\BibitemShut {NoStop}%
\bibitem [{\citenamefont {Head}(2025)}]{Head2025}%
  \BibitemOpen
  \bibfield  {author} {\bibinfo {author} {\bibfnamefont {D.~A.}\ \bibnamefont {Head}},\ }\bibfield  {title} {\bibinfo {title} {Predicting rigidity and connectivity percolation in disordered particulate networks using graph neural networks},\ }\href {https://doi.org/10.1103/PhysRevE.111.045411} {\bibfield  {journal} {\bibinfo  {journal} {Phys. Rev. E}\ }\textbf {\bibinfo {volume} {111}},\ \bibinfo {pages} {045411} (\bibinfo {year} {2025})}\BibitemShut {NoStop}%
\bibitem [{\citenamefont {Kim}\ and\ \citenamefont {Schwarz}(2024)}]{kim2024}%
  \BibitemOpen
  \bibfield  {author} {\bibinfo {author} {\bibfnamefont {K.}~\bibnamefont {Kim}}\ and\ \bibinfo {author} {\bibfnamefont {J.~M.}\ \bibnamefont {Schwarz}},\ }\href {https://arxiv.org/abs/2410.00317} {\bibinfo {title} {Rigidity condition for gluing two bar-joint rigid graphs embedded in $\mathbb{R}^d$}} (\bibinfo {year} {2024}),\ \Eprint {https://arxiv.org/abs/2410.00317} {arXiv:2410.00317 [cond-mat.dis-nn]} \BibitemShut {NoStop}%
\bibitem [{\citenamefont {Lu}\ \emph {et~al.}(2026)\citenamefont {Lu}, \citenamefont {Song}, \citenamefont {Shi}, \citenamefont {Li},\ and\ \citenamefont {Deng}}]{lu2026}%
  \BibitemOpen
  \bibfield  {author} {\bibinfo {author} {\bibfnamefont {M.}~\bibnamefont {Lu}}, \bibinfo {author} {\bibfnamefont {Y.}~\bibnamefont {Song}}, \bibinfo {author} {\bibfnamefont {Q.}~\bibnamefont {Shi}}, \bibinfo {author} {\bibfnamefont {M.}~\bibnamefont {Li}},\ and\ \bibinfo {author} {\bibfnamefont {Y.}~\bibnamefont {Deng}},\ }\href {https://arxiv.org/abs/2601.21399} {\bibinfo {title} {High-precision dynamic monte carlo study of rigidity percolation}} (\bibinfo {year} {2026}),\ \Eprint {https://arxiv.org/abs/2601.21399} {arXiv:2601.21399 [cond-mat.stat-mech]} \BibitemShut {NoStop}%
\bibitem [{\citenamefont {Vinutha}\ and\ \citenamefont {Sastry}(2019)}]{vinutha2019}%
  \BibitemOpen
  \bibfield  {author} {\bibinfo {author} {\bibfnamefont {H.~A.}\ \bibnamefont {Vinutha}}\ and\ \bibinfo {author} {\bibfnamefont {S.}~\bibnamefont {Sastry}},\ }\bibfield  {title} {\bibinfo {title} {Force networks and jamming in shear-deformed sphere packings},\ }\href {https://doi.org/10.1103/PhysRevE.99.012123} {\bibfield  {journal} {\bibinfo  {journal} {Phys. Rev. E}\ }\textbf {\bibinfo {volume} {99}},\ \bibinfo {pages} {012123} (\bibinfo {year} {2019})}\BibitemShut {NoStop}%
\bibitem [{\citenamefont {Gluck}(1975)}]{gluck}%
  \BibitemOpen
  \bibfield  {author} {\bibinfo {author} {\bibfnamefont {H.}~\bibnamefont {Gluck}},\ }\bibfield  {title} {\bibinfo {title} {Almost all simply connected closed surfaces are rigid},\ }in\ \href@noop {} {\emph {\bibinfo {booktitle} {Geometric Topology}}},\ \bibinfo {editor} {edited by\ \bibinfo {editor} {\bibfnamefont {L.~C.}\ \bibnamefont {Glaser}}\ and\ \bibinfo {editor} {\bibfnamefont {T.~B.}\ \bibnamefont {Rushing}}}\ (\bibinfo  {publisher} {Springer Berlin Heidelberg},\ \bibinfo {address} {Berlin, Heidelberg},\ \bibinfo {year} {1975})\ pp.\ \bibinfo {pages} {225--239}\BibitemShut {NoStop}%
\bibitem [{\citenamefont {Lee}\ and\ \citenamefont {Streinu}(2008)}]{Lee2008}%
  \BibitemOpen
  \bibfield  {author} {\bibinfo {author} {\bibfnamefont {A.}~\bibnamefont {Lee}}\ and\ \bibinfo {author} {\bibfnamefont {I.}~\bibnamefont {Streinu}},\ }\bibfield  {title} {\bibinfo {title} {Pebble game algorithms and sparse graphs},\ }\href {https://doi.org/https://doi.org/10.1016/j.disc.2007.07.104} {\bibfield  {journal} {\bibinfo  {journal} {Discrete Mathematics}\ }\textbf {\bibinfo {volume} {308}},\ \bibinfo {pages} {1425} (\bibinfo {year} {2008})},\ \bibinfo {note} {third European Conference on Combinatorics}\BibitemShut {NoStop}%
\bibitem [{\citenamefont {Lu}\ \emph {et~al.}(2024)\citenamefont {Lu}, \citenamefont {Song}, \citenamefont {Li},\ and\ \citenamefont {Deng}}]{lu2024}%
  \BibitemOpen
  \bibfield  {author} {\bibinfo {author} {\bibfnamefont {M.}~\bibnamefont {Lu}}, \bibinfo {author} {\bibfnamefont {Y.-F.}\ \bibnamefont {Song}}, \bibinfo {author} {\bibfnamefont {M.}~\bibnamefont {Li}},\ and\ \bibinfo {author} {\bibfnamefont {Y.}~\bibnamefont {Deng}},\ }\href {https://arxiv.org/abs/2411.04748} {\bibinfo {title} {Self-similar gap dynamics in percolation and rigidity percolation}} (\bibinfo {year} {2024}),\ \Eprint {https://arxiv.org/abs/2411.04748} {arXiv:2411.04748 [cond-mat.stat-mech]} \BibitemShut {NoStop}%
\bibitem [{\citenamefont {Thorpe}\ \emph {et~al.}(2000)\citenamefont {Thorpe}, \citenamefont {Jacobs}, \citenamefont {Chubynsky},\ and\ \citenamefont {Phillips}}]{Thorpe2000}%
  \BibitemOpen
  \bibfield  {author} {\bibinfo {author} {\bibfnamefont {M.}~\bibnamefont {Thorpe}}, \bibinfo {author} {\bibfnamefont {D.}~\bibnamefont {Jacobs}}, \bibinfo {author} {\bibfnamefont {M.}~\bibnamefont {Chubynsky}},\ and\ \bibinfo {author} {\bibfnamefont {J.}~\bibnamefont {Phillips}},\ }\bibfield  {title} {\bibinfo {title} {Self-organization in network glasses},\ }\href {https://doi.org/https://doi.org/10.1016/S0022-3093(99)00856-X} {\bibfield  {journal} {\bibinfo  {journal} {Journal of Non-Crystalline Solids}\ }\textbf {\bibinfo {volume} {266-269}},\ \bibinfo {pages} {859} (\bibinfo {year} {2000})}\BibitemShut {NoStop}%
\bibitem [{\citenamefont {Gurmessa}\ \emph {et~al.}(2019)\citenamefont {Gurmessa}, \citenamefont {Bitten}, \citenamefont {Nguyen}, \citenamefont {Saleh}, \citenamefont {Ross}, \citenamefont {Das},\ and\ \citenamefont {Robertson-Anderson}}]{Gurmessa2019}%
  \BibitemOpen
  \bibfield  {author} {\bibinfo {author} {\bibfnamefont {B.~J.}\ \bibnamefont {Gurmessa}}, \bibinfo {author} {\bibfnamefont {N.}~\bibnamefont {Bitten}}, \bibinfo {author} {\bibfnamefont {D.~T.}\ \bibnamefont {Nguyen}}, \bibinfo {author} {\bibfnamefont {O.~A.}\ \bibnamefont {Saleh}}, \bibinfo {author} {\bibfnamefont {J.~L.}\ \bibnamefont {Ross}}, \bibinfo {author} {\bibfnamefont {M.}~\bibnamefont {Das}},\ and\ \bibinfo {author} {\bibfnamefont {R.~M.}\ \bibnamefont {Robertson-Anderson}},\ }\bibfield  {title} {\bibinfo {title} {Triggered disassembly and reassembly of actin networks induces rigidity phase transitions},\ }\href {https://doi.org/10.1039/C8SM01912F} {\bibfield  {journal} {\bibinfo  {journal} {Soft Matter}\ }\textbf {\bibinfo {volume} {15}},\ \bibinfo {pages} {1335} (\bibinfo {year} {2019})}\BibitemShut {NoStop}%
\bibitem [{\citenamefont {Lenne}\ and\ \citenamefont {Trivedi}(2022{\natexlab{b}})}]{Lenne2022}%
  \BibitemOpen
  \bibfield  {author} {\bibinfo {author} {\bibfnamefont {P.-F.}\ \bibnamefont {Lenne}}\ and\ \bibinfo {author} {\bibfnamefont {V.}~\bibnamefont {Trivedi}},\ }\bibfield  {title} {\bibinfo {title} {Sculpting tissues by phase transitions},\ }\href {https://doi.org/10.1038/s41467-022-28151-9} {\bibfield  {journal} {\bibinfo  {journal} {Nature Communications}\ }\textbf {\bibinfo {volume} {13}},\ \bibinfo {pages} {664} (\bibinfo {year} {2022}{\natexlab{b}})}\BibitemShut {NoStop}%
\bibitem [{\citenamefont {Jackson}\ \emph {et~al.}(2022)\citenamefont {Jackson}, \citenamefont {Michel}, \citenamefont {Lwin}, \citenamefont {Fortier}, \citenamefont {Das}, \citenamefont {Bonassar},\ and\ \citenamefont {Cohen}}]{Jackson2022}%
  \BibitemOpen
  \bibfield  {author} {\bibinfo {author} {\bibfnamefont {T.~W.}\ \bibnamefont {Jackson}}, \bibinfo {author} {\bibfnamefont {J.}~\bibnamefont {Michel}}, \bibinfo {author} {\bibfnamefont {P.}~\bibnamefont {Lwin}}, \bibinfo {author} {\bibfnamefont {L.~A.}\ \bibnamefont {Fortier}}, \bibinfo {author} {\bibfnamefont {M.}~\bibnamefont {Das}}, \bibinfo {author} {\bibfnamefont {L.~J.}\ \bibnamefont {Bonassar}},\ and\ \bibinfo {author} {\bibfnamefont {I.}~\bibnamefont {Cohen}},\ }\bibfield  {title} {\bibinfo {title} {Structural origins of cartilage shear mechanics},\ }\href {https://doi.org/10.1126/sciadv.abk2805} {\bibfield  {journal} {\bibinfo  {journal} {Science Advances}\ }\textbf {\bibinfo {volume} {8}},\ \bibinfo {pages} {eabk2805} (\bibinfo {year} {2022})},\ \Eprint {https://arxiv.org/abs/https://www.science.org/doi/pdf/10.1126/sciadv.abk2805} {https://www.science.org/doi/pdf/10.1126/sciadv.abk2805} \BibitemShut {NoStop}%
\bibitem [{\citenamefont {Bi}\ \emph {et~al.}(2015)\citenamefont {Bi}, \citenamefont {Lopez}, \citenamefont {Schwarz},\ and\ \citenamefont {Manning}}]{Bi2015A}%
  \BibitemOpen
  \bibfield  {author} {\bibinfo {author} {\bibfnamefont {D.}~\bibnamefont {Bi}}, \bibinfo {author} {\bibfnamefont {J.~H.}\ \bibnamefont {Lopez}}, \bibinfo {author} {\bibfnamefont {J.~M.}\ \bibnamefont {Schwarz}},\ and\ \bibinfo {author} {\bibfnamefont {M.~L.}\ \bibnamefont {Manning}},\ }\bibfield  {title} {\bibinfo {title} {A density-independent rigidity transition in biological tissues},\ }\href {https://doi.org/10.1038/nphys3471} {\bibfield  {journal} {\bibinfo  {journal} {Nature Physics}\ }\textbf {\bibinfo {volume} {11}},\ \bibinfo {pages} {1074} (\bibinfo {year} {2015})}\BibitemShut {NoStop}%
\bibitem [{\citenamefont {Atia}\ \emph {et~al.}(2021)\citenamefont {Atia}, \citenamefont {Fredberg}, \citenamefont {Gov},\ and\ \citenamefont {Pegoraro}}]{Atia2021are}%
  \BibitemOpen
  \bibfield  {author} {\bibinfo {author} {\bibfnamefont {L.}~\bibnamefont {Atia}}, \bibinfo {author} {\bibfnamefont {J.~J.}\ \bibnamefont {Fredberg}}, \bibinfo {author} {\bibfnamefont {N.~S.}\ \bibnamefont {Gov}},\ and\ \bibinfo {author} {\bibfnamefont {A.~F.}\ \bibnamefont {Pegoraro}},\ }\bibfield  {title} {\bibinfo {title} {Are cell jamming and unjamming essential in tissue development?},\ }\href {https://doi.org/https://doi.org/10.1016/j.cdev.2021.203727} {\bibfield  {journal} {\bibinfo  {journal} {Cells \& Development}\ }\textbf {\bibinfo {volume} {168}},\ \bibinfo {pages} {203727} (\bibinfo {year} {2021})},\ \bibinfo {note} {quantitative Cell and Developmental Biology}\BibitemShut {NoStop}%
\bibitem [{\citenamefont {Lee}\ \emph {et~al.}(2005)\citenamefont {Lee}, \citenamefont {Streinu},\ and\ \citenamefont {Theran}}]{Lee2005Finding}%
  \BibitemOpen
  \bibfield  {author} {\bibinfo {author} {\bibfnamefont {A.}~\bibnamefont {Lee}}, \bibinfo {author} {\bibfnamefont {I.}~\bibnamefont {Streinu}},\ and\ \bibinfo {author} {\bibfnamefont {L.}~\bibnamefont {Theran}},\ }\bibfield  {title} {\bibinfo {title} {Finding and maintaining rigid components.}\ }(\bibinfo {year} {2005})\ pp.\ \bibinfo {pages} {219--222}\BibitemShut {NoStop}%
\bibitem [{\citenamefont {Lee}\ \emph {et~al.}(2008)\citenamefont {Lee}, \citenamefont {Streinu},\ and\ \citenamefont {Theran}}]{Lee2008Analyzing}%
  \BibitemOpen
  \bibfield  {author} {\bibinfo {author} {\bibfnamefont {A.}~\bibnamefont {Lee}}, \bibinfo {author} {\bibfnamefont {I.}~\bibnamefont {Streinu}},\ and\ \bibinfo {author} {\bibfnamefont {L.}~\bibnamefont {Theran}},\ }\bibfield  {title} {\bibinfo {title} {Analyzing rigidity with pebble games},\ }in\ \href {https://doi.org/10.1145/1377676.1377715} {\emph {\bibinfo {booktitle} {Proceedings of the Twenty-Fourth Annual Symposium on Computational Geometry}}},\ \bibinfo {series and number} {SCG '08}\ (\bibinfo  {publisher} {Association for Computing Machinery},\ \bibinfo {address} {New York, NY, USA},\ \bibinfo {year} {2008})\ p.\ \bibinfo {pages} {226–227}\BibitemShut {NoStop}%
\bibitem [{\citenamefont {Pollaczek-Geiringer}(1927)}]{Geiringer}%
  \BibitemOpen
  \bibfield  {author} {\bibinfo {author} {\bibfnamefont {H.}~\bibnamefont {Pollaczek-Geiringer}},\ }\bibfield  {title} {\bibinfo {title} {Über die gliederung ebener fachwerke},\ }\href {https://doi.org/https://doi.org/10.1002/zamm.19270070107} {\bibfield  {journal} {\bibinfo  {journal} {ZAMM - Journal of Applied Mathematics and Mechanics / Zeitschrift für Angewandte Mathematik und Mechanik}\ }\textbf {\bibinfo {volume} {7}},\ \bibinfo {pages} {58} (\bibinfo {year} {1927})},\ \Eprint {https://arxiv.org/abs/https://onlinelibrary.wiley.com/doi/pdf/10.1002/zamm.19270070107} {https://onlinelibrary.wiley.com/doi/pdf/10.1002/zamm.19270070107} \BibitemShut {NoStop}%
\end{thebibliography}%

\end{document}


\renewcommand{\theequation}{S\arabic{equation}}
\renewcommand{\thefigure}{S\arabic{figure}}
\renewcommand{\thesection}{Supplementary Note \arabic{section}}

\onecolumngrid

\section*{A fast algorithm for 2D Rigidity Percolation\\  Supplemental Material}

\begin{center}
    Nina Javerzat$^{\ast\star}$, Daniele Notarmuzi$^{\dag\star\star}$\\
    \textit{$^{\ast}$ Université Grenoble Alpes, CNRS, LIPhy, 38000 Grenoble, France} \\
    \textit{$^{\dag}$Institut f\"{u}r Theoretische Physik, TU Wien, Wiedner Hauptstraße 8-10, A-1040 Wien, Austria}
    $^{\star}$\small{nina.javerzat@univ-grenoble-alpes.fr}
    $^{\star\star}$\small{daniele.notarmuzi@tuwien.ac.at}
\end{center}

\par\noindent\rule{\textwidth}{0.4pt}
\section{Proofs of results in 2D rigidity theory}
\subsection{Preliminary results}
In the following we consider graphs $G=(V,E)$, with $n=|V|$ vertices and $m=|E|$ edges.
We refer to \cite{combinatorial_rigidity, holroyd98} for an introduction to graph's rigidity and the definition of the main concepts, and notably recall that the rigidity of the graph is independent from its embedding in $\mathbb{R}^d$ \cite{gluck}. We will need the following lemma (valid in any dimension):
\\\\
\textit{Lemma A -- } Given a rigid graph $G$, a rigid subgraph $g\subset G$, and an edge $e\in g$, if $g\setminus\{e\}$ is rigid then $G\setminus\{e\}$ is rigid.
\\\\
\textit{Proof -- } Denote $x,y$ the vertices of $e$, and suppose that $G\setminus\{e\}$ is not rigid, namely that there exists a motion (a differentiable family of embeddings) of $G\setminus\{e\}$ that does not preserve the distance between all pairs of vertices 
If this motion does not preserve the distance between $x$ and $y$, then $g\setminus\{e\}$ is not rigid, leading to a contradiction. Likewise, if the motion preserves the distance between $x$ and $y$, but does not preserve the distance between any other two vertices of $G\setminus\{e\}$, then $G$ cannot be rigid, leading to a contradiction as well. Therefore, $G\setminus\{e\}$ is rigid.
\\\\
For convenience we also recall Proposition 2, stated in the main text:\\\\
\textit{Proposition 2 (``$d$-pivots rule'')~\cite{combinatorial_rigidity} --} In $d$ dimensions, if two rigid graphs $G$ and $H$ share at least $d$ vertices (pivots), then $G\cup H$ is rigid.
\\\\\indent
In 2 dimensions,  a theorem due to Hilda Geiringer (later rediscovered by Gerard Laman), characterises rigid graphs \cite{Geiringer,Laman1970}:\\\\
\cite{Laman1970}:\\\\
\textit{Theorem (Geiringer-Laman) --} A graph with $2n- 3$ edges is rigid in two dimensions if and
only if no subgraph $G'$ has more than $2n' - 3$ edges.
\\\\
\textit{Proposition (Laman)} -- Every rigid graph $G=(V,E)$ has a rigid subgraph with $|V|$ vertices and $2|V|-3$ edges.
\\\\
We will call \textit{rigid components} (or rigid clusters) of a graph $G$ its maximal rigid subgraphs, i.e. the maximal sets of mutually rigid edges. Namely, we can write the decomposition of $G$ in its rigid components $G_j$ as $G = \bigcup_j\,G_j$. Note that the $G_j$ are edge-disjoint, but that vertices of $G$ may belong to $k$ rigid components. Such vertices are pivots, of pivotal class (cf. main text) $\pi = k$.
\\\\
We define an edge $e\in G=\bigcup_j\,G_j$ as \textit{redundant} if the rigid cluster decomposition of $G\setminus\{e\}$ is $G\setminus\{e\} = G_1\cup\cdots\cup G_e\setminus\{e\}\cup\cdots$, where $G_e$ is the rigid component of $G$ that contains $e$. 
From this definition follows that:\\\\
\textit{Proposition A --} $e$ is redundant for $G$ if and only if $G_e\setminus\{e\}$ is rigid.\\\\
Note that, if $e=\{x,y\}$ is redundant, its two vertices $x,y$ necessarily belong to $G_e\setminus\{e\}$, as a graph with isolated vertices cannot be rigid.
Edges that are not redundant are \textit{independent}. The number $m^R$ of redundant edges of $G$ is the number of edges such that $G\setminus\{e_1,\cdots,e_{m_R}\}$ has only independent edges. Note that the set of redundant edges $\{e_1,\cdots,e_{m_R}\}$ is not unique, while its cardinality $m_R$ is. The number of independent edges is denoted $m^I = m-m^R$. The number of \textit{floppy modes} (or degrees of freedom, see \cite{combinatorial_rigidity, JacobsThorpe1995}) of $G$ is $F = 2n-m^I\geq3$, with $F=3$ if and only if $G$ is rigid. This implies notably that $2n-3$ is the maximal number of independent edges for any given graph. With these definitions of redundant and independent edges we have:
\\\\
\textit{Theorem A -- } The edges of a graph $G=(V, E)$ are independent in
 two dimensions if and only if $G$ has no rigid subgraph $G'=(V', E')$ with $(2n'-3)+1$ edges.
 \\\\
 \textit{Proof:}\\
 $\Leftarrow$ Suppose $e\in G$ is redundant. $G_e\setminus\{e\}$ is rigid, so by Laman proposition it has a rigid spanning subgraph $\tilde{G}$ with $2 \tilde{n}-3$ edges. Then, the graph $G'\equiv \tilde{G}\cup\{e\}\subseteq G$ has $\tilde{n}$ vertices and $2\tilde{n}-3+1$ edges. It is rigid by Proposition 2, since $\tilde{G}$ is rigid and shares two pivots with the edge $e$.\\
  $\Rightarrow$ Suppose there exists a rigid $G'\subseteq G$ with $m'=(2n'-3)+1$. We take $G'$ as the minimal such subgraph, namely with the smallest number of vertices $n'$. All $G''\subset G'$ have $m''\leq 2n''-3$. In particular, for any edge $e$ of $G'$, the graph $G''=G'\setminus\{e\}$ has exactly $n'$ vertices\footnote{Indeed, both vertices of $e$ necessarily belong to $G''$: suppose instead that $n''=n'-1$ (i.e, one vertex of $e$ is not in $G'''$). Then $m''=2n'-3=2(n''+1)-3=2n''-1>2n''-3$, leading to a contradiction.} and $2n'-3$ edges. By minimality of $G'$, $G''$ satisfies the Geiringer-Laman theorem and is rigid. Therefore, any edge of $G'$ is redundant for $G'$.
  By Lemma A, this implies that any edge of $G'$ is redundant for $G$.
\\\\
A weaker version of Theorem A is given in Ref. \cite{Jacobs1997An} (Theorem 2.2, attributed to Laman). The minimal rigid graph $G'$ is said to be \textit{overconstrained} \cite{JacobsThorpe1995,Jacobs1997An}: any of its edges can be removed without losing rigidity. In the above proof we have indeed shown that any edge $e$ of $G'$ is redundant in the sense of Proposition A: $\forall e\in G'$, $G'\setminus \{e\}$ remains rigid. Note that the number of redundant edges $m^R$ of $G'$ is $m^R=1$: once an edge has been cut, no further edge of $G'\setminus\{e\} = G''$ can be removed without losing rigidity. Noting that any added edge connecting two vertices of $G'$ is also redundant for $G'$ (by Theorem A), an \textit{overconstrained region} \cite{JacobsThorpe1995} with $m^R\geq1$ redundant edges generally consists of such a minimal rigid graph $G'$ together with a set of edges $\{e_1,e_2,\cdots,e_{m^R-1}\}$ such that each $e_i$ has vertices in $G'$. Namely, an overconstrained region becomes floppy when (any) $m^R+1$ edges are cut.
\\\\
We can now generalize Laman's proposition to:
\\\\
\textit{Proposition B} -- Any graph $G = (V, E)$ with  $\left|V\right|$ vertices has a spanning subgraph $\tilde{G} = (V, \tilde{E})$ where $\tilde{E}$ is a set of independent edges. 
\\\\\
\textit{Proof:} We decompose $G$ in its rigid components as $G = \bigcup_j G_j$. Since each $G_j=(V_j,E_j)$ is rigid, by Laman proposition it has a spanning rigid subgraph $\tilde{G}_j = (V_j,\tilde{E}_j)$ with $2n_j-3$ edges. Let's consider $\tilde{G} = \bigcup_i \tilde{G}_j = (V,\tilde{E})$, a spanning subgraph of $G$. We want to show that the edges $\tilde{E}$ are independent. Let's suppose that they are not. By theorem A, there exists a rigid subgraph $\tilde{G}'\subset \tilde{G}$ with $2n'-2$ edges. Because it is rigid, $\tilde{G}'$ is necessarily contained in a unique rigid component $\tilde{G}_j$ of $\tilde{G}$. So $\tilde{G}_j$ has a subgraph $\tilde{G}'$ with more than $2n'-3$ edges, which, by the Geiringer-Laman theorem, contradicts the fact that it is rigid and has $2\tilde{n}_j-3$ edges.

\subsection{Theorems on single bond activation}
We are now ready to prove the three theorems on which we base our algorithm. In the following, we consider the graph $G=\bigcup_j G_j$, and activate a new edge $e$, with vertices $u$ and $v$, that we denote $uv$ for convenience. The three theorems below are concerned with the rigid cluster decomposition of $G^+\equiv G\cup\{uv\}$. Rigid and connectivity components in $G$ are respectively denoted $\mathcal{R}$ and $\mathcal{C}$, while rigid and connectivity components in $G^+$ are denoted $\mathcal{R}^+$ and $\mathcal{C}^+$.\\\\
\textit{Theorem 1 (Pivoting) -- }
Assume $u\in \mathcal{C}_u$ and $v\in\mathcal{C}_v$ with $\mathcal{C}_u\neq\mathcal{C}_v$.
Then,
\begin{enumerate}
    \item The bond $uv$ is independent.
    \item The activation of $uv$ creates a new rigid cluster composed of the bond $uv$ only.
    \item The pivotal class of $u$ and $v$ increases by 1. The pivotal class of any other node
     is unchanged.
\end{enumerate}
\textit{Proof:} 
\begin{enumerate}
\item 
Since $u$ and $v$ are in distinct connectivity clusters, there is no rigid component of $G = G^+\setminus\{uv\}$ that contains both $u$ and $v$, i.e. $\mathcal{R}^+_{uv}\setminus\{uv\}=\emptyset$, where $\mathcal{R}^+_{uv}$ is the
 rigid cluster of $G^+$ that contains $uv$.
 \item Let's consider $\mathcal{C}^+\equiv \mathcal{C}_u\cup\mathcal{C}_v\cup\{uv\} = \bigcup_j \mathcal{R}^+_j$ where $\mathcal{R}^+_j$ are the rigid components of $\mathcal{C}^+$. We want to show that  $\mathcal{R}^+_{uv}=\{uv\}$. Let's suppose instead that $\mathcal{R}^+_{uv} = \mathcal{R}^+_u\cup\{uv\}\cup \mathcal{R}^+_v$ 
where $\mathcal{R}^+_u = \mathcal{C}_u\bigcap \mathcal{R}^+_{uv}$ has $n_u$ vertices and $m_u$ edges, and $\mathcal{R}^+_v = \mathcal{C}_v\bigcap \mathcal{R}^+_{uv}$
has $n_v$ vertices and $m_v$ edges, and $m_u+m_v>0$. By Proposition B, $\mathcal{R}^+_{uv}$ has a (rigid) spanning subgraph $\tilde{\mathcal{R}}^+_{uv} =\tilde{\mathcal{R}}^+_u\cup\{uv\}\cup \tilde{\mathcal{R}}^+_v$. $\tilde{\mathcal{R}}^+_{uv}$ has $n_u+n_v$ vertices and $\tilde{m}_u+\tilde{m}_v+1$ independent edges, so that it has $F=2(n_u+n_v)- \tilde{m}_u-\tilde{m}_v-1$ floppy modes. Since $\tilde{\mathcal{R}}_u$ and $\tilde{\mathcal{R}}_v$ are rigid, $2 n_u-\tilde{m}_u = 2n_v-\tilde{m}_v = 3$. Therefore, $F = 5>3$, contradicting the assumption that $\tilde{\mathcal{R}}_{uv}$ is rigid. Thus it cannot be that $\mathcal{R}^+_{uv} = \mathcal{R}^+_u\cup\{uv\}\cup \mathcal{R}^+_v$ with non-empty $\mathcal{R}^+_u$ and $\mathcal{R}^+_v$, and so $\mathcal{R}_{uv}^+=\{uv\}$.
\item Follows from 2.
\end{enumerate}
\mynewline{blabla}
\\
\textit{Theorem 2 (Overconstraining) -- } Assume that there exists a rigid cluster $\mathcal{R}$ such that $u,v\in\mathcal{R}$ before the activation of $uv$. Then,
\begin{enumerate}
    \item $uv$ is a redundant bond.
    
    \item   Upon activation of $uv$, $\mathcal{R}$ becomes $\mathcal{R}\cup uv$; all other rigid clusters remain identical.
   
     \item The pivotal classes of all nodes are unchanged.
\end{enumerate}
\textit{Proof:} 
\begin{enumerate}
    \item[]1. and 2. Follow from the definition of redundant bonds with $\mathcal{R}^+_{uv}\setminus\{uv\} = \mathcal{R}$.
    \item[]3. Follows from 2.
\end{enumerate}
\mynewline{blabla}
\\
\textit{Theorem 3 (Rigidification) -- }
Assume $\mathcal{C}_u = \mathcal{C}_v=\mathcal{C}$ and that, before the activation of $uv$, there is no rigid cluster that contains both $u$ and $v$. Then,
\begin{enumerate}
    \item $uv$ is an independent bond.
    \item The activation of $uv$ triggers a rigidification process resulting in the coalescence of $k+1$ rigid clusters $\mathcal{R}_1,\,\mathcal{R}_2,\cdots,\mathcal{R}_k$ and $\mathcal{R}_{k+1}=\{uv\}$ into a single rigid cluster $\mathcal{R}^+=\mathcal{R}_1\cup\mathcal{R}_2\cup\cdots\cup\mathcal{R}_k\cup \{uv\}$. The clusters $\mathcal{R}_1,\,\mathcal{R}_2,\cdots,\mathcal{R}_{k}$ all belong to $\mathcal{C}$, and each of them is connected to at least two other ones via at least two distinct pivots. 
    \item The pivotal class of these pivots decreases by at least one.
\end{enumerate}

\textit{Proof:}
\begin{enumerate}
    \item Since there is no rigid component of $G$ containing both $u$ and $v$, $\mathcal{R}_{uv}^+\setminus\{uv\}=\emptyset$.
    \item That $\mathcal{R}_1,\,\mathcal{R}_2,\cdots,\mathcal{R}_{k}$ all belong to $\mathcal{C}$ is obvious. 
    That the activation of $uv$ triggers a rigidification process that leads to $k \geq 0$ can be simply seen
    by an example as, e.g., the activation of the third edge of a triangle.
    Let's now consider any $\mathcal{R}_i$ among $\mathcal{R}_1,\,\mathcal{R}_2,\cdots,\mathcal{R}_{k+1}$, and show that it must have at least two pivots with $\mathcal{R}^+\setminus\mathcal{R}_i=\mathcal{R}_1\cup\cdots\cup\mathcal{R}_{i-1}\cup\mathcal{R}_{i+1}\cup\cdots\cup\mathcal{R}_{k+1}$. That  $\mathcal{R}_i$ 
    shares zero pivot with $\mathcal{R}^+\setminus\mathcal{R}_i$ is excluded since $\mathcal{R}^+$ must be connected.
    $\mathcal{R}_i$ and $\mathcal{R}^+\setminus\mathcal{R}_i$ cannot share only one pivot: if they would ($\mathcal{R}_i$ would be a ``dangling-end'' of $\mathcal{R}^+$)  the number of floppy modes of $\mathcal{R}^+$ would be (using Proposition B) $F_{\mathcal{R}^+}=2\left(n_{\mathcal{R}_i}+n_{\mathcal{R}^+\setminus\mathcal{R}_i}-1\right)-m_{\mathcal{R}_i}^I-m_{\mathcal{R}^+\setminus\mathcal{R}_i}^I = F_{\mathcal{R}_i}+F_{\mathcal{R}^+\setminus\mathcal{R}_i}-2\geq4$ and $\mathcal{R}^+$ would not be rigid.
    Therefore, any $\mathcal{R}_i\in\mathcal{R}^+$ shares at least two pivots with $\mathcal{R}^+\setminus\mathcal{R}_i$.\\
   Moreover, it cannot be that $\mathcal{R}_i$ is connected to only one of the other $k$ coalescing clusters through at least two pivots, or, by Proposition 2, they would not be two distinct rigid clusters before the activation of $uv$. 

   \item Upon rigidification, any pivot $p$ shared by rigidifying clusters $\mathcal{R}_i$ and $\mathcal{R}_j$ ceases to be a pivot between $\mathcal{R}_i$ and $\mathcal{R}_j$, so its pivotal class decreases by one. If $p$ belongs to other rigidifying clusters, its pivotal class decreases by more than one.
  
\end{enumerate}
\section{Cost of each event}
We first note that, for any system in which the
number of bonds is linear in the number of nodes, i.e., any system where the average degree
of a node is independent on the system size, the number of bonds to be activated over any fixed range  of bond concentration $\Delta p$ is linear in $N$. 

\begin{figure}[h]
    \includegraphics[scale=.6]{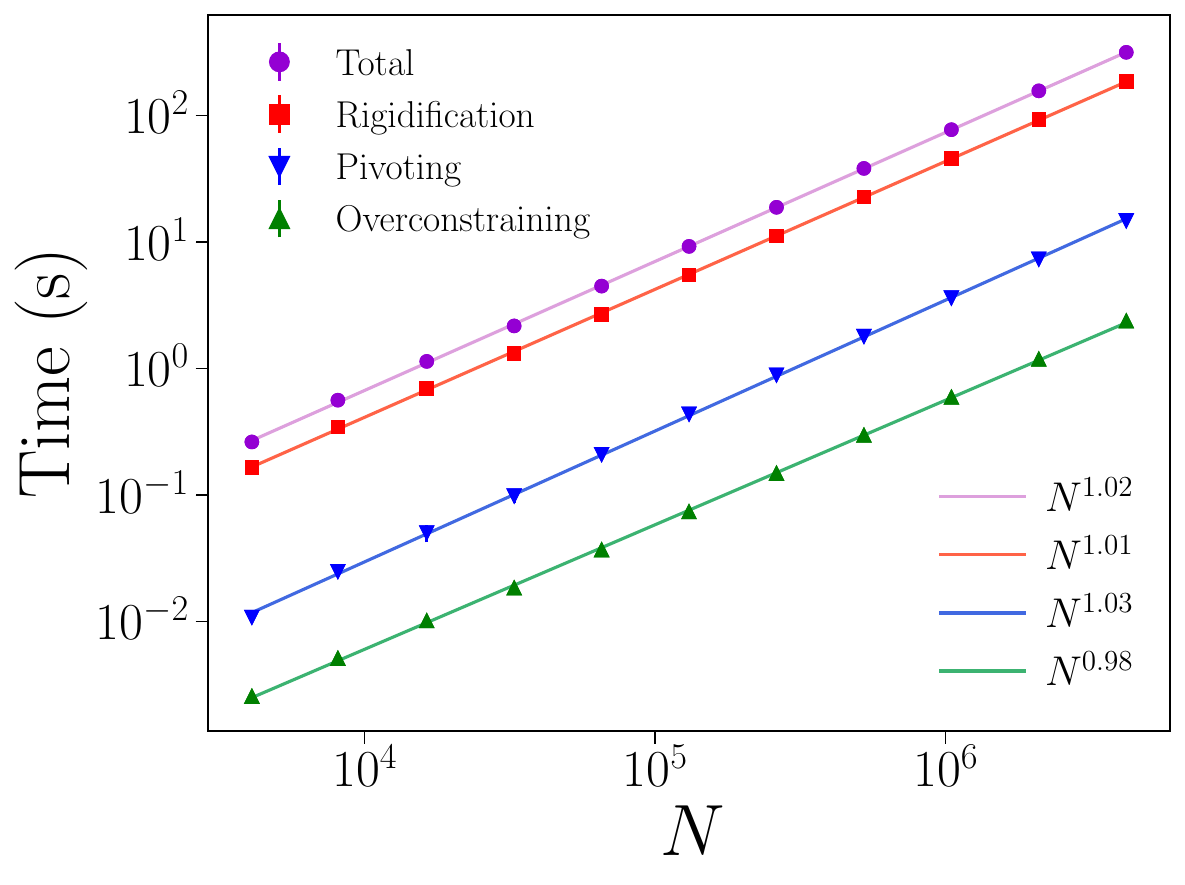}
    \caption{Scaling of the computing time for each type of event with the system size. The data represents the total
    cost of each event in a single simulation, i.e., the aggregated cost of all events of a given type.}
    \label{fig:theorems_cost}
\end{figure}

From Fig.~2 in the main text, each given type of event (pivoting, rigidification, overconstraining) occurs a number of times that is linear in $N$, for all the three events. 
Note that overconstraining events, which follow exclusively from the activation of redundant bonds, happen exactly $M-B^I = N+3$ times. Figure~\ref{fig:theorems_cost} shows the average computing time of, respectively, all the pivoting, rigidification
and overconstraining steps in a single simulation, as a function of $N$. The linear scaling of the cost of overconstraining events shows that each such event has constant cost.
The cost of each rigidification and pivoting event, instead, is not constant but rather
depends on the number of pebble searches needed to perform the step, and on their cost. As such,
the cost depends on the concentration $p$. For example, in pivoting events,
the cost of the (at most one) pebble search of type I is much higher near the CP transition than below it.
Finally, note that the major cost of the algorithm is clearly due to rigidification events. 
\\


\section{Non optimal rigidification strategies}
Here we give  details on possible legitimate but non-optimal rigidification strategies referred to in the main text.

Any rigidification stategy that we attempt begins by first gathering four
pebbles at $uv$ (using pebble searches of type I). One pebble is used to cover the edge and to update the
pebble graph, the remaining three are frozen at their location. Furthermore, a new $\mathcal{T}_{\mathcal{R}}$
tree is initialized with root $uv$ and the bonds that belong to it are identified. To this aim,
$u$ and $v$ are marked rigid, with all the other nodes initially unmarked.

\subsection{Exploiting the JH approach}

\begin{figure}[h]
    \includegraphics[width=\textwidth]{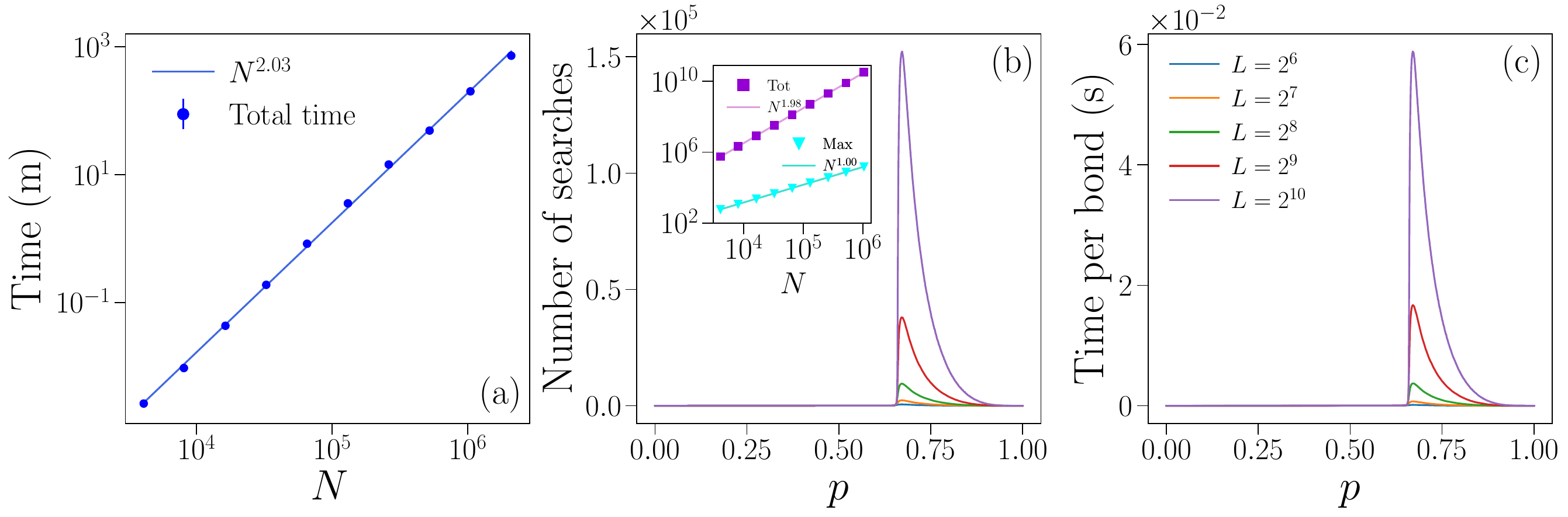}
    \caption{Performance of the algorithm implementing the rigidification protocol based
    on the JH algorithm.    
    (a) The average time (in minutes) to complete a simulation as a function of the system size.
    The blue solid line is the best fit to the log-log of the data and shows the scaling $N^{2.03}$.
    (b) Average number of pebble searches (of any type) needed at each bond activation, as function of the bond concentration $p$, for different system sizes $N$. 
    Inset: Scaling of the total and maximum number of pebbles searches as a function of $N$, 
    i.e., integral and peak of the curves in the main panel as a function of $N$. 
    (c) Average time (in seconds) of each bond activation as function of $p$, for different system sizes $N$.}
    \label{fig:NZ_JH_mix}
\end{figure}

As stated in the main text, the most obvious approach to build the rigid cluster
that forms upon rigidification is to build it following the approach of the JH algorithm.
Namely, to perform a Breadth First Search starting at $uv$, over neighbours of nodes marked rigid. Results are shown in Fig.~\ref{fig:NZ_JH_mix}.
As stated in the main text, the algorithm is roughly quadratic.
This can be understood by
looking at the average number of pebble searches (panel (b) in Fig.~\ref{fig:NZ_JH_mix}), and the average time  needed to activate a bond and update the system's state accordingly (panel (c)),
as a function of the bond concentration $p$. Pebble searches are all of type II, except at most four. Their average number is hence
clearly dominated by the type II searches. The time per bond function (panel (c)) is substantially identical
 to the number of searches (panel (b)), hence suggesting that the time per bond is a constant 
--the average cost of
one search of type II-- times the number of searches. The integral of the latter, shown in the inset of panel (b), scales
nearly quadratically with the system size, hence explaining the overall 
quadratic scaling of the algorithm.
The quadratic scaling of the integral can be understood by noting that the maximum number 
of searches scales linearly with the system size (inset), meaning that there is at least
one bond activation that leads to the execution of $\mathcal{O}(N)$ searches. However, the
region of the concentration range where the number of searches scales linearly with $N$
is roughly constant, implying it contains $\mathcal{O}(N)$ bonds, all of which require $\mathcal{O}(N)$
searches, explaining the total scaling of this algorithm.

This approach is the simplest and certainly the most obvious to obtain a Newman-Ziff algorithm for RP, and is taken as a benchmark:
a more complex rigidification protocol is meaningful if it results in an algorithm that scales 
significantly better than $N^2$.

\subsection{Exploration of the whole connectivity cluster}

A straightforward strategy to build the rigid cluster that emerges during a rigidification
step is to exploit the transitivity of rigidity (Proposition 1), and play the Pebble Game only on each root bond of the rigid clusters contained inside the connectivity
cluster $\mathcal{C}_u = \mathcal{C}_v$. To adopt this strategy, it is required to have a set
containing the roots of rigid clusters contained in each connectivity cluster. 
When two connectivity clusters coalesce,
which is a pivoting step, the two sets are merged and the new rigid cluster created, $uv$, 
is added to resulting set. Overconstraining events do not affect these sets. 
Finally, during a rigidification step, the mutual rigidity of the new activated bond with respect to each element of the set (namely each root contained in $\mathcal{C}_u$)
is explicitly
checked using pebble searches of type II. This approach results in a quadratic algorithm, for
a very simple reason. The average number of bonds that result in a rigidification step
is $\mathcal{O}(N)$ (see Fig. 2 in the main text).
These $\mathcal{O}(N)$ events happen
almost always inside the macroscopic connectivity cluster, which contains a very large
number of small rigid clusters (especially just above $p_c^{CP}$), each of which
has to be explicitly checked using pebbles searches of type II. 
These $\mathcal{O}(N)$ rigidification steps have
a cost that is $ \approx \mathcal{O}(N)$ 
hence leading to a roughly quadratic algorithm.


\subsection{Exploitation of the ``Two pivots rule''}

We have also envisaged to exploit further rigidity theory to reduce even more the number of pebble searches. In particular the ``Two pivots rule'' (Proposition 2), can be used in principle to identify rigid clusters that have two pivots
with the forming one, so as to mark them rigid without performing any pebble search of type II. 

In that case, the eighth step of the algorithm presented in the main text (section VI) is replaced by:
\\\\
8. If $p'$ is not labeled as enqueued, then
\begin{myenum}
    \item Insert it in the queue of pivots and label it as enqueued.
    This is absolutely crucial to prevent inserting the same pivot in the queue more than once. 
    \item Iterate over all the bonds of $p'$ and, if they are active, find the root $r''$ of the rigid cluster
    they belong to.
    \item If the rigid cluster rooted by $r''$ is $\mathcal{R}_{r_{\rm large}}$ or
    $\mathcal{R}_{r_{\rm small}}$, go back to step 8b and process the next root of $p'$. 
    If $r''$ is already marked (floppy or rigid) go back to step 8b and process the next root of $p'$.
    Otherwise go to step 8d.
    \item Iterate over all the pivots $p''$ of $\mathcal{R}_{r''}$ and start counting how many
    of them are in $\mathcal{P}_{\rm r_{large}}$. If the count reaches two, break the loop
    and go to step 8e, otherwise go back to step 8b and process the next root of $p'$.
    \item Mark $r''$ as rigid.
\end{myenum}

In this way $r''$ is marked rigid without using pebble searches but rather by checking its mutual rigidity with respect to the new bond $uv$ exploiting Proposition 2.
However,
quickly testing this approach revealed it to be extremely slow. The reason is twofold. First,
the above approach does not allow to mark $r''$ as floppy. Hence,
as the rigid cluster gets build and the set of pivots of $\mathcal{R}_{\rm large}$ changes,
this approach might require to check the same root $r''$ more than once, as new rigid clusters
coalesce into the forming rigid cluster (until $r''$ gets marked rigid, if ever, or until
it gets marked floppy by pebble searches.) 
Second, every root $r''$ that enters step 8d, results in an iteration over $\mathcal{P}_{\rm r''}$.
These are the third, innermost loops inside the process (outer loop is over the queue of 
pivots, inner loop is over the pivots of $\mathcal{R}_{r_{\rm small}}$), making this approach highly inconvenient.

We stress that one might therefore not want to avoid playing pebble searches at all cost, as their use can be more efficient (numerically) than applying directly rigidity theory. Our final algorithm represents an efficient balance between the cost of pebble searches and the cost of applying rigidity theorems.

\subsection{A different implementation of the pivot network}

\begin{figure}[h]
    \includegraphics[scale=.25]{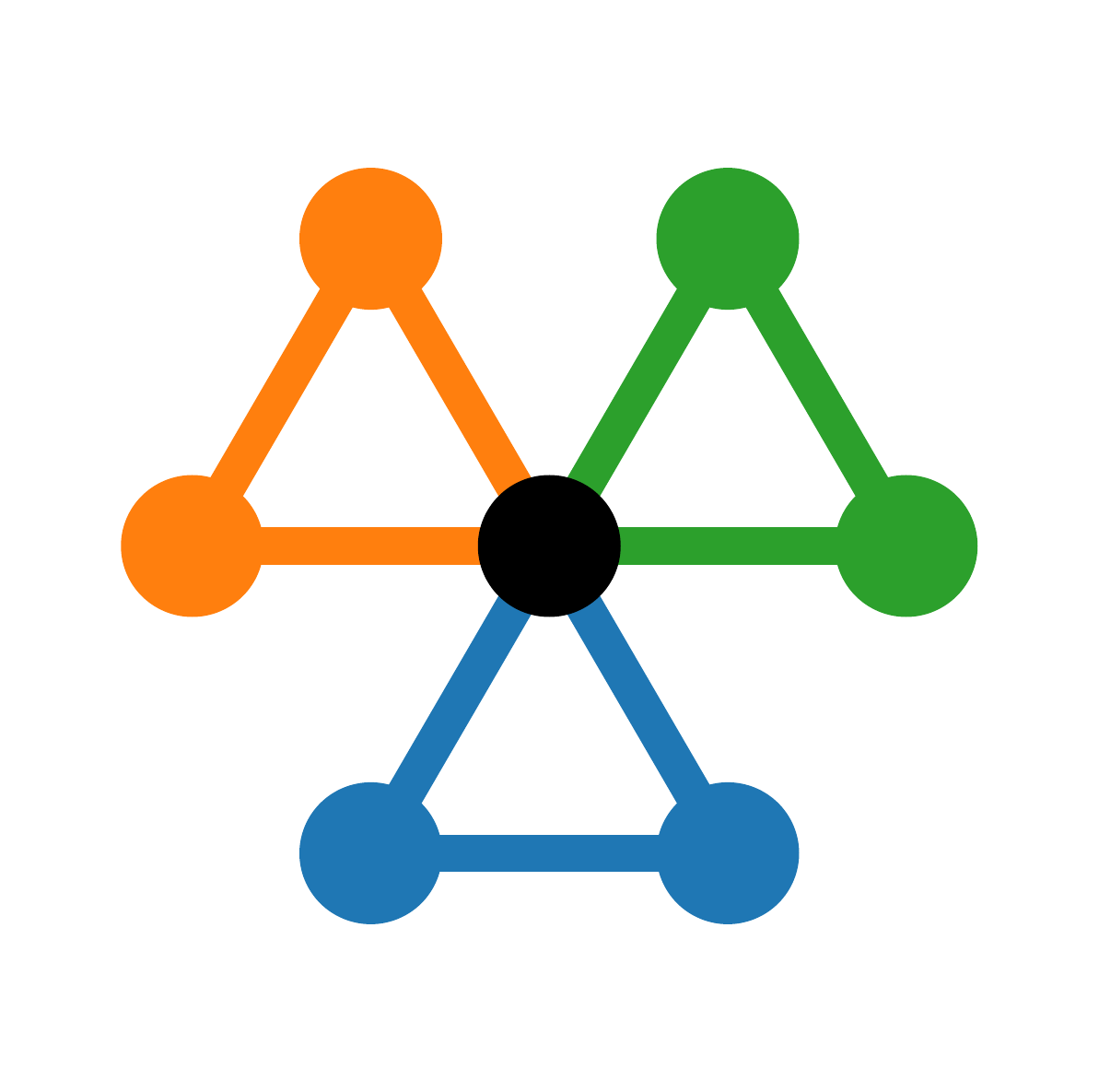}
    \caption{Three rigid clusters, identified by the colors of the bonds, pivoted
    by one node colored in black.}
    \label{fig:RBnet}
\end{figure}

To represent the Pivot network we defined $\mathcal{P}$ as an array of 
sets such that, if the entry $b$  of the array is a  non empty set, it contains the pivots of the rigid
cluster rooted by bond $b$ (see section VI). A somewhat more natural approach would be 
to directly store the roots of the rigid clusters to which $b$ is connected to in the Pivot network,
rather than its pivots. 
This is more natural  as one directly stores the neighbor-neighbor
pairs that exist in the pivot network.
This approach turns out to have two main
limitations: first and most importantly, it does not allow to easily reconstruct the paths needed
to detect wrapping (cf. section VII), as they pass through pivots. Second, it is more expensive to keep
in memory and to update. Consider Figure~\ref{fig:RBnet}.
Storing the network of rigid clusters in terms of neighbor-neighbor relationships 
results in the entry associated to 
the orange cluster to have two elements, i.e., $\rm{Net[orange]} = \{blue, green\}$ and
analogously for the entry of the blue and of the green clusters, for a total
of six values. The network $\mathcal{P}$ that we defined contains exactly half
of this data, as $\mathcal{P}[\rm{orange}] = p$, where $p$ is the black pivot node shared by all the three rigid clusters. The same holds for the entry
in $\mathcal{P}$ of the blue and of the green clusters. 
For these two reasons, we have implemented $\mathcal{P}$ in the way described in the main
text. 
The cost associated to this choice is that to reconstruct the neighbor-neighbor
relation, like orange-green and orange-blue, we have to iterate over the bonds of $p$ and
perform one find operation on each. The find operation, however, has constant cost
and we found it to be more efficient.


\section{Details of our C++ implementation}

The number of pebbles per node and the pivotal class of each node
are two arrays of length $N$ each. Note that, at a little efficiency cost, both these arrays
can be dropped. The number of pebbles of node $i$ is two minus the number of positive
entries in the two cells of its pebble graph array; the pivotal class of $i$ is the number of rigid clusters it belongs to, which can be computed by applying the find operation on its bonds
Our experiments reveal that the highest efficiency is reached by introducing
seven more arrays of length $N$. Two are needed to store nodes visited during pebble searches
of type II. Referring to our code at Ref.~\cite{GitProject}, these are the arrays
named \texttt{visited} and \texttt{visited\_indices} respectively.
Both are initialized with all negative entries and, when a node is visited, the
corresponding entry becomes positive.
Indices of the positive entries of \texttt{visited} are quickly recovered
by using \texttt{visited\_indices}: its first entry is assigned the
index of the first visited node and so on. Iterating over the non-negative values of \texttt{visited\_indices}
allows to quickly reinitialize both, avoiding the need of an 
$\mathcal{O}(N)$ iteration for each initialization.
The second pair of arrays is used in the same way, but stores the marks (rigid, floppy, unmarked)
of the nodes visited during repeated pebble searches of type II. The array \texttt{marks}, where the mark are stored,
is reinitialized to all nodes being unmarked by iterating over non-negative entries of \texttt{marks\_indices}.
The third pair works again in the same way and is used to keep track of which pivots are enqueued
in the queue of pivots and which are not. This allows to prevent inserting the same pivot more
than once. Referring to the code we make available at~\cite{GitProject}, these are the arrays
named \texttt{enqueued} and \texttt{enqueued\_indices} respectively.
The seventh array of length $N$ is used to store the paths that are created by pebble searches
and is named \texttt{searched\_bonds} in our code.
If node $i$ leads to node $j$, then the entry of node $j$ is set to $i$. In this way we can reconstruct
the path from a pebble (or from a floppy node) to the node that started the search and flip
the direction of the corresponding pebble edges (in case of searches of type I) 
or mark the corresponding nodes as  floppy (in case of searches of type II). 
Most of these statically allocated arrays could be replaced by
sets with dynamically handled memory to save some space, but we have observed the performance to degrade
with this approach, due to a significant amount of allocations and de-allocations.

The Breadth First Searches performed by the pebble searches (of both type)
are implemented using a stack that we allocate statically with $10^7$ entries, which turned
out to be sufficient even for $N \gg 10^7$. The queue of pivots is also allocated statically
to contain $10^6$ elements, which also turned out to be always sufficient.

\section{Performance of the code and time complexity of the algorithm}
\begin{figure}[h]
    \includegraphics[width=\textwidth]{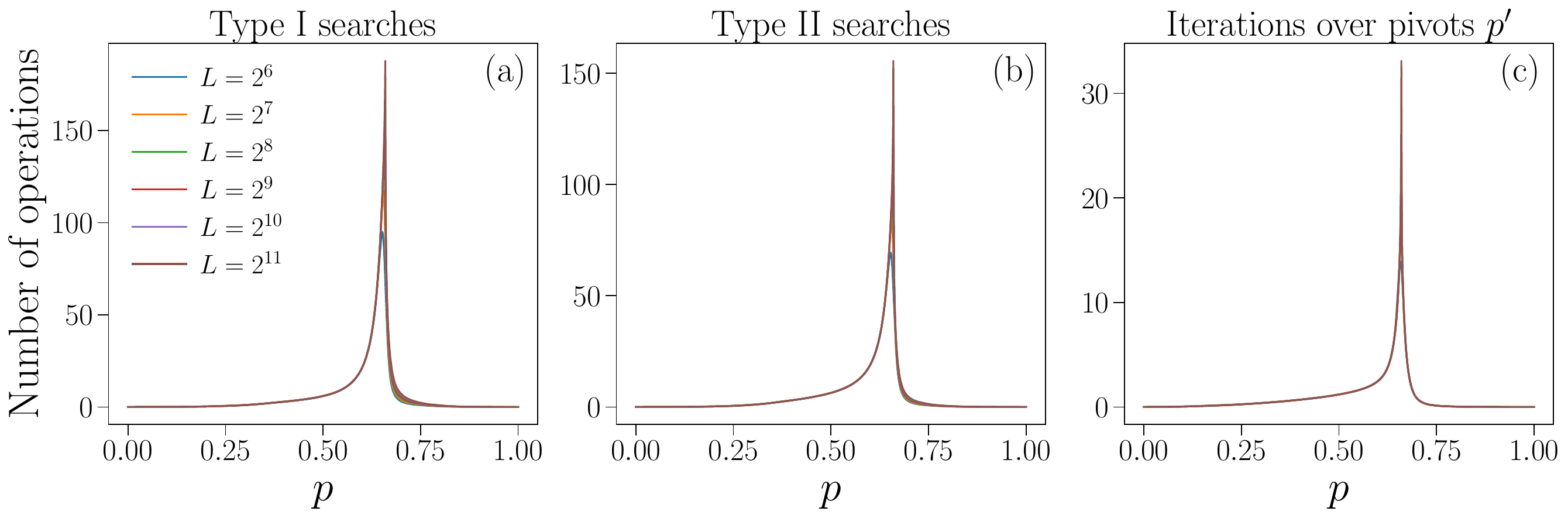}
    \caption{Average number of operations as a function of the bond concentration.
    (a) Number of nodes visited during type I searches.
    (b) Number of nodes visited during type II searches.
    (c) Number of pivots over which we iterate during step 7 of the algorithm.}
    \label{fig:comp_complex_curves}
\end{figure}

The time complexity of an algorithm is a theoretical quantity that, at least in principle,
can be computed analytically. For example, the Newman-Ziff algorithm for connectivity percolation
has been proven to scale linearly with the system size~\cite{Newman2000Efficient} (by showing that
path compression keeps the cost of find operations constant). In our work, we have not 
estimated analytically the time complexity of our algorithm. 
Rather, we have measured
the performance of a code that implements it. This implies that we have put an upper bound
on the complexity of the algorithm: it is clearly impossible to write a code that
performs better than the algorithmic complexity of the algorithm it implements. Hence, as discussed in the Results section, the computational complexity of our algorithm is at most 
$\mathcal{O}(N^{1.02})$. 
Furthermore, we have shown that measuring the number of operations performed in the most
expensive steps of the algorithm, the number of visited nodes during type I pebble searches 
 scales as $N^{1.02}$ as well,
suggesting that the performance observed is indeed a good estimator of the computational complexity
of the algorithm. Here we furthermore show that the relationship between the performance --as measured by 
the time--, and the computational complexity --as measured by the number of visited nodes
(in both types of pebble searches) and by the number of pivots $p'$ over which we iterate-- is well grounded.
Fig.~\ref{fig:comp_complex_curves} shows the average number of nodes visited in type I and II searches
and the number of pivots $p'$ over which we iterate during step 7 of the algorithm, as a function
of the bond concentration $p$. The curves, whose integral we have shown to scale with exponents
$1.02$, $1.01$ and $1.00$ respectively (see Fig.~5 of the main text) all behave like 
the time per bond function (Fig. 4 (c) in the main text), further strengthening the hypothesis that the performance (time)
can be understood by considering these quantities as proxies of the computational
complexity (number of operations). 

\begin{figure}[h]
    \includegraphics[scale=.6]{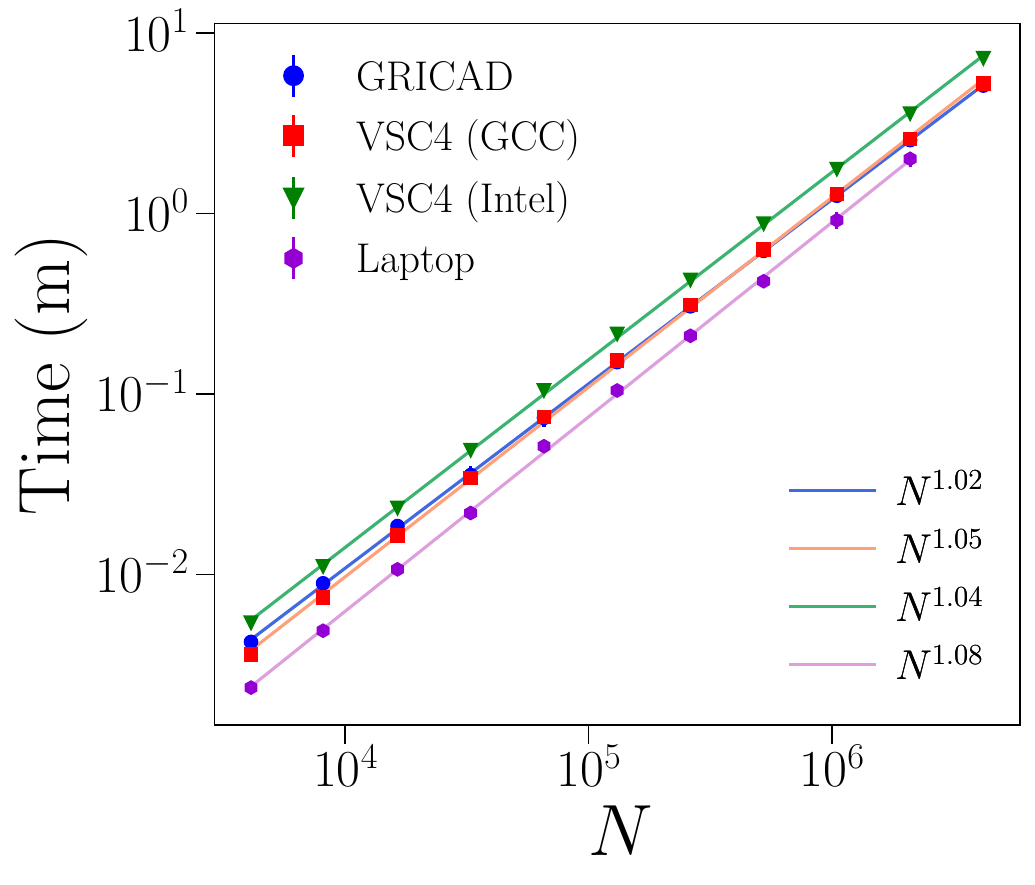}
    \caption{Performance of the code as measured on different machines and using different
    compilers. }
    \label{fig:reproducibility}
\end{figure}

Regarding the performance, we note that programs that are implemented using arrays only 
(such as NZ \cite{Newman2000Efficient} or JH \cite{Jacobs1997An})
have performances that are typically stable and reproducible across different machines. 
C++ sets, however, are expected to have performance that (weakly) depends on several 
factors: the implementation of the C++ standard library can vary between different compilers as, e.g., the 
strategy to keep the trees balanced might vary. 
Furthermore, the allocator that handles memory allocations of sets is also compiler dependent. 
Finally, sets are also expected to perform in a 
hardware dependent way as how the containers access to the cache usually depends on the machine.
Hence, it is not surprising that we did observe the scaling of the performance 
of our code to weakly depend on the compiler and on the machine used. 
The results shown in the main text have been obtained running the code available at~\cite{GitProject} on the 
Grenoble Alpes Research Center for Advanced Computing
(GRICAD), which uses Intel(R) Xeon(R) Gold 5218 CPUs, compiling it with the GNU Compiler Collection (GCC)
version 10.2.1. In Fig.~\ref{fig:reproducibility}
we compare the results shown in the main text with three different datasets. 
Two datasets have are obtained on the Vienna Scientific Computing 4 (VSC 4), 
which uses Intel(R) Xeon(R) Platinum 8174 CPUs, compiling the code with Intel C++ Compiler (ICC) 
version 19.1.3 for one dataset and with GCC version 10.2.0 for the other.
Lastly, a dataset has been obtained using one of our laptops, which uses
11th Gen Intel(R) Core(TM) i7, compiling with GCC version 11.4.0.

We observe the performance on VSC 4 to scale with 
estimated exponents that are only slightly larger than the one obtained on GRICAD
(the results shown in the main paper). Finally, the scaling observed with the laptop is
sufficiently higher for the difference to be appreciable by the naked eye.
Note that no optimization flags have been used to assess the performance of our code.
The data shown has been obtained by launching one series of $T$ trials for each size, 
with $T=10^5$ for the smallest sizes and $T\sim600$ for the largest one.
The following table summarizes the results obtained.

\begin{table}[h]
\centering
\begin{tabular}{l|c|c|c} 
\hline
\textbf{Machine} & \textbf{CPU}  & \textbf{Compiler} & \textbf{Scaling exponent} \\
\hline
GRICAD & Intel(R) Xeon(R) Gold 5218 & GNU Compiler Collection 10.2.1 & 1.0210(26)  \\
VSC 4 & Intel(R) Xeon(R) Platinum 8174 & Intel C++ Compiler 19.1.3 & 1.051(6) \\
VSC 4 & Intel(R) Xeon(R) Platinum 8174 & GNU Compiler Collection 10.2.0 & 1.0392(43)  \\
Laptop & 11th Gen Intel(R) Core(TM) i7 & GNU Compiler Collection 11.4.0 & 1.077(6)\\
\hline
\end{tabular}
\caption{Performance of our C++ code on different machines and on different compilers. Each row corresponds
to a performance measure. From left to right we show the machine name, the CPU of the machine, the
compiler used and the performance measured.}
\label{tab:performance}
\end{table}

\section{Estimation of the critical exponents and critical threshold}

In this section we provide more details about the measurements and scaling analysis to determine the critical exponents. Table \ref{tab:measurements} summarizes the measurement
we made: we started from $L=2^6$ and multiplied it by a factor $\sqrt{2}$ until we reached
$L=2^{14.5}$. For each value of $L$ and hence of $M$ we chose the number of realizations $\mathcal{N}$ such that the total number of activated bonds, $\mathcal{N}\times M$, is of order $10^{12}$.

\begin{table}[h]
\centering
\begin{tabular}{r | r | r | r | r}
\hline
$L$ & $N$ & $M$ & $\mathcal{N}$ & total number of activated bonds \\
\hline
64    & 4 096       & 12 288        & 96 000 000 & 1 179 648 000 000 \\
90    & 8 100       & 24 300        & 42 000 000 & 1 020 600 000 000 \\
128   & 16 384      & 49 152        & 24 000 000 & 1 179 648 000 000 \\
181   & 32 761      & 98 283        & 12 000 000 & 1 179 396 000 000 \\
256   & 65 536      & 196 608       & 5 760 000  & 1 132 462 080 000 \\
362   & 131 044     & 393 132       & 1 800 970  & 708 018 938 040   \\
512   & 262 144     & 786 432       & 1 440 000  & 1 132 462 080 000 \\
724   & 524 176     & 1 572 528     & 720 000    & 1 132 220 160 000 \\
1024  & 1 048 576   & 3 145 728     & 336 000    & 1 056 964 608 000 \\
1448  & 2 096 704   & 6 290 112     & 161 000    & 1 012 708 032 000 \\
2048  & 4 194 304   & 12 582 912    & 100 000    & 1 258 291 200 000 \\
2896  & 8 386 816   & 25 160 448    & 45 000     & 1 132 220 160 000 \\
4096  & 16 777 216  & 50 331 648    & 25 000     & 1 258 291 200 000 \\
5792  & 33 547 264  & 100 641 792   & 8 728      & 878 401 560 576   \\
8192  & 67 108 864  & 201 326 592   & 5 000      & 1 006 632 960 000 \\
11585 & 134 212 225 & 402 636 675   & 5 250      & 2 113 842 543 750 \\
16384 & 268 435 456 & 805 306 368   & 1 420      & 1 143 535 042 560 \\
23170 & 536 848 900 & 1 610 546 700 & 498        & 802 052 256 600 
\end{tabular}
\label{tab:measurements}
\caption{Simulation settings. From left to right we report
the lattice linear size $L$, the number of nodes $N=L^2$, the number of
activated bonds $M=3N$, the number $\mathcal{N}$ of simulations performed, which results in a
nearly constant total number of activated bonds, $\mathcal{N}\times M\sim 10^{12}$.}
\end{table}

\subsection{Computation of the transition widths $\Delta(L)$ and critical thresholds $p_c(L)$}
The wrapping probabilities $R_x$ and $R_{xy}$ are measured as functions of the number of active bonds $m$ (see main text), and convoluted with the binomial distribution
to obtain them as functions of the bond concentration $p$~\cite{NZ2001}. The wrapping probabilities $R_e(p)$ and $R_1(p)$ are computed from $R_x(p)$
and $R_{xy}(p)$ as 
\begin{align}
    R_e &= 2R_x-R_{xy}\\
    R_1 &= R_x - R_{xy}.
\end{align}

\begin{figure}\label{fig:SM_wrapping_probs}
\centering
\includegraphics[width=\linewidth]{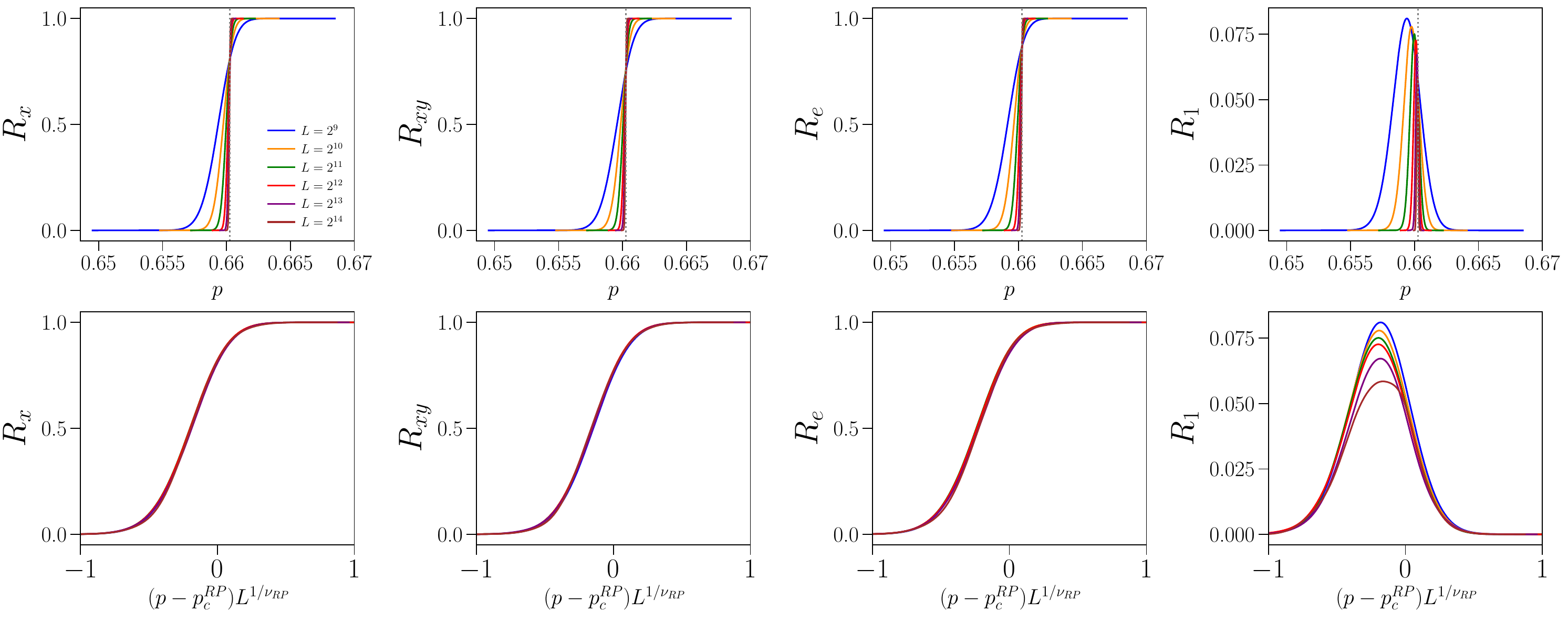}
\caption{ Upper row: The four wrapping probabilities as a function of the bond concentration $p$. The critical point at $p_c^{\rm RP}\approx 0.66027 $ is indicated as a dashed line. Bottom row: Collapse of the wrapping probabilities when plotted as functions of $(p-p_c^{\rm RP})\,L^{1/\nu^{\rm RP}}$, using the values of $p_c^{\rm RP}$ and $\nu^{\rm RP}$ given in Table 1 of the main text.}
\end{figure}

From the four wrapping probabilities, shown in Fig.~\ref{fig:SM_wrapping_probs},  we compute four transitions widths $\Delta(L)$, and four  thresholds functions $p_c(L)$, as
\begin{align}
    &p_c(L) = \sum_i p_i\; \text{Prob}[p_i] = \langle p \rangle\\
    &\Delta(L) = \sqrt{\sum_i \Big(p_i-p_c(L)\Big)^2\;\text{Prob}[p_i]} = \sqrt{\langle p^2 \rangle - \langle p \rangle^2}
\end{align}
For each wrapping type, the probability distribution Prob$[p_i]$ is obtained from the derivative
of the corresponding wrapping probability, except for $R_1$ where Prob$[p_i]$ is proportional to $R_1$ itself.

\subsection*{Supplementary References}

\bibliography{biblio}